\documentclass{aa}  

\usepackage{graphicx}
\usepackage{txfonts}
\usepackage{placeins}
\usepackage{hyperref}
\hypersetup{colorlinks=true,citecolor=blue,linkcolor=black, urlcolor=black, linktoc=all}
\newcommand{\dcop}[1]{DCO$^+$} 
\newcommand{\ccd}[1]{CCD}  
\newcommand{\docp}[1]{DOC$^+$}
\newcommand{\xcch}[1]{$^{13}$CCH} 
\newcommand{\hcxop}[1]{HC$^{18}$O$^+$} 
\newcommand{\cxch}[1]{C$^{13}$CH} 
\newcommand{\hoxcp}[1]{HO$^{13}$C$^+$} 
\newcommand{\hxcn}[1]{H$^{13}$CN}      
\newcommand{\hxcop}[1]{H$^{13}$CO$^+$} 
\newcommand{\hnxc}[1]{HN$^{13}$C} 
\newcommand{\hcop}[1]{HCO$^+$} 
\newcommand{\hocp}[1]{HOC$^+$} 

\begin{document}

   \title{Similar levels of deuteration in the pre-stellar core L1544 and the protostellar core HH211}

   \author{K. Giers
          \inst{1},
          S. Spezzano\inst{1}, 
          P. Caselli\inst{1},
          E. Wirström\inst{2},
          O. Sipilä\inst{1},
          J. E. Pineda\inst{1},
          E. Redaelli\inst{1},
          C. T. Bop\inst{3},
          F. Lique\inst{3}
          }

   \institute{Max-Planck-Institute for Extraterrestrial Physics, Giessenbachstrasse 1, D-85748 Garching, Germany\\
              \email{kgiers@mpe.mpg.de}
         \and 
         Department of Space, Earth, and Environment, Chalmers University of Technology, Onsala Space Observatory, 439 92 Onsala, Sweden
         \and 
         Univ. Rennes, CNRS, IPR (Institut de Physique de Rennes) – UMR 6251, 35000 Rennes, France
             }



 
  \abstract
   { In the centre of pre-stellar cores, deuterium fractionation is enhanced due to low temperatures and high densities. 
   Therefore, the chemistry of deuterated molecules can be used to probe the evolution and the kinematics in the earliest stages of star formation. }
   {We analyse the deuterium fractionation of simple molecules, comparing the level of deuteration in the envelopes of the prototypical pre-stellar core L1544 in Taurus and the young protostellar core HH211 in Perseus.} 
   {We used single-dish observations of CCH, HCN, HNC, and \hcop, and their $^{13}$C-, $^{18}$O,- and D-bearing isotopologues, detected with the 20\,m telescope at the Onsala Space Observatory. We derived the column densities, and subsequently the carbon isotopic ratios and deuterium fractions of the molecules. Additionally, we used radiative transfer simulations and results from chemical modelling to reproduce the observed molecular lines. We used new collisional rate coefficients for HNC, \hnxc\,, DNC, and DCN that consider the hyperfine structure of these molecules.}
   {
   For CCH, we find high levels of deuteration (10\%) in both sources, consistent with other carbon chains. 
   We find moderate deuteration of HCN (5-7\%), with a slight enhancement towards the protostellar core. Equal levels of deuteration for HNC towards both cores ($\sim$8\%) indicate that HNC is tracing slightly different layers compared to HCN.
   We find that the deuterium fraction of \hcop\, is enhanced towards HH211, most likely caused by isotope-selective photodissociation of C$^{18}$O.
   With radiative transfer, we were able to reproduce the observed lines of CCH, HCN, \hxcn\,, HNC, \hnxc\,, and DNC towards L1544 as well as CCH, \hxcn\,, \hnxc\,, DNC, \hxcop\,, \hcxop\,, and \dcop\, towards HH211.  
   }
   { 
   Similar levels of deuteration show that the deuterium fractionation is most probably equally efficient towards both cores, suggesting that the protostellar envelope still retains the chemical composition of the original pre-stellar core. 
   The fact that the two cores are embedded in different molecular clouds also suggests that environmental conditions do not have a significant effect on the deuterium fractionation within dense cores.
   Our results highlight the uncertainties when dealing with $^{13}$C isotopologues and the influence of the applied carbon isotopic ratio.
   Radiative transfer modelling shows that it is crucial to include the effects of the hyperfine structure to reproduce the observed line shapes. In addition, to correctly model emission lines from pre-stellar cores, it is necessary to include the outer layers of the core to consider the effects of extended structures. In addition to \hcop\, observations, HCN observations towards L1544 also require the presence of an outer diffuse layer where the molecules are relatively abundant.}

               
\titlerunning{Similar levels of deuteration in L1544 and HH211}
\authorrunning{Giers et al.}

   \maketitle

\section{Introduction}
Deuterated molecules are important diagnostic tools of the earliest phases of star formation, allowing astronomers to study the central regions of pre-stellar cores in detail \citep[e.g.][]{Caselli2002b,Pineda2022arXiv}, where CO and other CO-bearing molecules are heavily frozen onto dust grains \citep[e.g.][]{Caselli1999,BerginTafalla2007}. High levels of deuteration are also found in the cold envelopes of newly formed protostars \citep{VanDishoeck1995,Hatchell1998,Roberts2002,Parise2004,Emprechtinger2009}, bringing up the question of how much of this material can survive during the star and planet formation process.
High levels of deuteration have been found in planet-forming disks \citep[e.g.][]{Mathews2013}, comets \citep[e.g.][]{Altwegg2015}, and carbonaceous chondrites \citep[e.g.][]{Robert2003,Busemann2006,Ceccarelli2014}. Our oceans are also enriched in heavy water, partially inherited from the solar pre-stellar phase \citep{Cleeves2014}. 
It is therefore important to understand the chemical processes that regulate deuterium fractionation in the early phases of star formation in detail and investigate possible differences between various evolutionary stages. In particular, observational constraints are needed for the chemical models to discriminate among the processes of deuteration happening in the gas phase and on the surface of dust grains. 

At low temperatures, deuterium fractionation is driven by the gas phase reaction
\begin{equation}
    \mathrm{H}_3^+ + \mathrm{HD} \rightleftharpoons \mathrm{H}_2\mathrm{D}^+ + \mathrm{H}_2 + 232\,\mathrm{K}.
\end{equation}
Due to its exothermicity, the reaction proceeds more efficiently from left to right with decreasing temperature \citep[assuming a low ortho-to-para H$_2$ ratio; e.g. ][]{Pagani1992}. 
This leads to an enhancement of the abundance of $\rm H_2D^+$ and an associated increase in the abundances of $\rm D_2H^+$ and $\rm D_3^+$, and subsequently to efficient deuteration of other molecules.
In the cold and dense conditions of the earliest stages of star formation, the main destructor of $\rm H_2D^+$, CO, is highly depleted from the gas phase and frozen out onto the surfaces of dust grains \citep[e.g.][]{Caselli1999}, and this further increases the deuterium fractionation \citep{DalgarnoLepp1984}.
The deuterium fraction of a molecular tracer is measured by dividing the column density of the deuterated molecule by the column density of the hydrogenated molecule.

In this paper, we focus on deuterated molecules present in the well-known pre-stellar core L1544 in Taurus, and in one of the youngest (and most highly deuterated) Class 0 sources, HH211 in Perseus.
L1544 is located at a distance of 170\,pc \citep{Galli2019}, with clear evidence of gravitational contraction \citep{Caselli2012}. Its central temperature approaches 6\,K \citep{Crapsi2007} and its central density reaches $10^7$\,cm$^{-3}$ \citep{Keto2010,Caselli2019}. Towards its centre, L1544 exhibits a high level of deuteration \citep{Crapsi2005,Redaelli2019}.
Within the central 2000\,au, recent Atacama Large Millimeter/submillimeter Array observations have been found consistent with almost complete (99.999\%) depletions of elements heavier than He \citep{Caselli2022}.
L1544 is on the verge of star formation, thus shedding light on the initial conditions in the process of star formation. 
HH211 is a newly born Class 0 protostar \citep[e.g.][]{Enoch2006,LeeLi2018} at a distance of 321\,pc \citep{OrtizLeon2018}, which presents high degrees of deuterium fractionation \citep{Emprechtinger2009}. 
It hosts a jet-driven molecular outflow \citep{McCaughrean1994,GuethGuilloteau1999}, and has a Keplerian disk \citep{SeguraCox2016}.
The surrounding envelope is elongated in a direction roughly perpendicular to the jet and outflow axis \citep{GuethGuilloteau1999}.
In \cite{Chantzos2018}, it is shown that the deuteration of cyclopropenylidene, c-C$_3$H$_2$, is more efficient towards HH211 (N(c-C$_3$HD)/N(c-C$_3$H$_2$)=20\%) than in L1544 (N(c-C$_3$HD)/N(c-C$_3$H$_2$)=10\%). 
Gas phase deuteration processes are sufficient to reproduce the N(c-C$_3$HD)/N(c-C$_3$H$_2$) ratio observed in L1544, so the enhancement of deuteration in HH211 is assumed to be due to efficient deuteration happening on the surface of dust grains in the pre-stellar phase, followed by the release from the grains in the protostellar stage.

We present a survey of ground state-rotational lines of deuterated molecules towards L1544 and HH211, observed with the single-dish 20\,m radio telescope at the Onsala Space Observatory. 
Our observations cover CCH, HCN, HNC, and HCO$^+$, and their deuterated isotopologues. 
To avoid optical depth limitations with the main isotopologues, we also observed \xcch\,, 
\cxch\,, \hxcn\,, \hnxc\,, \hxcop\,, and \hcxop\,. 
This data set allows for a unique comparison between the deuteration in a very dynamically evolved pre-stellar core on the verge of forming a low-mass star, and a young Class 0 protostellar core, and thus, to study the influence of the evolutionary stage on the deuterium fractionation. Such a comparison is important to understand how the deuterated molecules are inherited in the earlier stages of formation of a low-mass star, and eventually in its planetary system.

In Sect.~\ref{observations}, we describe the observations, followed by the results in Sect.~\ref{results}. 
The analysis in Sect.~\ref{analysis} covers the derivation of the column densities and deuterium fraction assuming local thermal equilibrium (LTE) conditions.
In Sect.~\ref{modeling}, we use the non-LTE radiative transfer code LOC to model our observations. 
We discuss the results of the LTE and non-LTE analysis in Sect.~\ref{discussion} and present our conclusions in Sect.~\ref{conclusion}.

\section{Observations}\label{observations}

The data presented in this paper were obtained with the 20\,m radio telescope at the Onsala Space Observatory. 
The observed spectra are centred on the dust emission peaks of the sources (L1544: $\alpha_{2000}=05^\mathrm{h}04^\mathrm{m}17^\mathrm{s}.21$, $\delta_{2000}=+25^\circ10'42''.8$; 
HH211: $\alpha_{2000}=03^\mathrm{h}43^\mathrm{m}56^\mathrm{s}.80$, $\delta_{2000}=+32^\circ00'50''.0$). 
The first part of the observations was done in May 2020 (project ID: O2019b-04). A follow-up project observed the second part from January to May 2021 (project ID: O2020b-05).
For the observations, we used the 3\,mm \citep{OSO3mm} and the 4\,mm receiver \citep{OSO4mm} in combination with the OSA spectrometer at 19\,kHz resolution, with 
625\,MHz bandwidth and dual polarisation.
We applied the dual beam-switched mode, with the off position shifted by 11.8\,arcmin in the azimuth direction. The intensity calibration was done using standard Dicke-switched type calibration, switching between a hot load and a sky direction close to the source. The $1\sigma$ pointing accuracy is estimated to be 3\,arcsec.
The frequency resolution of the data corresponds to a velocity resolution of about 0.07\,km\,s$^{-1}$ at 84\,GHz. 
The weather conditions during the observations were variable, with the system temperature ranging between $120-300$\,K.

\begin{figure*}
   \centering
   \includegraphics[width=.99\textwidth]{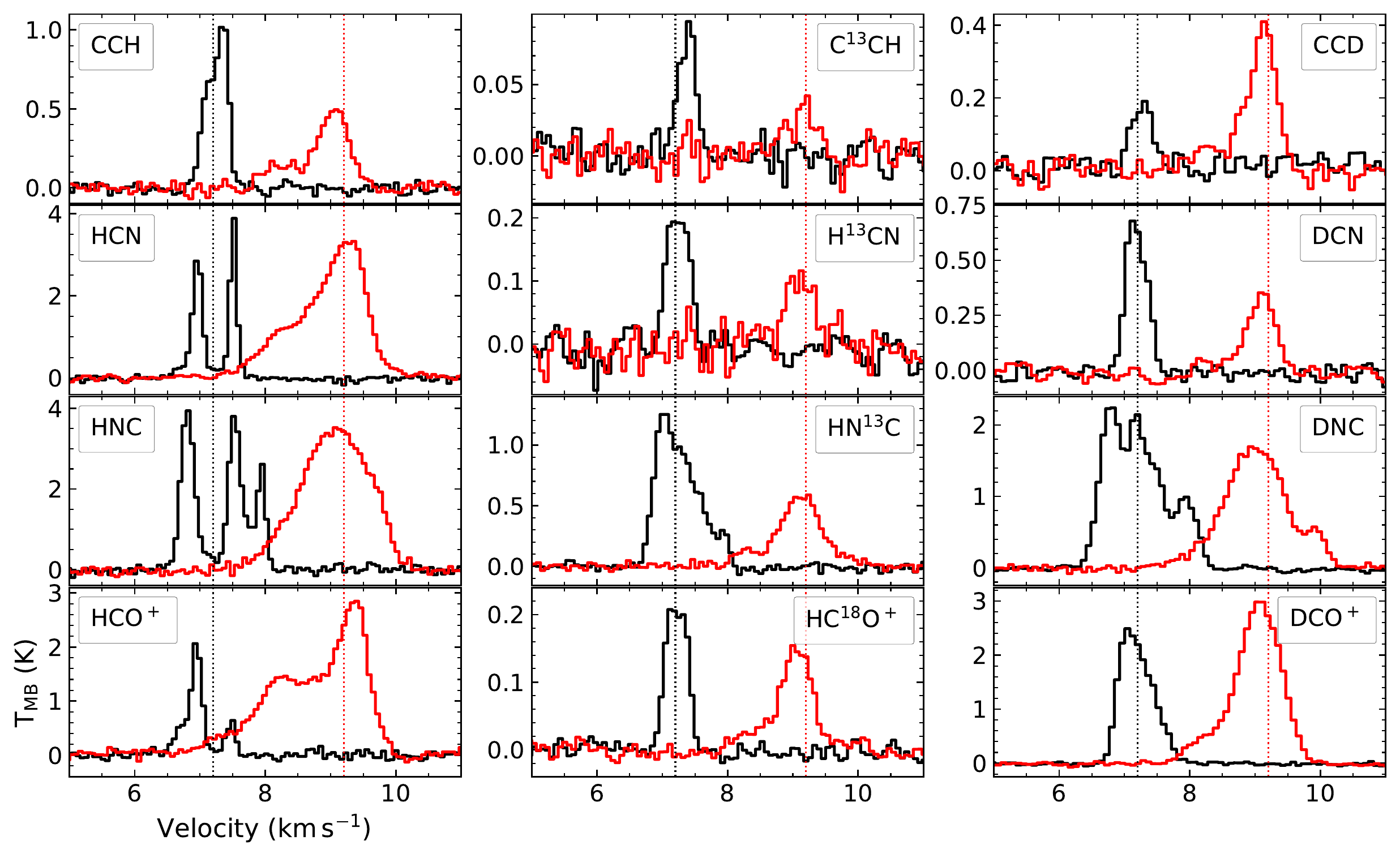}
   \caption{Selection of the spectra observed towards the pre-stellar core L1544 (black) and the protostellar core HH211 (red). The dotted vertical lines indicate the respective rest velocities of the cores. For HCN, the $F=2-1$ transition is shown. For CCH, \cxch\,, \ccd\,, \hxcn\,, and DCN, the transitions listed in Table~\ref{tab:observedlines} are shown.}
              \label{Fig:SelectedSpectra}
\end{figure*}

The observed transitions are summarised in Table~\ref{Tab:AllObervedLines}. The corresponding spectra towards L1544 and HH211 are presented in Fig.~\ref{Fig:L1544all} and Fig.~\ref{Fig:HH211all}, respectively. A selection of the observed transitions is given in Table~\ref{tab:observedlines} and Fig.~\ref{Fig:SelectedSpectra}.
The average beam size of the observations is about 50\,arcsec, which corresponds to a physical size of 8500\,au for L1544 and 16000\,au for HH211.
 
The data processing was done using the GILDAS software \citep{Pety2005} and the Python package \textsc{pyspeckit} \citep{pyspeckit,pyspeckit2}. 
The antenna temperature $T_A^*$ was converted to the main beam temperature $T_\mathrm{mb}$ using the relation $T_\mathrm{mb}=T_A^*/B_\mathrm{eff} $. The corresponding values for the main beam efficiencies ($B_\mathrm{eff}$) of the 20\,m telescope are given in Table~\ref{tab:observedlines}.

\begin{table*}[h]
    \centering
    \caption{Properties of the selected observed lines. Numbers in brackets give the uncertainty on the last digit.}
    \scalebox{0.9}{
    \begin{tabular}{l l l c l r c l l}
    \hline\hline 
    \noalign{\smallskip}
     Species & Transition & Frequency\tablefootmark{a} & $B_\mathrm{eff}$ & $T_\mathrm{mb,peak}$ & rms & $V_\mathrm{LSR}$\tablefootmark{b} & $\Delta v$\tablefootmark{b} & $\int T_\mathrm{mb}\mathrm{d}v$ \\
      &  & (MHz) &  & (K) & (mK) & (km\,s$^{-1}$) & (km\,s$^{-1}$) & (K\,km\,s$^{-1}$) \\
    \noalign{\smallskip}
    \hline
    \noalign{\smallskip}
    \multicolumn{9}{c}{\textbf{L1544}} \\
    CCH & $N$=1-0, $J$=1/2-1/2, $F$=1-0  & 87446.47(1)   & 0.52 & 0.96(3) & 27 & 7.24 & 0.39 & 0.40(2) \\
    \xcch\,  & $N$=1-0, $J$=3/2-1/2, $F_1$=2-1, $F$=5/2-3/2 & 84119.33(2)  & 0.53 & 0.039(5) & 7 & 7.31 & 0.32 & 0.013(2)   \\
    \cxch\,  & $N$=1-0, $J$=3/2-1/2, $F_1$=2-1, $F$=5/2-3/2 & 85229.335(4)   & 0.53 & 0.089(6) & 9 & 7.35 & 0.32 & 0.030(3) \\
    \ccd\,  & $N$=1-0, $J$=1-0, $F_1$=3/2-1/2, $F$=3/2-3/2  & 72101.811(5)  & 0.54 & 0.19(1) & 24 & 7.23 & 0.39 & 0.078(9)  \\
    \hxcn\,  & $J$=1-0, $F$=0-1 & 86342.2543(3) & 0.53 & 0.22(1) & 18 & 7.21 & 0.39 & 0.091(7) \\
    DCN  & $J$=1-0, $F$=0-1 & 72417.03(1) & 0.54 & 0.69(2) & 23 & 7.14 & 0.37 & 0.27(1) \\
    \hxcop\, & $J$=1-0 & 86754.288(5)   & 0.53 & 1.38(4) & 20 & 7.13 & 0.52 & 0.77(4) \\
    \hcxop\, & $J$=1-0 & 85162.223(5)   & 0.53 & 0.229(6) & 9 & 7.18 & 0.38 & 0.093(4)  \\
    \dcop\, & $J$=1-0 & 72039.3124(8)  & 0.54 & 2.44(5) & 25 & 7.12 & 0.60 & 1.55(4)  \\
    \hnxc\, & $J$=1-0, $F_1$=0-1, $F_2$=1-2, $F$=1-2 & 87090.675(3) & 0.52 & 0.27(3) & 25 & 7.92 & 0.24 & 0.07(2)\\
    DNC   & $J$,$I$,$F$=1,2,1-0,2,2; 1,2,1-0,0,0 & 76305.513(1) & 0.54 & 0.99(2) & 26 & 7.90 & 0.40 & 0.42(2) \\
    \noalign{\smallskip}
    \hline 
    \noalign{\smallskip}
    \multicolumn{9}{c}{\textbf{HH211}} \\
    CCH  & $N$=1-0, $J$=3/2-1/2, $F$=1-1  & 87284.11(1)   & 0.52 & 0.50(1) & 27 & 9.02 & 0.54 & 0.29(1) \\
    \xcch\,  & $N$=1-0, $J$=3/2-1/2, $F_1$=1-0, $F$=3/2-1/2 & 84153.31(2)  & 0.53 & 0.016(4) & 6 & 9.07 & 0.4 & 0.007(2) \\
    \cxch\,  & $N$=1-0, $J$=3/2-1/2, $F_1$=2-1, $F$=5/2-3/2  & 85229.335(4)   & 0.53 & 0.032(4) & 8 & 9.10 & 0.53 & 0.018(4) \\
    \ccd\,  & $N$=1-0, $J$=1-0, $F_1$=3/2-1/2, $F$=5/2-3/2 & 72107.721(3)   & 0.54 & 0.37(5) & 28 & 9.10 & 0.45 & 0.18(3)\\
    \hxcn\,  & $J$=1-0, $F$=2-1 & 86340.1666(1)  & 0.53 & 0.45(2) & 23 & 9.05 & 0.61 & 0.29(2)\\
    DCN  & $J$=1-0, $F$=0-1 & 72417.03(1)  & 0.54 & 0.35(2) & 22 & 9.06 & 0.50 & 0.18(4)\\
    \hxcop\, & $J$=1-0 & 86754.288(5)  & 0.53 & 1.95(1) & 24 & 8.97 & 0.60 & 1.24(2)\\
    \hcxop\, & $J$=1-0 & 85162.223(5)  & 0.53 & 0.153(4) & 8 & 9.03 & 0.53 & 0.087(5)\\
    \dcop\, & $J$=1-0 & 72039.3124(8)  & 0.54 & 3.01(1) & 22 & 9.04 & 0.74 & 2.38(2)\\
    \hnxc\, & $J$=1-0 & 87090.825(4) & 0.52 & 0.58(1) & 27 & 9.09 & 0.71 & 0.46(1)\\
    \hnxc\, & $J$=1-0, $F_1$=0-1 $F_2$=1-2, $F$=1-2 & 87090.675(3) & 0.52 & 0.090(5) & 27 & 9.78 & 0.35 & 0.034(7)\\
    DNC   & $J$,$I$,$F$=1,2,1-0,2,2; 1,2,1-0,0,0 & 76305.513(1) & 0.54 & 0.56(3) & 31 & 9.90 & 0.38 & 0.22(4) \\
    \noalign{\smallskip}
    \hline 
    \noalign{\smallskip}
    \end{tabular}}
    \label{tab:observedlines}
    \tablefoot{
    \tablefoottext{a}{Extracted from Cologne Database for Molecular Spectroscopy \citep{Mueller2001}.}
    \tablefoottext{b}{The uncertainties of the fits are smaller than the velocity resolution. Therefore, the error on $V_\mathrm{LSR}$ and $\Delta$v are given by the observed spectral resolution, $0.07$\,km\,s$^{-1}$. }
    }
\end{table*}

%
   
\section{Results}\label{results}

Many species observed in this survey show either resolved (CCH, \xcch\,, \cxch\,, \ccd\,, HCN, \hxcn\,, DCN), or unresolved (HNC, \hnxc\,, DNC, \hxcop\,, \dcop\,) hyperfine structure. 
Due to the amount of hyperfine lines in the resolved case, we decided to select one hyperfine component per species for the LTE analysis, based on optical depth and signal-to-noise ratio (S/N).
The properties of the selected lines are given in Table~\ref{tab:observedlines}, determined by fitting a Gaussian profile to the lines (see Fig.~\ref{Fig:L1544all} and Fig.~\ref{Fig:HH211all}). The corresponding spectra are shown in Fig.~\ref{Fig:SelectedSpectra}, where lines observed towards L1544 and HH211 are plotted in black and red, respectively.
In the following, we describe the observed spectra and discuss the line selection.

\subsection{L1544}

The radical CCH and its isotopologues present lines with multiple hyperfine (hf) components. In L1544, we detected all six hf components of CCH, where four out of six likely show a dip due to self-absorption. 
This is caused by absorption in the outer layers of the core and dilutes the signal from the high-density central regions. 
We selected the line with the lowest optical depth ($N$=1-0, $J$=1/2-1/2, $F$=1-0), to minimise any issues with optical depth effects. 
In L1544, we successfully observed the components of the $^{13}$C isotopologues of CCH: 
For \xcch\,, we detected three of the eight hyperfine components with a signal-to-noise ratio $>3$; \cxch\, shows five of the seven components with S/N\,$>4$. For both, we selected the transition with the highest S/N, ($N$=1-0, $J$=3/2-1/2, $F_1$=2-1, $F$=5/2-3/2).
The deuterated isotopologue, \ccd\,, is much brighter, and seven of the nine hyperfine components are detected (S/N\,$>8$). Some of them show self-absorption, thus, we selected the line with the lowest S/N ($\approx$8), $N$=1-0, $J$=3/2-1/2, $F$=5/2-3/2, to avoid optical depth issues. 
The Gaussian fit parameters of the \xcch\, and \cxch\, hf components show that the lines are shifted by approximately 0.1\,km\,s$^{-1}$ ($V_\mathrm{LSR}=$7.3\,km\,s$^{-1}$) compared to CCH and \ccd\, (7.2-7.25\,km\,s$^{-1}$), which are located at the typical system velocity observed for L1544 (7.2\,km\,s$^{-1}$). However, this shift is close to the spectral resolution of the data (0.07\,km\,s$^{-1}$) and might therefore not be significant.

HCN is optically thick, showing a double-peaked line profile in each hyperfine component with a dip reaching down to almost zero level, caused by self-absorption in the outer layers of the core. 
Due to its high optical depth, the line profile provides information on the kinematics in the outer layers of the core. 
A blue asymmetry in a self-absorbed peak is a typical signature of infall motions - extended inward velocities in the outer regions of a core cause enhanced self-absorption at redshifted velocities \citep[e.g.][]{Evans1999}.
The intensity ratio between the two peaks can give information on the infall rates itself \citep[e.g.][]{Keto2015}, because with increasing infall motions, the redder peak decreases in intensity and evolves to a shoulder of the blue peak (like it is seen with \hcop\, in Fig.~\ref{Fig:SelectedSpectra} and for example in \citealt{Redaelli2022}).
However, the peak of the central HCN hf component shows a red asymmetry. Following the argumentation above, this is caused by an expansion motion of the core in its outer regions, more specifically, in the layers traced by the HCN transition.

\begin{figure}
   \centering
   \includegraphics[width=\hsize]{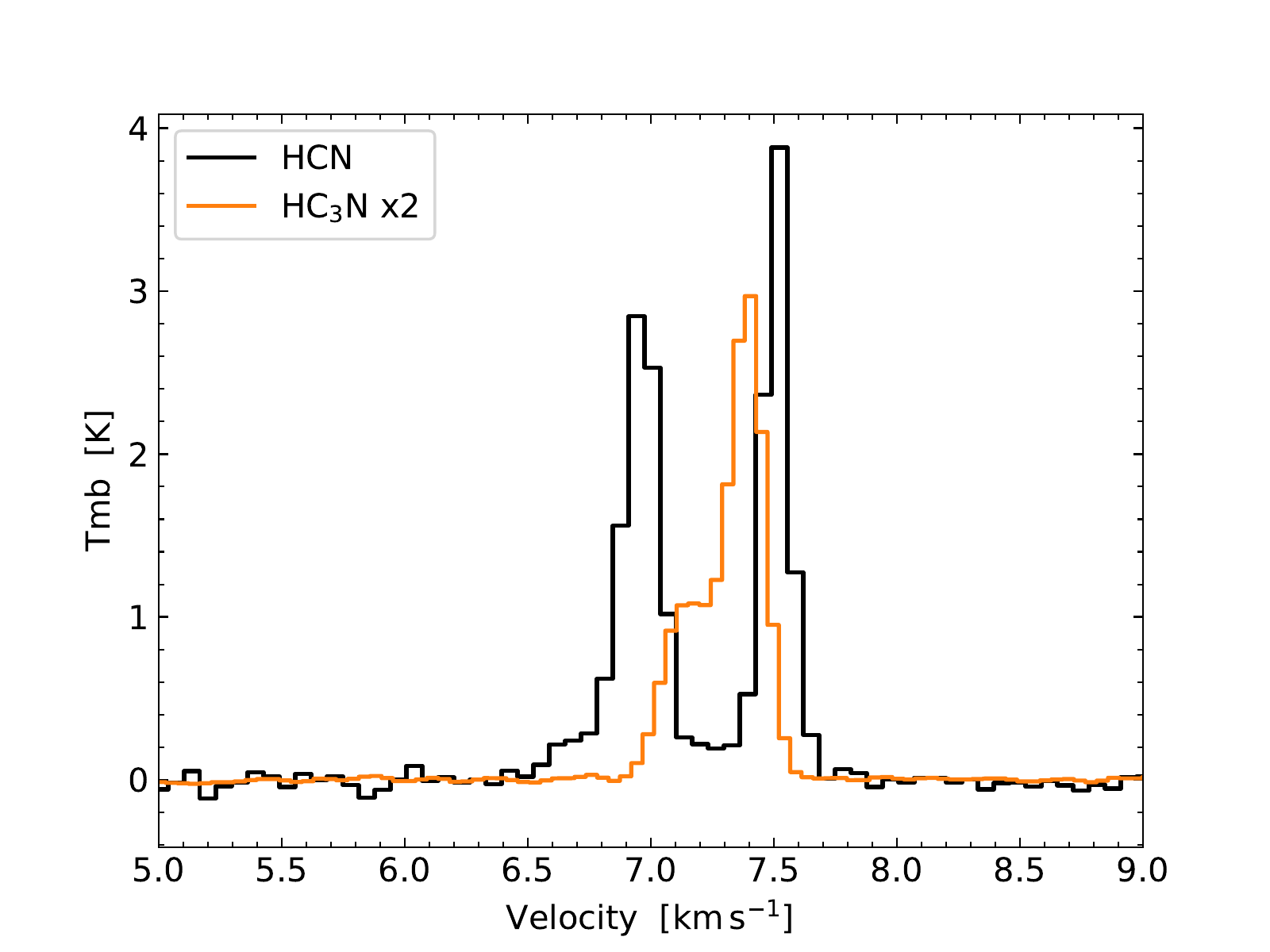}
   \caption{Central hyperfine ($F=2-1$) of the ($1-0$) transitions of HCN (black) and HC$_3$N (orange), observed towards L1544. The data for HC$_3$N are taken from \cite{Bianchi2023}}
              \label{Fig:overplotHCNHC3N}
\end{figure}

It has to be noted that the red asymmetry of HCN has a different origin than the spectra of HC$_3$N (1-0) towards L1544 observed by \cite{Bianchi2023}.
In the case of the optically thin HC$_3$N, the red asymmetry is caused by the molecule tracing the southern region of the core, which is redshifted with respect to the rest velocity of L1544 (see the $V_\mathrm{LSR}$ map in Fig.~B.1 in \citealt{Spezzano2016b}).
By overplotting the central hyperfines of HCN and HC$_3$N, Fig.~\ref{Fig:overplotHCNHC3N} displays the large line width of HCN. This additionally shows the optical thickness of the transition, as in L1544 the line width observed for optically thin species is approximately 0.4\,km\,s$^{-1}$ \citep[e.g.][]{Caselli2002a}.
The red asymmetry is not observed for the other two hyperfine components of HCN. They show a rather symmetric double-peak profile, possibly tracing a more static or non-moving layer, which was also observed in CS (2-1) by \cite{Tafalla1998}. This shows that the hyperfine components clearly trace different layers, depending on their optical depth.
In Sect.~\ref{modeling}, we analyse this behaviour by applying non-LTE radiative transfer modelling. 
Due to its optical thickness, the main species cannot be used to derive the column density and deuterium fraction of HCN, so we focus instead on the $^{13}$C isotopologue.

The hyperfine structures of both the $^{13}$C- and the deuterated isotopologue of HCN show some self-absorption.
A hyperfine structure fitting, using the HFS method provided by the GILDAS package CLASS, reveals that in both cases the weakest component ($F$=0-1) is moderately optically thin, with an optical depth of  $\tau=0.48(9)$ and $\tau=0.9(2)$ for \hxcn\, and DCN, respectively (numbers in brackets give the uncertainty on the last digit). Therefore, we selected these lines for the further analysis.

\begin{figure}
   \centering
   \includegraphics[width=\hsize]{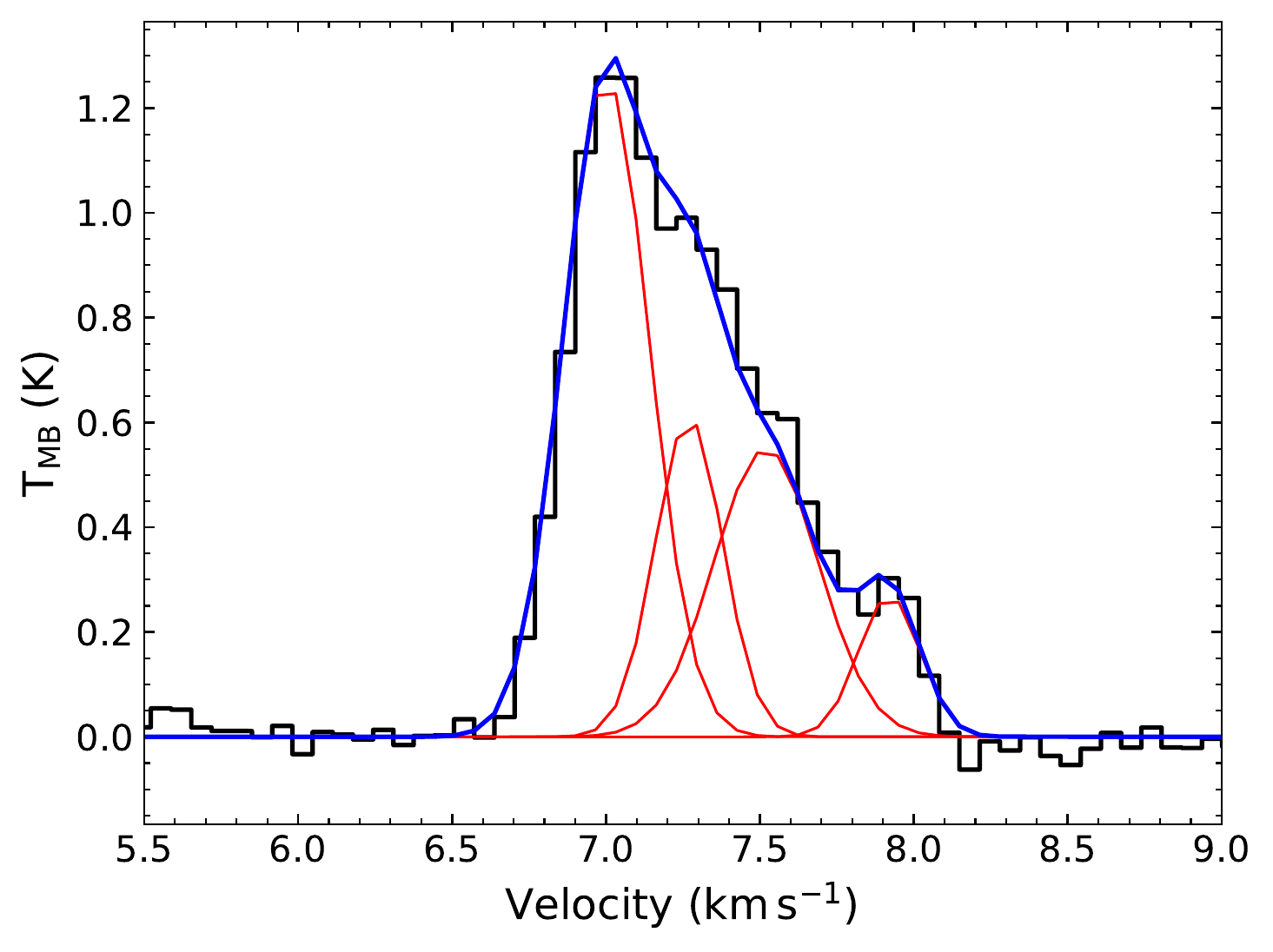}
   \caption{Multi-component Gaussian fit of the four effective hyperfine components found by \cite{VanDerTak2009} to the spectrum of \hnxc\, observed towards L1544.}
              \label{Fig:L1544hn13c4gaussians}
\end{figure}

Our observations cannot resolve the hyperfine structure of HNC, but only show blended components. In addition, the lines are expected to be heavily self-absorbed like in the case of HCN, which makes it impossible to reconstruct the components. 
Therefore, we included this molecule in the non-LTE analysis in Sect.~\ref{modeling} but do not consider it in the derivation of column densities.
The lines of the isotopologues, \hnxc\, and DNC, are split into multiple hyperfine components, mainly due to the non-zero nuclear spins of the D and N nuclei.
For \hnxc\,, \cite{VanDerTak2009} identified four effective hyperfine components, with relative intensities of 1.00, 4.04, 6.63, and 3.66, from low to high frequency.
In Fig.~\ref{Fig:L1544hn13c4gaussians} we fitted the four effective components of \hnxc\, to our observed spectrum by applying a multi-component Gaussian. The amplitudes of the fits result in relative intensities of 1.00, 2.18, 2.41, 4.95 (from high to low velocity components), corresponding to optical depths of 0.3, 0.7, 0.8, 4.5, respectively.
Therefore, we conclude that all components of \hnxc\,, except for the weakest component at the reddest velocity, are optically thick and most probably self-absorbed.
Hence, we only used the hf component at 87090.675\,MHz to derive the column density of \hnxc\,. 
For DNC, \cite{VanDerTak2009} identified six effective hyperfine components, reporting relative intensities of 0.33, 2.67, 1.67, 2.33, 0.99, 1.00, from low to high frequency. 
A CLASS HFS fitting gives out optical depths of 0.24, 1.94, 1.21, 1.69, 0.72, 0.72, from low to high frequency.
Therefore, we selected the lowest frequency component of the DNC hyperfine structure for our further analysis.

The $J$=1-0 transition of \hcop\, shows strong self-absorption in L1544, with a blue asymmetry caused by infall motions in the outer layers of the core. Therefore, we excluded it in the LTE analysis. A detailed discussion and non-LTE modelling of this line can be found in \cite{Redaelli2022}.
The $^{13}$C isotopologue of \hcop\, also shows asymmetric self-absorption. However, we used the line to derive a lower limit for the integrated intensity and column density of \hcop\,. 
The spectrum of the $^{18}$O isotopologue shows a flattened top.
With higher spectral resolution (20\,kHz) and a smaller telescope beam (31$"$), a small dip is visible \citep[e.g.][]{Caselli2002a,Redaelli2019} that is caused by kinematics and CO depletion in the central regions of the core. Due to the lower spectral resolution and larger telescope beam of our data, however, the emission is diluted. The line does not show any asymmetry, as the abundance of the molecule in the outer envelope is too low to be affected by self-absorption.
The deuterated isotopologue consists of three blended hyperfine components that we cannot resolve with our spectral resolution. Therefore, we treated the line as one component.

\subsection{HH211}

The spectra observed towards HH211 show an additional velocity component centred at around 8.22\,km\,s$^{-1}$, which is significantly blue-shifted with respect to the main emission peak ($\sim$9.2\,km\,s$^{-1}$).
This emission likely originates from the outskirts of a lower velocity plateau which might be part of or influenced by the outflow. 
It is located to the south-east of the core (Pineda \& Friesen et al. in prep.), which is partly covered by the telescope beam (40-60$"$, depending on the frequency of the transition) of the Onsala 20\,m telescope.
To account for the extra emission of the additional velocity component, we apply a two-component Gaussian fit to the spectral lines observed towards HH211, wherever possible. In the analysis, we used the fit parameters of the main component emission to derive the column density and deuterium fraction of the molecular tracers.
The only exception are the two $^{13}$C isotopologues of CCH, where the signal-to-noise ratio is too low to resolve the second component.
In this case, a one-component Gaussian fit is sufficient to obtain the line properties.
In Table~\ref{tab:HH2112Gaussians}, we list the best-fit parameters and the derived column densities for the additional velocity component. 
There is no clear trend visible in the FWHM of the additional component compared to the main component. For CCH, \ccd\, and \hcop\,, the second component appears to be broader than the main component, for the rest it is narrower. However, the line width has a rather large error and in some cases, the two line widths are, within the error bars, the same.

The main emission components of the observed lines all show a Gaussian line shape and do not exhibit any dips or asymmetries caused by self-absorption. 
This is possibly caused by a combination of infall motions and turbulent motions induced by the protostellar outflow in the less dense regions of the core, which is enlarging the velocity range in the core and thus, decreasing the possibility for self-absorption. In addition, the lines are also tracing outflow material, for example \hcop\, or HNC \citep[e.g.][]{Arce2004}.
In general, the increased level of turbulence in the core causes a broadening of the line widths
(on average 0.2\,km\,s$^{-1}$ larger than in L1544).

For CCH, we detected all six hf components, and selected the transition with the lowest optical depth to calculate the column density ($N=1-0$, $J=3/2-1/2$, $F=1/1$).
Towards the protostellar core, the $^{13}$C isotopologues of CCH are less bright than towards L1544, and we detected only two hyperfine components with S/N\,$>3$ for each of them. We chose the components with the higher S/N, $N$=1-0, $J$=3/2-1/2, $F_1$=1-0, $F$=3/2-1/2 for \xcch\, and $N$=1-0, $J$=3/2-1/2, $F_1$=2-1, $F$=5/2-3/2 for \cxch\,. 
For the deuterated isotopologue of CCH, we detected seven out of nine hyperfine transitions. Among them, the weaker lines do not show the additional velocity component. We selected the transition with the highest signal-to-noise ratio (S/N\,$>13$), which is $N$=1-0, $J$=3/2-1/2, $F$=5/2-3/2.
The lines of the main isotopologue, CCH, seem to be shifted to lower velocity ($\sim$9.0\,km\,s$^{-1}$), while the lines of the rarer isotopologues are located around the rest velocity of the system ($\sim$9.1\,km\,s$^{-1}$). 

For HCN, \hxcn\, and DCN, we applied the hyperfine-fitting method of CLASS to check the optical depth. As in L1544, all components of HCN are optically thick. We therefore excluded the molecule from the LTE analysis, and only considered it in the radiative transfer modelling in Sect.~\ref{modeling}. The optical depths of DCN are similar to those towards L1544, only the weakest hyperfine is optically thin, with $\tau=0.35$. For \hxcn\,, however, all hyperfine components turn out to be optically thin, with $\tau=0.08-0.4$. Therefore, we selected the brightest hyperfine (F=$2-1$) for the further analysis. 

In the spectra of HNC and isotopologues, the hyperfine components are completely blended. There is no clear evidence of self-absorption as it was seen in L1544. However, the main species is expected to be optically thick, thus, we focus on the $^{13}$C isotopologue. 
Towards HH211, the hf components of \hnxc\, and DNC are less distinct than towards L1544, and the reddest component is less pronounced. Due to this and the additional velocity component at lower velocity ($\sim8.2$\,km\,s$^{-1}$), we were not able to fit the four effective components of \hnxc\, and the six effective components of DNC reported by \cite{VanDerTak2009}. Therefore, we applied a three-component Gaussian fit to the observed spectrum to account for the additional velocity component and the high velocity hf component separately from the remaining, optically thick emission.

The observed emission of \hcop\, shows a very pronounced additional velocity component, with almost half of the intensity of the main emission. 
To avoid any optical depth issues, we did not use \hcop\, for the LTE analysis.
The main velocity components of the $^{13}$C- and D-bearing isotopologues have an optical depth of 0.9, and therefore have to be interpreted with caution.

\section{Analysis}\label{analysis}

\subsection{Excitation temperature and column density}\label{columndensity}
In this work, all column densities were calculated under the assumption of optically thin emission, using the formula presented in \cite{Mangum2015}. Furthermore, we applied the approximation of a constant excitation temperature throughout the core (CTex), following \cite{Caselli2002b} and \cite{Redaelli2019}:
\begin{equation}\label{EquColDensThin}
    N_\mathrm{tot}^\mathrm{thin}=\frac{8\pi\nu^3}{c^3}\frac{Q_\mathrm{rot}(T_\mathrm{ex})}{g_uA_\mathrm{ul}}\left[J_\nu(T_\mathrm{ex})-J_\nu(T_\mathrm{bg})\right]^{-1}\frac{\mathrm{e}^{\frac{E_u}{kT_\mathrm{ex}}}}{\mathrm{e}^{\frac{h\nu}{kT_\mathrm{ex}}}-1}\int T_\mathrm{mb}\mathrm{d}v,
\end{equation}
where $Q_\mathrm{rot}(T_\mathrm{ex})$ is the partition function of the molecule at an excitation temperature $T_\mathrm{ex}$, $g_u$ and $E_u$ are the degeneracy and energy of the upper level of the transition, respectively, $A_\mathrm{ul}$ the Einstein coefficient for spontaneous emission, $T_\mathrm{bg}=2.73$\,K the temperature of the cosmic microwave background, $J(T)$ the Rayleigh-Jeans equivalent temperature and $T_\mathrm{mb}$ the main beam temperature. Assuming a Gaussian line profile, the integrated main beam temperature is calculated by $\frac{1}{2}\sqrt{\pi/\ln2}\cdot\Delta V \cdot T_\mathrm{mb,peak}$.
The corresponding parameters for each transition are listed in Table~\ref{tab:observedlines} and Table~\ref{tab:ColumnDensityParameters}. We derived the partition function from the energies and degeneracies of the rotational levels \citep[see][]{Mangum2015}, taken from the Cologne Database for Molecular Spectroscopy \citep[CDMS][]{Mueller2001}.

However, as some of our transitions are only moderately optically thin, we applied the optical depth correction factor \citep{GoldsmithLanger1999} to all of our lines to derive the total column density:
\begin{equation}\label{EquColDensTot}
    N_\mathrm{tot}=N_\mathrm{tot}^\mathrm{thin}\frac{\tau}{1-\mathrm{e}^{-\tau}}.
\end{equation}
The optical depth is derived using:
\begin{equation}\label{equ:opticaldepth}
    \tau_\nu = -\ln \left[1 - \frac{T_\mathrm{mb}}{f\left[J(T_\mathrm{ex})-J(T_\mathrm{bg})\right]}\right],
\end{equation}
where we assumed a filling factor $f=1$.

For CCH and isotopologues, we assumed an excitation temperature of 6\,K in both sources. This temperature has been applied to similar carbon chain molecules such as c-C$_3$H$_2$ in \cite{Giers2022} and \cite{Chantzos2018}, following \cite{Crapsi2005} and \cite{Emprechtinger2009}. It is also consistent with \cite{Taniguchi2019} who derive $T_\mathrm{ex}\approx6$\,K for CCH in the starless cores L1521B and L134N.
To estimate the excitation temperature of \hxcn\,, DCN, \hnxc\, and DNC, we used the results of the hyperfine-fitting done earlier with the HFS method in CLASS. 
The derived values are listed in Table~\ref{tab:ColumnDensityParameters} and are consistent with excitation temperatures used by \cite{Padovani2011}, \cite{HilyBlant2013} and \cite{Quenard2017}.
For \hcxop\, and \dcop\,, we used the excitation temperatures derived by \cite{Redaelli2019} using radiative transfer. 
For \hxcop\,, we assumed an excitation temperature of 6\,K, which is an intermediate value and similar to what we used for CCH and isotopologues. 
In general, we assumed the same or similar excitation temperatures for both the pre-stellar and the protostellar core, as our observations mainly trace the cold outer envelopes of the two cores, where similar conditions hold. Indeed, the effect of the excitation temperature on the derived column density was found to be moderately small. Changing $T_\mathrm{ex}$ by one Kelvin introduces an error of about 20\,\%.

The derived column densities of all species observed towards the two sources are presented in Table~\ref{tab:ColumnDensities}. To analyse the nature of the additional velocity component in HH211, we derived the column density of it separately. The results are presented in the last column of Table~\ref{tab:HH2112Gaussians}.
The uncertainties for the column densities were derived by propagating the 1$\sigma$ errors on the integrated intensity and the optical depth, and adding an additional error of 15\% to account for the uncertainties in the flux calibration.

For CCH, the derived column densities are consistent with previous measurements within a factor of two \citep{Sakai2008,ZhangWuLiu2021}.
The $^{13}$C isotopologues of CCH are very weak and difficult to observe. Therefore, they have only been detected towards the sources L1521B, TMC-1, L134N/L183, L483 and L1527 so far \citep{Taniguchi2019, SakaiSaruwatari2010, Agundez2019, Yoshida2019}. 

HCN, HNC and isotopologues have been observed multiple times towards L1544 \citep[e.g.][]{Hirota2003, HilyBlant2010, Quenard2017, Spezzano2022} and HH211 \citep{Imai2018} in previous studies, reporting results which are consistent with our measurements.
Previous measurements towards HH211 by \cite{Roberts2002} reported column densities for \hxcn\, and DCN that are smaller by a factor of two compared to our results. 
These differences are likely due to different approximations used in the derivation of the column densities, as well as a different telescope beam.

By assuming the carbon isotopic ratio for the local interstellar medium, $\rm^{12}C/^{13}C=68$ \citep{Milam2005}, we derived a HNC/HCN ratio of 0.3(1) and 0.4(1) for L1544 and HH211, respectively. Previous measurements of this ratio towards L1544 and other cores have revealed a value revolving around 1 (L1544: 1.0(5), \citealt{Quenard2017}; L1498: 0.9(1), L1521E: 1.1(1), TMC-2: 1.1(1), \citealt{Padovani2011}), suggesting that HNC and HCN have similar abundances. However, as this is derived using the $^{13}$C isotopologues, it implies that the fractionation of the two molecules is the same. In \cite{Colzi2020} it is shown that the carbon fractionation of HCN and the carbon fractionation of HNC do behave similarly with a ratio close to 68. However, most of the time, the isotopic ratio for HNC is slightly larger, which can change the resulting HNC/HCN ratio by up to 40\%. Therefore, using the same $^{12}$C/$^{13}$C ratio for the two molecules might not be appropriate. 

In L1544, \hcop\, and its deuterated and non-deuterated isotopologues have been the subject of multiple studies \citep[e.g.][]{Caselli2002b,Jorgensen2004,Vastel2006,Redaelli2019}.
Recently, \cite{Redaelli2019} analysed high-sensitivity maps of \hcxop\, and \dcop\,, where the peak column densities are consistent with our measurements, taking into account that our observations are taken towards the dust peak and are averaged over the whole beam.
To our knowledge, there is no literature data of \hcop\, and isotopologues towards HH211 so far.

\begin{table*}[h]
    \centering
    \caption{Parameters used in the derivation of the column densities.}
    \begin{tabular}{l l c c c c c c l}
    \hline\hline 
    \noalign{\smallskip}
     Molecule & Frequency & $T_\mathrm{ex}$ (K) & $Q(T_\mathrm{ex})$ & $E_\mathrm{u}$\,\tablefootmark{a} (K) & $A$\,\tablefootmark{a} (s$^{-1})$ & $n_\mathrm{crit}$ (cm$^{-3}$) & $g_\mathrm{u}$\,\tablefootmark{a} & $\tau$ \\
    \noalign{\smallskip}
    \hline
    \noalign{\smallskip}
    \multicolumn{8}{l}{\textbf{L1544}} \\
    CCH     & 87446.47(1) & 6 & 12.88 & 4.2 & 2.61e-7 & 9e+4 & 3 & 0.348(8)\\
    \xcch\,  & 84119.33(2) & 6 & 26.40 & 4.08 & 1.37e-6 & 5e+5 & 6 & 0.012(1)\\
    \cxch\,  & 85229.335(4) & 6 & 26.30 & 4.1 & 1.42e-6 & 5e+5 & 6 & 0.028(2)\\
    \ccd\,   & 72101.811(5) & 6 & 22.92 & 3.5 & 3.28e-7 & 1e+5 & 4 & 0.058(9)\\
    \hxcn\, & 86342.2543(3) & 3.3 & 5.92 & 4.1 & 2.22e-5 & 7e+5 & 1 & 0.48(9)\\
    DCN    & 72417.03(1) & 3.7 & 7.49 & 3.5 & 1.32e-5 & 4e+5 & 1 & 0.9(2)\\
    \hxcop\, & 86754.288(5) & 6 & 3.24 & 4.2 & 3.85e-5 & 2e+5 & 3 & 0.55(1)\\
    \hcxop\, & 85162.223(5) & 5.5 & 3.05 & 4.1 & 3.65e-5 & 1e+5 & 3 & 0.086(2)\\
    \dcop\,  & 72039.3124(8) & 7.8 & 4.86 & 3.5 & 2.21e-5 & 1e+5 & 3 & 0.657(7)\\
    \hnxc\,  & 87090.675(3) & 4 & 27.49 & 4.2 & 1.58e-5 & 1e+5 & 7 & 0.24(3)\\
    DNC    & 76305.513(1) & 5 & 27.81 & 3.7 & 1.60e-5 & 2e+5 & 6 & 0.57(2)\\
    \multicolumn{8}{l}{\textbf{HH211}} \\
    CCH    & 87284.11(1) & 6 & 12.88 & 4.2 & 2.60e-7 & 9e+4 & 3 & 0.165(5)\\
    \xcch\, & 84153.31(2) & 6 & 26.40 & 4.08 & 1.37e-6 & 5e+5 & 4 & 0.005(1)\\
    \cxch\, & 85229.335(4) & 6 & 26.30 & 4.1 & 1.42e-6 & 5e+5 & 6 & 0.010(1)\\
    \ccd\,  & 72107.721(3) & 6 & 22.92 & 3.5 & 8.60e-7 & 3e+5 & 6 & 0.12(2)\\
    \hxcn\,& 86340.1666(1) & 4.1 & 7.05 & 4.1 & 2.22e-5 & 7e+5 & 5 & 0.39(2)\\
    DCN  & 72417.03(1) & 3.7 & 7.49 & 3.5 & 1.32e-5 & 4e+5 & 1 & 0.44(7)\\
    \hxcop\, & 86754.288(5) & 6 & 3.24 & 4.2 & 3.85e-5 & 2e+5 & 3 & 0.91(1)\\
    \hcxop\, & 85162.223(5) & 5.5 & 3.05 & 4.1 & 3.65e-5 & 1e+5 & 3 & 0.057(2)\\
    \dcop\,  & 72039.3124(8) & 7.8 & 4.86 & 3.5 & 2.21e-5 & 1e+5 & 3 & 0.899(7)\\
    \hnxc\, & 87090.675(3) & 4 & 27.49 & 4.2 & 1.58e-5 & 1e+5 & 7 & 0.073(4)\\
    DNC    & 76305.513(1) & 5 & 27.81 & 3.7 & 1.60e-5 & 2e+5 & 6 & 0.28(2)\\
    \noalign{\smallskip}
    \hline 
    \noalign{\smallskip}
    \end{tabular}
    \label{tab:ColumnDensityParameters}
    \tablefoot{
    \tablefoottext{a}{Extracted from the Cologne Database for Molecular Spectroscopy \citep{Mueller2001}}}
\end{table*}

\begin{table}[h]
    \centering
    \caption{Derived column densities of both sources, given in units of $\times10^{12}$\,cm$^{-2}$.}
    \begin{tabular}{l c c}
    \hline\hline 
    \noalign{\smallskip}
     Molecule & L1544       & HH211 \\
    \noalign{\smallskip}
    \hline
    \noalign{\smallskip}
    CCH        & 320(40)  & 210(30) \\
    \xcch\,    & 1.6(4)  & 1.3(5) \\
    \cxch\,    & 3.6(7)  & 2.2(5) \\
    \ccd\,     & 37(10)  & 22(5) \\
    \hxcn\,    & 5(1)    & 1.7(2) \\
    DCN        & 18(4)  & 9(2) \\
    \hxcop\,   & 1.2(1) & 2.2(2) \\
    \hcxop\,   & 0.12(2) & 0.11(2) \\
    \dcop\,    & 3.1(3) & 5.4(5) \\
    \hnxc\,    &  1.6(4) & 0.7(2) \\
    DNC        & 9(1)  & 4(1) \\
    \noalign{\smallskip}
    \hline 
    \noalign{\smallskip}
    \end{tabular}
    \label{tab:ColumnDensities}
\end{table}

\subsection{Deuterium fraction}\label{deuteriumfraction}

The deuterium fractions are derived by dividing the column density of the deuterated isotopologues by the column density of the main species.
As the main species of all observed molecules, except CCH, are optically thick, we used the $^{13}$C and $^{18}$ isotopologues to estimate their column densities. For this, we multiplied the column densities of the $^{13}$C isotopologues by $^{12}\mathrm{C}/^{13}\mathrm{C}=68$ \citep{Milam2005}. 
Similarly, the column density of \hcxop\, was multiplied by 557 \citep{Wilson1999} to obtain an estimate for $N$(\hcop\,). The resulting deuterium fractions are summarised in Table~\ref{tab:DeuterationLevel}. 
Furthermore, we derived the deuteration level in the additional velocity component of HH211. The results are presented in Table~\ref{tab:HH2112GaussiansDeuteration}. 
Despite the high uncertainties in some cases, the additional component and the main emission generally show quite similar deuteration levels. 
This likely suggests that they both trace the same gas, with similar properties along the line-of-sight.

\cite{Colzi2020} show that in the local interstellar medium the carbon isotopic ratio is molecule-dependent and varies with time, volume density, and temperature, and can deviate significantly from 68. For example, for a fixed density and temperature, it may change by a factor of two between 10$^5$ and 10$^6$ years of core evolution, depending on the molecule.
In this work, we applied the canonical value, taking note of the uncertainties it might introduce.

When interpreting the deuterium fraction, one has to keep in mind that, first, with single-dish telescopes the observations of deuterated and non-deuterated species correspond to a different beam size and therefore can cover different regions of the core.  
In our case, the beam sizes for the deuterated isotopologues on average are generally 10$''$ larger than the telescope beams of the normal and $^{13}$C isotopologues (see Table~\ref{Tab:AllObervedLines}). 
This might dilute the real deuterium fraction of a molecule and cause systematic errors in the ratios.
Moreover, the species themselves are abundant in different regions or shells of the cores. 
In pre-stellar cores, deuterated species are expected to trace higher-density regions close to the centre of the core, where the deuteration process is most efficient \citep[e.g.][]{Giers2022}. 
Non-deuterated isotopologues, on the other hand, are additionally also abundant in the outer parts of a core. In L1544, this is observed for example in \hcop\,, CH$_3$OH, H$_2$CO, and c-C$_3$H$_2$ \citep{Redaelli2019,ChaconTanarro2019,Giers2022}.
Towards protostellar cores, various single dish observations have shown high levels of deuteration \citep[e.g.]{VanDishoeck1995,Parise2004}. 
However, \cite{Persson2018} show that with interferometric observations the D/H ratio is actually lower in the inner regions, as they are less affected by optical depth effects and beam dilution. These uncertainties are important to keep in mind when interpreting single dish observations of the inner warm regions of protostellar cores.
However, due to the size of the telescope beam, these regions are very much diluted, and with our data, we are mainly sensitive to the outer (and cold) protostellar envelope.

\begin{table}[h]
    \centering
    \caption{Column density ratios in both cores.}
    \begin{tabular}{l l l }
    \hline\hline 
    \noalign{\smallskip}
       & L1544 & HH211  \\
    \noalign{\smallskip}
    \hline
    \noalign{\smallskip}
    $N$(CCH)/$N$(\xcch\,) & 200(50) & 160(60) \\
    $N$(CCH)/$N$(\cxch\,) & 90(20) & 100(30) \\
    $N$(\cxch\,)/$N$(\xcch\,)\, & 2.2(7)   & 1.7(8)  \\
    $N$(\hnxc\,)/$N$(\hxcn\,) & 0.3(1) & 0.4(1) \\
    $N$(\hxcop\,)/$N$(\hcxop\,) & 10(1) & 20(4) \\
    \noalign{\smallskip}
    \hline
    \noalign{\smallskip}
    $N$(\ccd\,)/N(CCH)               & 0.11(3)  & 0.10(3) \\
    $N$(\ccd\,)/[$N$(\xcch\,)$\times$68]   & 0.3(1)  & 0.2(1) \\
    $N$(\ccd\,)/[$N$(\cxch\,)$\times$68]   & 0.15(5)  & 0.14(5) \\
    $N$(DCN)/[$N$(\hxcn\,)$\times$68]    & 0.05(2)  & 0.07(2)\\
    $N$(\dcop\,)/[$N$(\hxcop\,)$\times$68] & 0.040(7) & 0.037(5) \\
    $N$(\dcop\,)/[$N$(\hcxop\,)$\times$557] & 0.048(9) & 0.09(2) \\
    $N$(DNC)/[$N$(\hnxc\,)$\times$68]    & 0.08(3) & 0.08(3) \\
    \noalign{\smallskip}
    \hline 
    \noalign{\smallskip}
    \end{tabular}
    \label{tab:DeuterationLevel}
\end{table}

\section{Radiative transfer modelling}\label{modeling}
In this section, we use non-LTE radiative transfer simulations to model the observed molecular lines.
We applied the software LOC \citep[Line Transfer with OpenCl,][]{Juvela2020}, which is based on ray tracing.
To reduce the complexity of the simulations, we made the simplifying assumption of a one dimensional, spherically symmetric grid for both of our cores.

For CCH, HCN, \hxcn\,,  \hcop\, and \dcop\,, we used the latest collision rate coefficients available on the Leiden Atomic and Molecular Database \citep[LAMDA,][]{Schoier2005,VanDerTak2020,Dagdigian2018,hernandez2017rotational,DenisAlpizar2020,Pagani2012}. In the case of HCN, \hxcn\,, and \dcop\,, we applied the hyperfine-resolved rate coefficients. 
For the collisional coefficients of \hxcop\, and \hcxop\,, we used the rates of \hcop\, from LAMDA and scaled them with the corresponding reduced mass of the isotopologues.
For the $^{13}$C- and the deuterated species of CCH, a scaling of the CCH rates is not possible, as the hyperfine structures are different due to the different nuclear spins.
Therefore, we excluded these molecules from the non-LTE modelling.

To the best of our knowledge, hyperfine resolved rate coefficients for DCN, HNC, HN$^{13}$C and DNC do not exist\footnote{Hyperfine resolved rate coefficients for HNC do not exist but data of reasonable accuracy can be obtained from pure rotational rate coefficients as described in \cite{Goicoechea:22}}.
A simple approach to simulating lines of these species using radiative transfer would be to use the rate coefficients of the dominant isotopologues (HCN and HNC), neglecting the effect of isotopic substitution.
However, it has been shown recently \citep{navarro2022linking} that isotope effects can be important in the case of hydrogen cyanides and isocyanides. Indeed, the comparison of the different sets of collisional data revealed that isotopic substitution can lead to substantial changes (up to $\sim$50\%) at 10\,K (the typical temperature of cold cores).

Accurate hyperfine-resolved rate coefficients for the excitation of DCN, HNC, HN$^{13}$C and DNC induced by collisions with H$_2$ are computed in this work. The scattering calculations are based on the HCN-H$_2$ \citep{denis2013new} and HNC-H$_2$ \citep{dumouchel2011rotational} interaction potentials corrected to consider the effect of isotopic substitution. The quantum mechanical close-coupling approach \citep{green1975rotational} and the almost exact recoupling method \citep{lanza2014hyperfine} was used to compute the collisional data for temperature ranging between 5 and 30\,K. In the calculations, only the excitation by para-H$_2$, the strongly dominant form of H$_2$ in cold ISM, has been considered. Details on the scattering calculations are provided in Appendix~\ref{collisionrates}.

The data underlying this article will be made available through the EMAA\footnote{https://emaa.osug.fr/}, LAMDA \citep{Schoier2005,VanDerTak2020}, and BASECOL \citep{Dubernet2013} data bases. They are also available on request.

\subsection{Physical structure of L1544}
\begin{figure}
   \centering
    \includegraphics[width=\hsize]{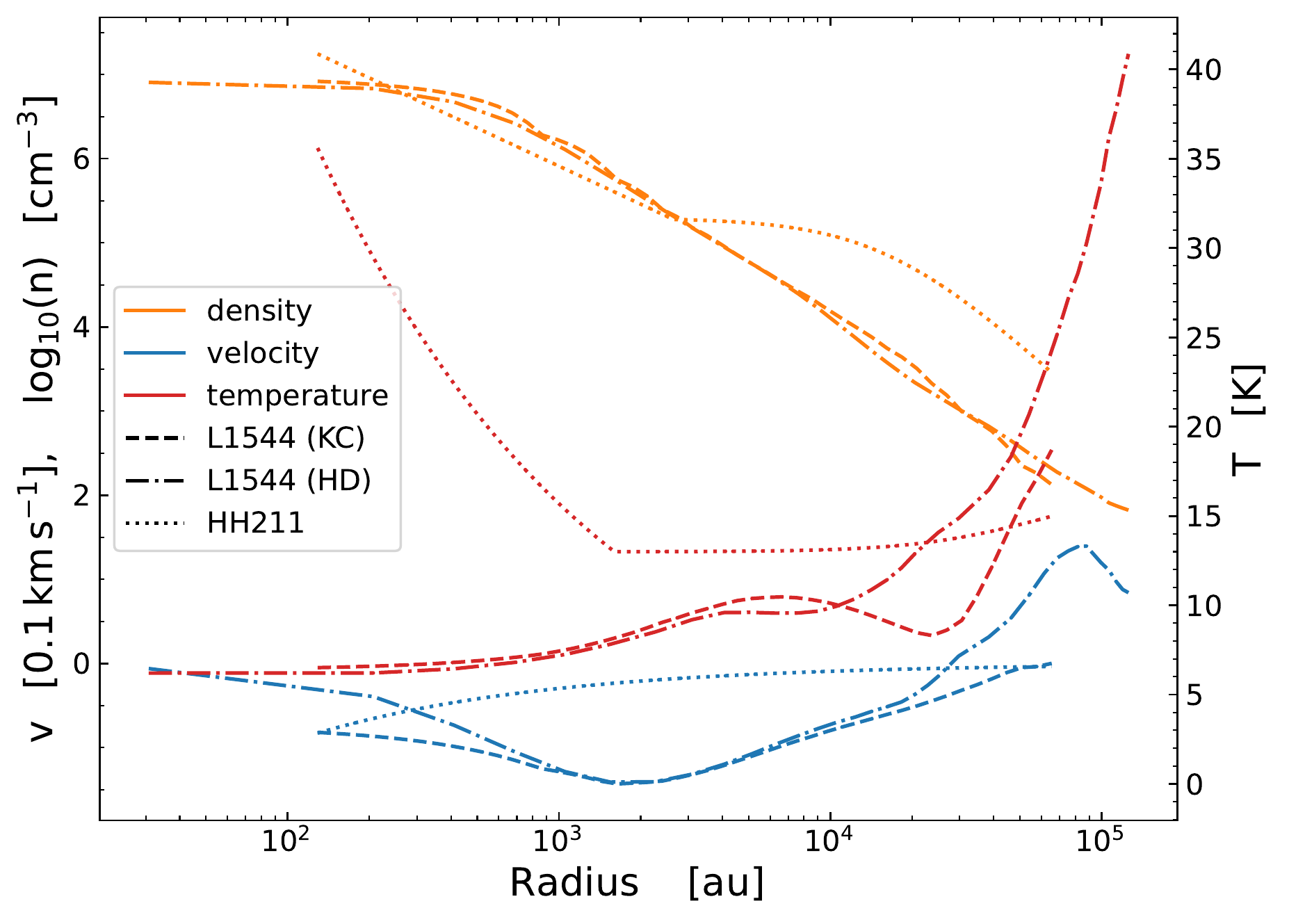}
   \caption{Profiles of the gas temperature (red), H$_2$ number density (orange, in logarithmic scale) and infall velocity (blue, in units of 0.1\,km\,s$^{-1}$) for the Keto-Caselli model \citep[dashed, KC][]{Keto2015}, the HDCRT model \citep[dotdashed, HD][]{Sipila2022} of L1544, and the physical model of HH211 (dotted) derived in this work.}
              \label{Fig:physicalmodels}
\end{figure}

The physical structure of L1544 is well-studied \citep[e.g.][and references therein]{Keto2015}.
To model the physical structure of the core, we applied the physical model presented by \citet[][hereafter Keto-Caselli model]{Keto2015}. It describes an unstable quasi-equilibrium Bonnor-Ebert sphere with a peak central H$_2$ volume density of $n_0\approx 10^7$\,cm$^{-3}$ and a central gas temperature of 6\,K. The model provides the infall velocity, density and gas temperature structure as a function of distance from the centre of the core (see Fig.~\ref{Fig:physicalmodels}). 

Recent studies prove the Keto-Caselli model to be unable to explain double-peaked profiles of optically thin lines \citep[e.g.][]{Redaelli2019,FerrerAsensio2022}. On top of that, \cite{Redaelli2022} show that to reproduce the blue asymmetry and high level of self-absorption of \hcop\, (1-0), it is necessary to add a low-density and contracting envelope around the Keto-Caselli model.
To test the effect of velocity variations in the outer core, we additionally used the HydroDynamics with Chemistry and Radiative Transfer model introduced by \citet[][hereafter HDCRT model]{Sipila2022}. The corresponding infall velocity, density and gas temperature structure are shown alongside the Keto-Caselli model in Fig.~\ref{Fig:physicalmodels}.
The HDCRT model was designed to recreate the conditions of the Keto-Caselli model in the inner regions of the core where the gas-dust thermal coupling is strong.
Unlike the Keto-Caselli model, the HDCRT model allows for expansion motions of the gas in the outer regions, caused by photoelectric heating.

\subsection{Physical structure of HH211}
There is no model of the physical structure of HH211 available in the literature. 
We computed it using data from the \textit{Herschel} Science Archive (HSA). Following \cite{Harju2017}, we used the archive \textit{Herschel}/SPIRE dust continuum emission maps to derive the temperature and column density maps. From those, we extracted radial temperature and volume density profiles for the outer regions of the core. For the inner regions, we used power-laws following \cite{Crimier2010} and \cite{MotteAndre2001}.
For the velocity, we assumed free fall with the literature mass of the protostar \citep{LeeLi2018}.
A detailed derivation of the physical structure of HH211 can be found in Appendix~\ref{AppPhysModelHH211}.

\subsection{Molecular abundances}
We used the state-of-the-art gas-grain chemical model of \cite{Sipila2019} to obtain an estimate of the fractional molecular abundances in the two cores. We applied the two-phase model of the chemical code, where the gas phase chemistry and the entire ice on the grains are active. 

To account for the fact that L1544 is embedded in a molecular cloud, we assumed an external visual extinction of $A_V=2$\,mag. We employed the same initial abundances and chemical networks as in \cite{Sipila2019} and the same standard values for various model parameters as in \cite{Giers2022}, not recounted here for brevity.
The abundance profiles are obtained by using the Keto-Caselli model  and the HDCRT model as a static physical model of the core. 
We ran the chemical simulation, extracting the abundance profiles for various evolutionary times.
In addition, we tested the abundances at the time of the best match between the Keto-Caselli model and the fiducial hydrodynamical simulation, as presented in \cite{Sipila2022}.

To derive the molecular abundances for HH211, we first adopted the physical conditions that roughly correspond to the edge of the HH211 model ($n_H=5\times10^3\,\mathrm{cm}^{-3}$, $T=15$\,K, $A_V=6$\,mag) and let the chemistry evolve over 10$^6$\,yr in those conditions, extracting the abundances at that time to use as the initial abundances for the protostellar core model.

The deuterium chemical model does not consider the isotope chemistry of carbon, therefore we scaled the abundance profiles by the carbon isotopic ratio of the local ISM \citep[68,][]{Milam2005} to obtain estimates for the $^{13}$C-species. The same applies for the $^{18}$O isotopologue of \hcop\, \citep[557,][]{Wilson1999}.

In addition to the molecular abundance profiles derived by chemical modelling, we applied a simple approach by assuming a constant abundance of the molecules throughout the core. 
This simplification allows to estimate the abundance levels of the molecules, especially in the case of the HDCRT model, and can be used for parameter-space explorations in future studies.
However, in the case of L1544, \cite{Caselli2022} find evidence of almost complete freeze out within the central 1800\,au. We took this into account by setting the abundance to zero for radii within 1500\,au.

\subsection{Results}
For the comparison with the observed spectra, the synthetic ones were convolved with the respective corresponding telescope beam size (see Table~\ref{Tab:AllObervedLines}).
To find the best fits to the observed spectra, we tested the different timesteps of the abundance profiles (10 logarithmically spaced steps between 10$^5$ and 10$^6$ years of chemical evolution) and run through a range of constant abundances in log scale.
The best fits were determined by optimising the line shape, line width and peak intensity of the synthetic spectra in accordance with the observations.
A selection of the best-fit synthetic spectra derived with radiative transfer modelling using LOC are presented in Figs.~\ref{Fig:LOC-CCH}-\ref{Fig:LOC-HCOp}. The respective labelling of the applied models is explained in Table~\ref{tab:SummaryOfModels}.

For L1544, the results are split between the Keto-Caselli model and the HDCRT model. The labels KC (red solid line) and HD (blue solid line) denote the results of a constant abundance applied to the respective model, while STKC (cyan dashed line) and STHD (orange dashed line) refer to the abundance profiles derived with chemical modelling. The results produced by the abundance profile from the best match between Keto-Caselli and fiducial hydrodynamical model are described by the label HD0 (green dotdashed line). For HH211, we plot the best-fit results of applying a constant abundance (blue solid line) and an abundance profile derived with the two-phase chemical model (red dotdashed line).
The observed spectra are overplotted in black.
The corresponding best-fitting abundances are plotted in Fig.~\ref{Fig:LOC-abuprofilesL1544} and Fig.~\ref{Fig:LOC-abuprofilesHH211} for L1544 and HH211, respectively.

To account for the level of turbulence in the cores, we used the average of the non-thermal components of the observed velocity dispersions. The values that best reproduce the line widths of the optically thin species are  
$\sigma_\mathrm{turb}=0.1$\,km\,s$^{-1}$ (L1544, Keto-Caselli model), $\sigma_\mathrm{turb}=0.17$\,km\,s$^{-1}$ (L1544, HDCRT model) and $\sigma_\mathrm{turb}=0.23$\,km\,s$^{-1}$ (HH211). Separate values for the two different physical models of L1544 are necessary, because the outer radius of the HDCRT model is a factor of two larger compared to the Keto-Caselli model. Therefore, the model covers more turbulent gas. 

To account for the additional velocity component at 8.22\,km\,s$^{-1}$ observed towards HH211, we approximated it by a static Gaussian that is added to the modelling results afterwards. For this, we used the results of the two-component Gaussian fits listed in Table~\ref{tab:HH2112Gaussians}. In the case of HNC, where a two-component Gaussian fit was not successful, we estimated the static Gaussian by eye to improve the readability of the modelled spectra.

\subsubsection{L1544}
The synthetic spectra of CCH are shown in Fig.~\ref{Fig:LOC-CCH}, with each hyperfine component presented in a separate plot (numbered from low to high frequency).
Towards L1544, the constant abundances applied to the two different physical models are able to reproduce the intensities of the more optically thin hyperfines. 
The line profiles of the more optically thick components are not reproduced by the Keto-Caselli model. 
The observed lines show rather symmetric double-peaks, whereas the synthetic lines display a blue asymmetry. 
This behaviour was also observed for CS ($2-1$) by \cite{Tafalla1998} and might be an indication that these transitions trace a static envelope. Using the HDCRT model and a constant abundance, we are indeed able to reproduce the symmetric double-peak profiles, except for hyperfine component 5. 
The abundance profiles derived with chemical modelling largely underestimate the amount of CCH, and therefore fail to reproduce the lines in both the STKC and the STHD case.
To reproduce the same intensities as in the KC case, the STKC abundance profile has to be multiplied by a factor of 20. The discrepancy between the simulated and observed CCH abundances is most likely related to the uncertainties in the chemistry of CCH, though the use of a simple 1D physical model may contribute as well. Future studies of the formation and destruction pathways of CCH are required to improve the chemical simulations, as well as the adoption of 3D models \citep{Jensen2023arXiv}.

The synthetic spectra of HCN, \hxcn\,, and DCN are shown in Fig.~\ref{Fig:LOC-HCN} and Fig.~\ref{Fig:LOC-DCN}.
The Keto-Caselli model fails at reproducing HCN with both a constant abundance and an abundance profile. The synthetic spectra show blue asymmetries in all hyperfines, indicating infall motion, and do not recreate the red asymmetry observed in the central hyperfine. This shows that the applied abundance profiles overestimate the amount of HCN in regions of infall and indicates that HCN might have a higher abundance in the low-density regions of the core. 
In addition, the modelled dip caused by the self-absorption does not reach down to zero level. This might be solved by adding an extended low-density envelope to the Keto-Caselli model, such as in the case of \hcop\, \citep{Redaelli2022}. 
The HDCRT model, on the other hand, is able to produce the red asymmetry profiles for HCN by applying the STHD abundance profile. However, it results in red asymmetries for all hyperfines, reproducing the intensity of two hyperfines, but underestimating the weakest component by a factor of two.
Applying a constant abundance results in a rather symmetric line profile for all components. The self-absorption dip of the synthetic spectrum goes down to almost zero level, as observed. 
A comparison of the STHD and HD0 abundance profiles (see Fig.~\ref{Fig:LOC-abuprofilesL1544}) shows that, in the case of HD0, HCN is more abundant at larger radii than in the case of STHD. This results in a deeper self-absorption dip.
The distance between the two peaks is not reproduced (see Fig.~\ref{Fig:LOC-HCN}), indicating that for HCN the turbulence level of 0.17\,km\,s$^{-1}$ is overestimated in the hydrodynamical case.

In the case of \hxcn\, (see Fig.~\ref{Fig:LOC-HCN}), the Keto-Caselli model works very well and is able to reproduce the spectrum with both the KC and STKC abundance profile. 
The HDCRT model, however, struggles to reproduce the correct shapes of the self-absorbed hyperfines. 
Both the HD and STHD abundance profiles result in symmetric rather than blue asymmetric lines, which might hint at an overestimation of \hxcn\, in the outer regions and/or a radially changing carbon isotopic ratio.
In the case of HD0, the spectrum is severely underestimated. This might be a hint that at this specific timestep chosen by \cite{Sipila2022}, the carbon isotopic ratio is actually below 68.
For DCN (see Fig.~\ref{Fig:LOC-DCN}), the Keto-Caselli model can reproduce the intensities of the hyperfines within a factor of two. However, it fails to reproduce the shapes of the lines. Both the KC and the STKC abundance profiles predict a strong red asymmetry that is not observed. In addition, the synthetic spectra show higher self-absorption than the observed spectrum. 
This shows that the abundance profiles are actually overestimating the amount of DCN and that in reality there is less material available to trace the infall motions. 
The HDCRT model cannot reproduce the spectrum, neither the line shapes nor the intensities match the observations.

The synthetic spectra of HNC, \hnxc\,, and DNC  are shown in Fig.~\ref{Fig:LOC-HNC}.
The collision rate coefficients for \hnxc\, and DNC only consider the spin of the $^{14}$N for the hyperfine splitting. Therefore, the modelling fails to reproduce the lowest frequency component of these two molecules. However, work is underway to address this issue, where the collisional rate coefficients are computed by also including the effect of $^{13}$C and D, respectively, which will allow to reproduce the full spectrum.
Towards L1544, HNC can be reproduced quite well, though the Keto-Caselli model seems to perform better than the HDCRT model at reproducing the observed intensity peaks. 
However, due to the greater physical size of the structure in the HDCRT model, the dip at 7.2\,km\,s$^{-1}$ is better reproduced than in the case of the Keto-Caselli model.
At high velocity there seems to be an issue with the hyperfine component that has the lowest statistical weight. This might be caused by an underestimation of the general abundance of the molecule and subsequently of the depth of self-absorption in the lower velocity hyperfine components. 

In the case of \hnxc\,, the Keto-Caselli model is able to reproduce the spectrum with both constant abundance and abundance profile, except for the high velocity component. The HDCRT model, however, results in a stronger self-absorption and therefore underestimates the observed intensity.
In the case of DNC, the STKC abundance profile, applied to the Keto-Caselli model, results in a good match with the observed line shape and intensity (excluding the high velocity component). The synthetic spectrum produced by the constant abundance largely overestimates the amount of DNC in the outer regions of the model, resulting in a deep dip caused by self-absorption of the strongest hyperfines that is not observed.
The HDCRT model is not able to reproduce DNC, it underestimates both intensity and line width. A comparison of the abundance profiles (see Fig.~\ref{Fig:LOC-abuprofilesL1544}) shows that the peaks of the STHD and the HD0 profiles are at least one magnitude lower than for the STKC profile, which leads to an underestimation of the molecular abundance.

The synthetic spectra of \hcop\, are shown in Fig.~\ref{Fig:LOC-HCOp}.
Towards L1544, our models fail to reproduce the strong self-absorption and asymmetry of \hcop\,. However, the case of this molecule is extensively studied in \cite{Redaelli2022}. By adding a low-density (27\,cm$^{-3}$), contracting envelope extending out to 1\,pc, they manage to reproduce the line shape of the transition.
In the case of \hxcop\,, a constant abundance profile applied to the Keto-Caselli model can reproduce the observed spectrum. 
The modelled line profile shows a stronger blueshift and a higher intensity than observed. However, a constant abundance of \hxcop\, with complete freeze out in the central 1500\,au is clearly too simplistic for this molecule, present not only in the core, but also in the surrounding cloud.
In the case of the HDCRT model, a constant abundance of \hxcop\, results in an almost symmetric double-peak profile.
The abundance profiles derived by chemical modelling severely underestimate the observed intensity and cannot reproduce the spectrum. This might indicate that in this case the carbon isotopic ratio is less than 68. 

For \hcxop\,, the line cannot be reproduced by either of the models. The HD abundance profile roughly reproduces the line width, but fails at the flattened top caused by CO depletion in the central regions of the core.
Like in the case of \hxcop\,, the synthetic spectra derived with the abundance profiles underestimate the intensity of the line and indicate that the assumption of the isotopic ratio to be 557 might be not correct in this case.  
For \dcop\,, the observed line shape cannot be reproduced. The simulation is considering the overlapping hyperfine structure of the molecule and the modelled spectra show a rather flat peak. This is likely caused by individual, self-absorbed hyperfine components.

\subsubsection{HH211}
\begin{figure*}
   \centering
   \includegraphics[width=0.99\textwidth]{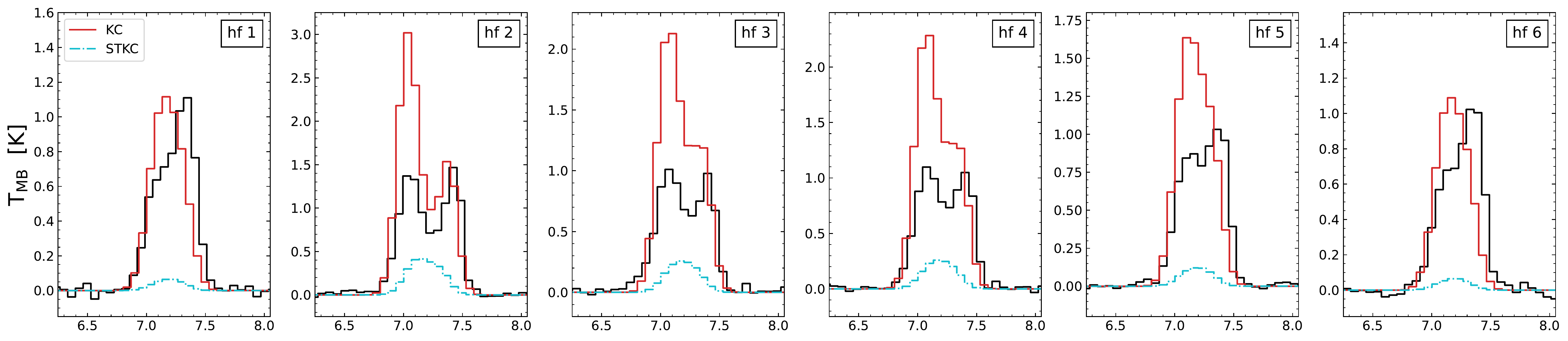}
   \includegraphics[width=0.99\textwidth]{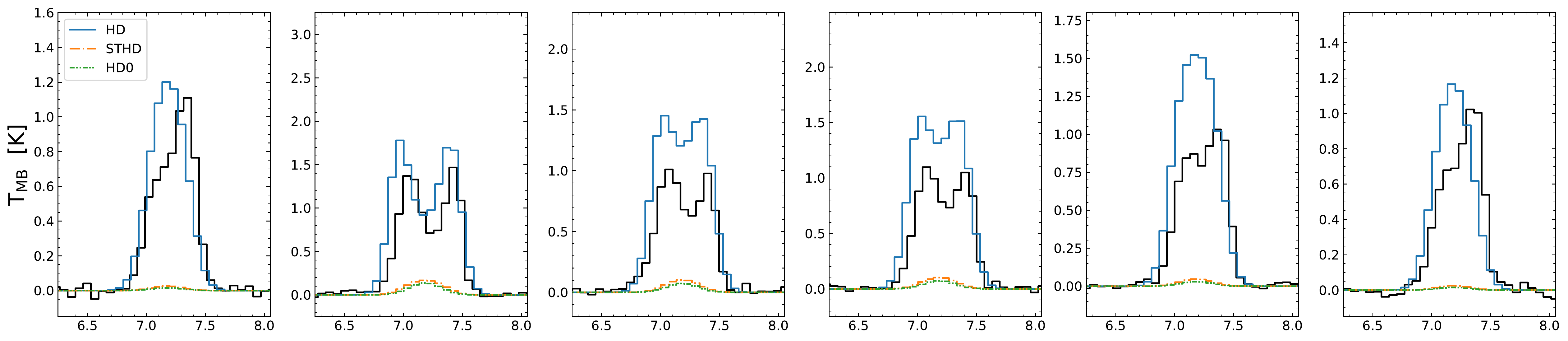}
   \includegraphics[width=0.99\textwidth]{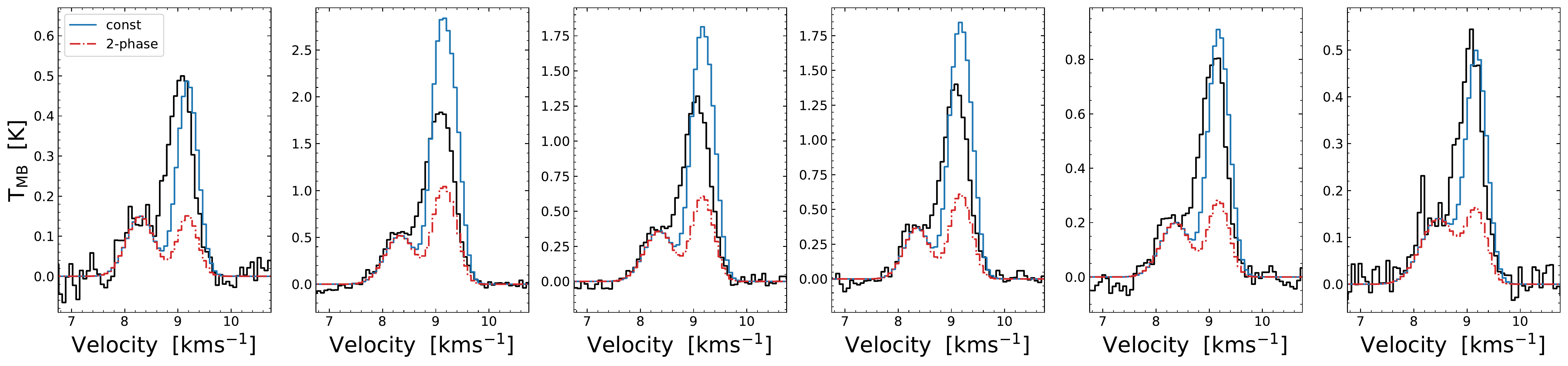}
   \caption{Synthetic spectra of CCH obtained with LOC (top rows: L1544, bottom row: HH211). Each hyperfine line is plotted separately, numbered from low to high frequency. For L1544, the results are splitted between using the Keto-Caselli model (top) and the HDCRT model (centre) as approximation for the physical structure of the core. Towards L1544, the best-fitting abundances and timesteps are $2\times10^{-8}$ (KC), 1e5\,yr (STKC), $3\times10^{-8}$ (HD), 1e5\,yr (STHD). Towards HH211, the best-fitting abundances and timesteps are $2\times10^{-9}$ (const), 1e6\,yr (2-phase).}
              \label{Fig:LOC-CCH}
\end{figure*}

The synthetic spectra of CCH are shown in Fig.~\ref{Fig:LOC-CCH}, with each hyperfine component presented in a separate plot (numbered from low to high frequency).
The observed spectrum can be reproduced assuming a constant molecular abundance of $2\times10^{-9}$ with respect to H$_2$ throughout the core. The intensity ratios between the hyperfine components are reproduced within a factor of two in the abundance. 
The synthetic spectra resulting from abundance profiles derived with the 2-phase chemical model, where gas phase and grain chemistry are considered, underestimate the observed lines by a factor of two to five. This supports the need to constrain the reactions involved in the chemistry of CCH to improve the chemical simulations. 

\begin{figure*}
   \centering
   \includegraphics[width=0.49\textwidth]{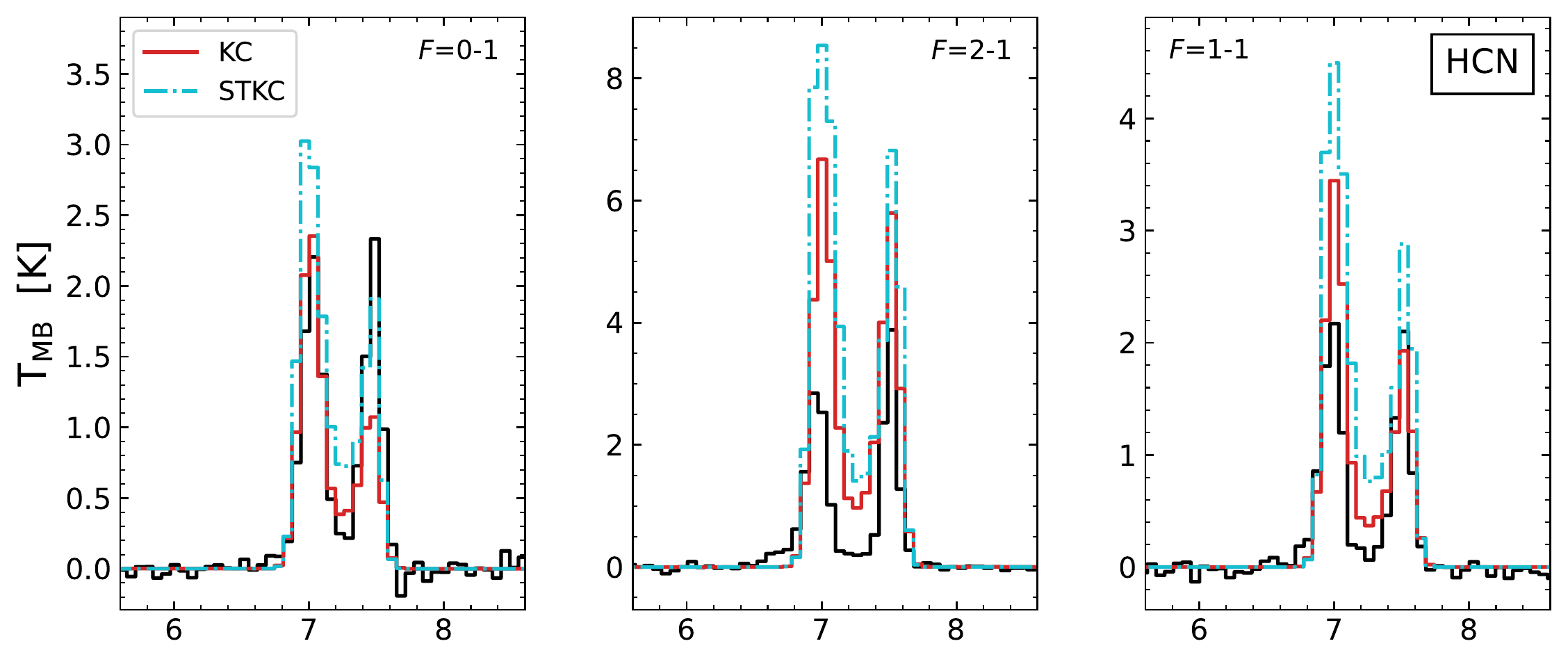}
   \includegraphics[width=0.49\textwidth]{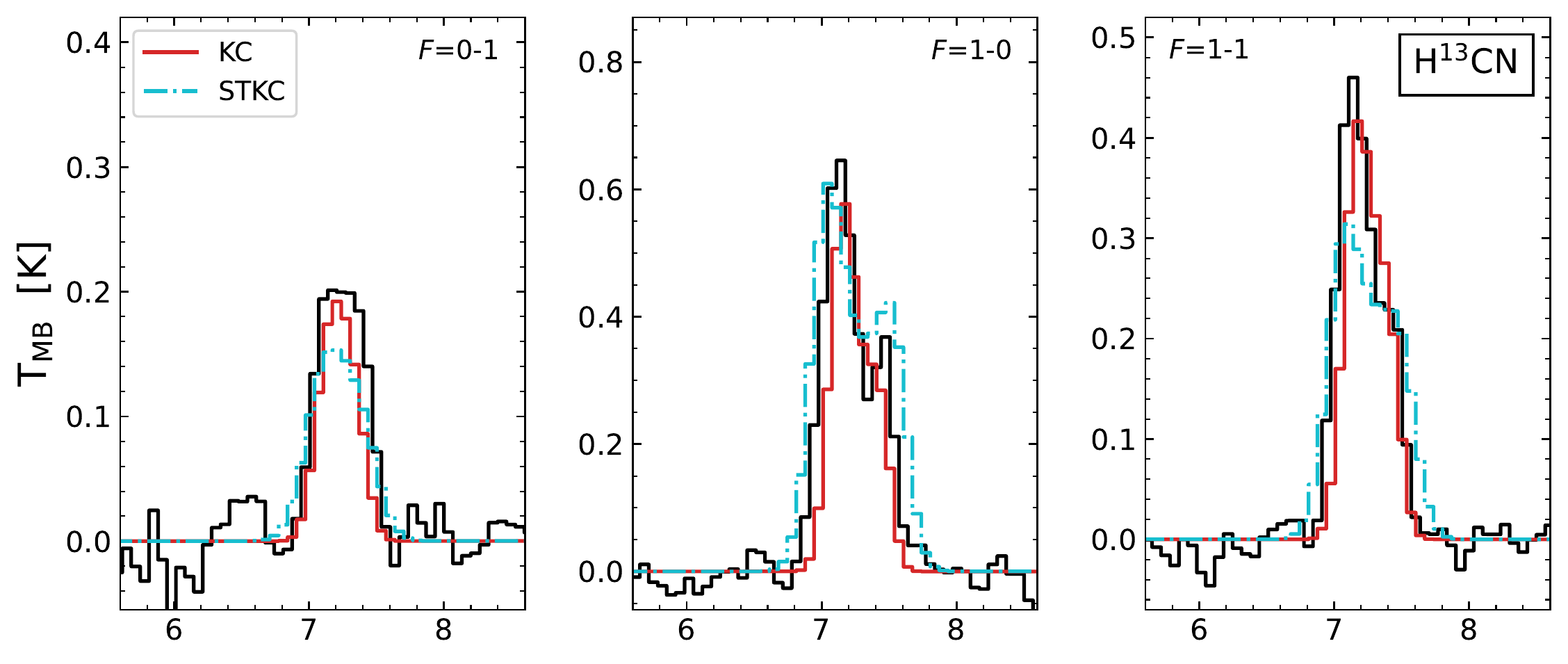}
   \includegraphics[width=0.49\textwidth]{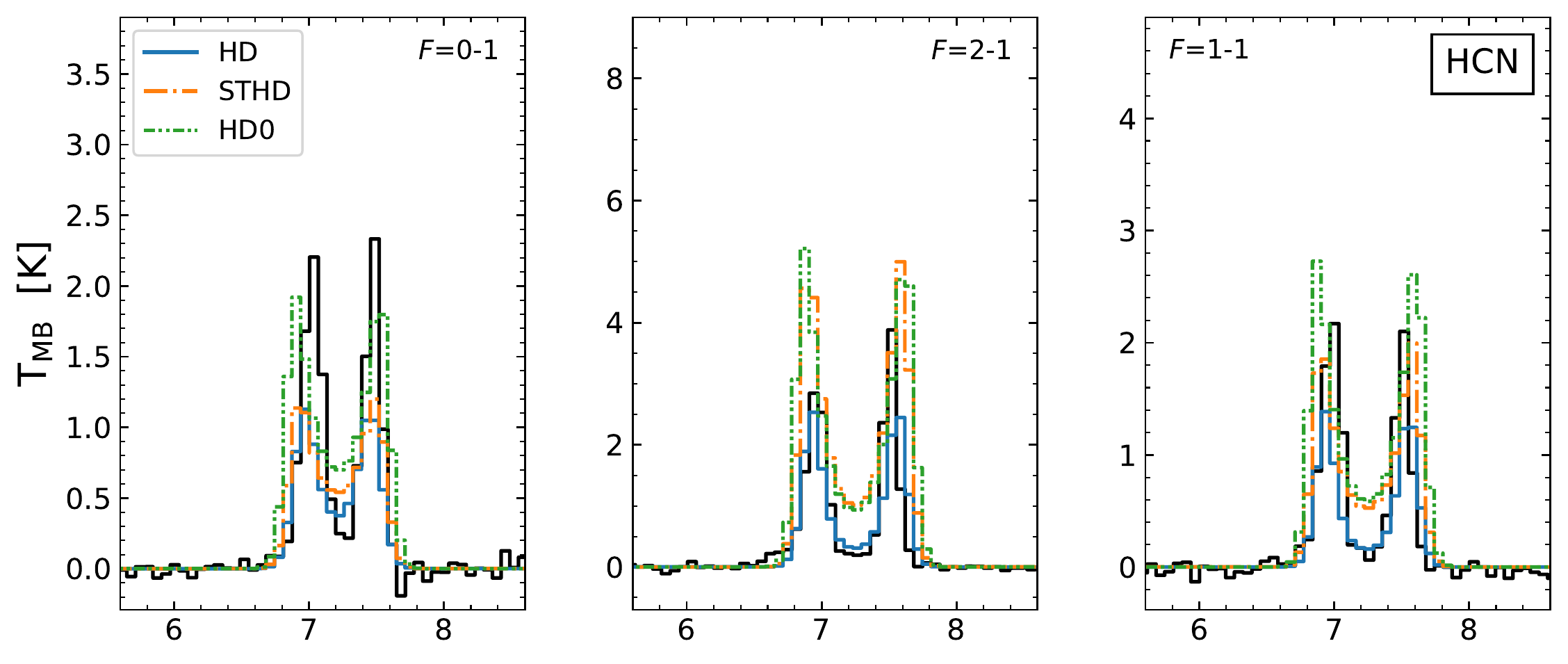}
   \includegraphics[width=0.49\textwidth]{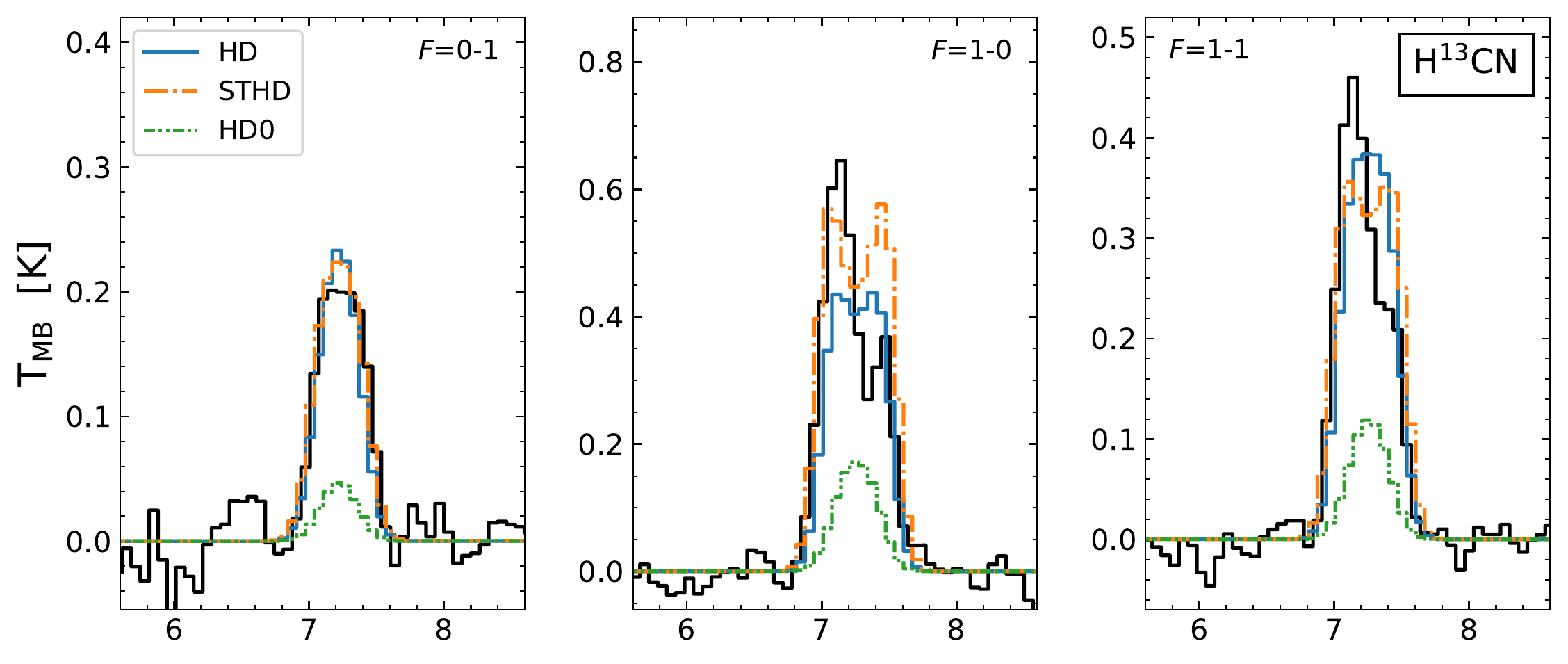}
   \includegraphics[width=0.49\textwidth]{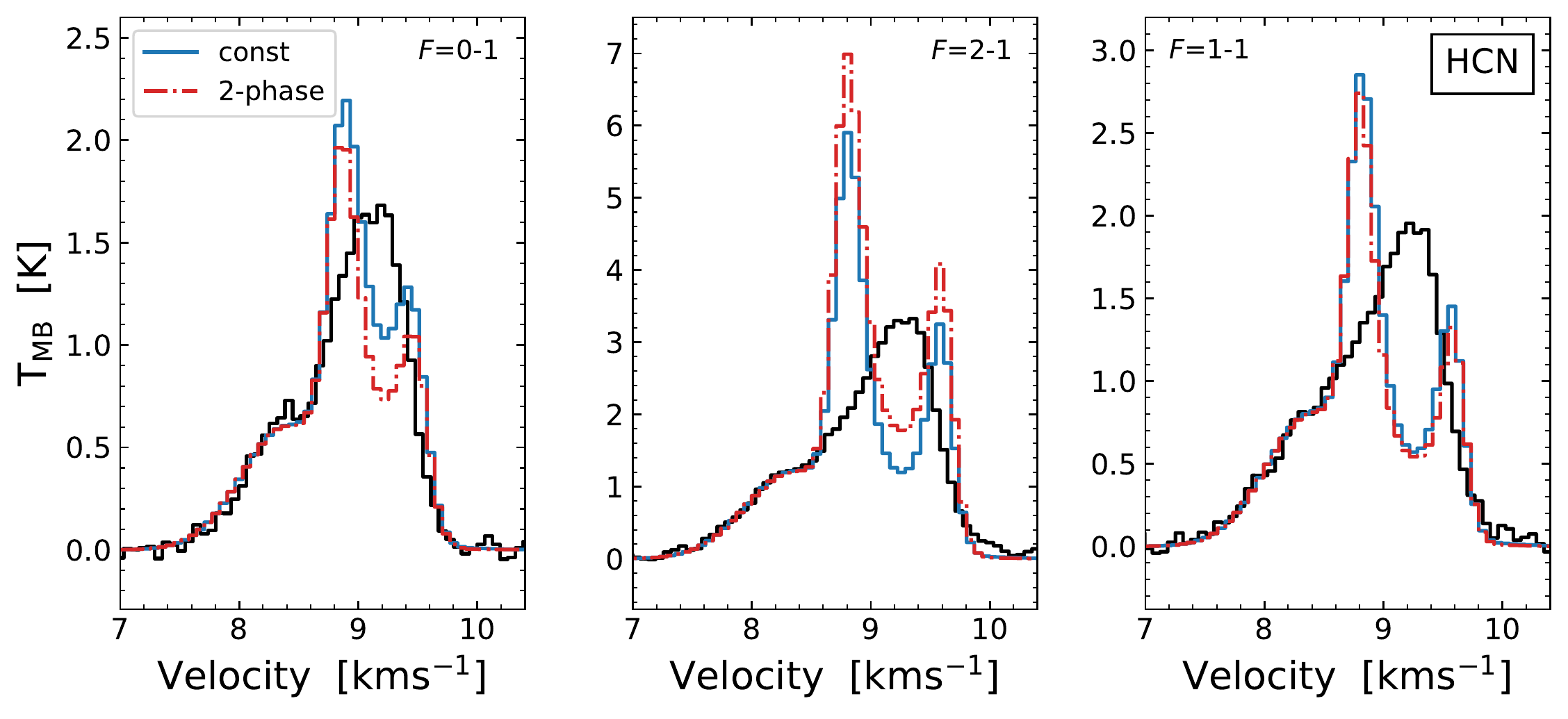}
   \includegraphics[width=0.49\textwidth]{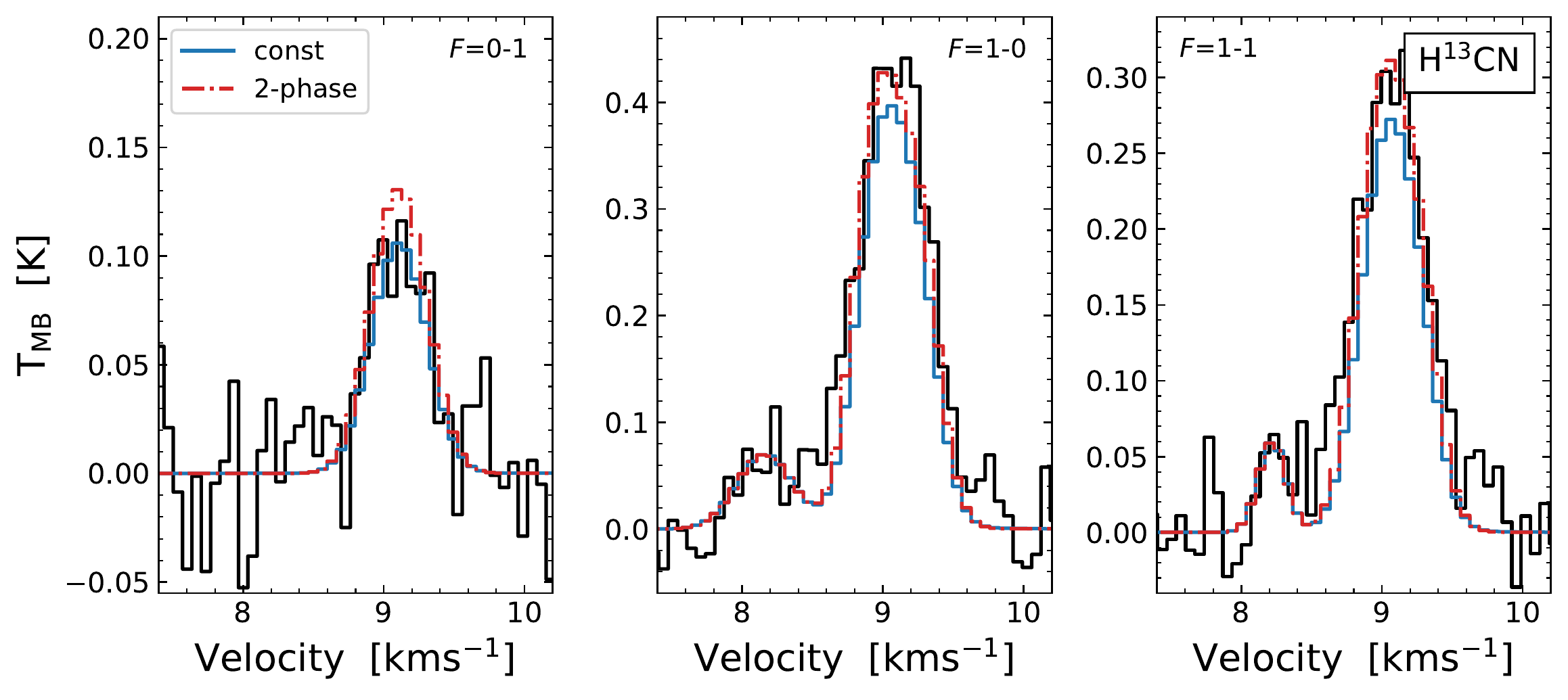}
   \caption{Synthetic spectra of HCN (left) and $\rm H^{13}CN$ (right) obtained with LOC (top rows: L1544, bottom row: HH211). For L1544, the results are splitted between using the Keto-Caselli model (top) and the HDCRT model (centre) as approximation for the physical structure of the core. Towards L1544, the best-fitting abundances and timesteps are $2\times10^{-8}$ (KC), 7.9e5\,yr (STKC), $1\times10^{-8}$ (HD), 1.3e5\,yr (STHD) for HCN; $2\times10^{-10}$ (KC), 6e5\,yr (STKC), $3\times10^{-10}$ (HD), 1e5\,yr (STHD) for $\rm H^{13}CN$. Towards HH211, the best-fitting abundances and timesteps are $2\times10^{-9}$ (const), 1e5\,yr (2-phase) for HCN; $3\times10^{-11}$ (const), 1.3e5\,yr (2-phase) for $\rm H^{13}CN$.}
              \label{Fig:LOC-HCN}
\end{figure*}

\begin{figure}
   \centering
   \includegraphics[width=0.49\textwidth]{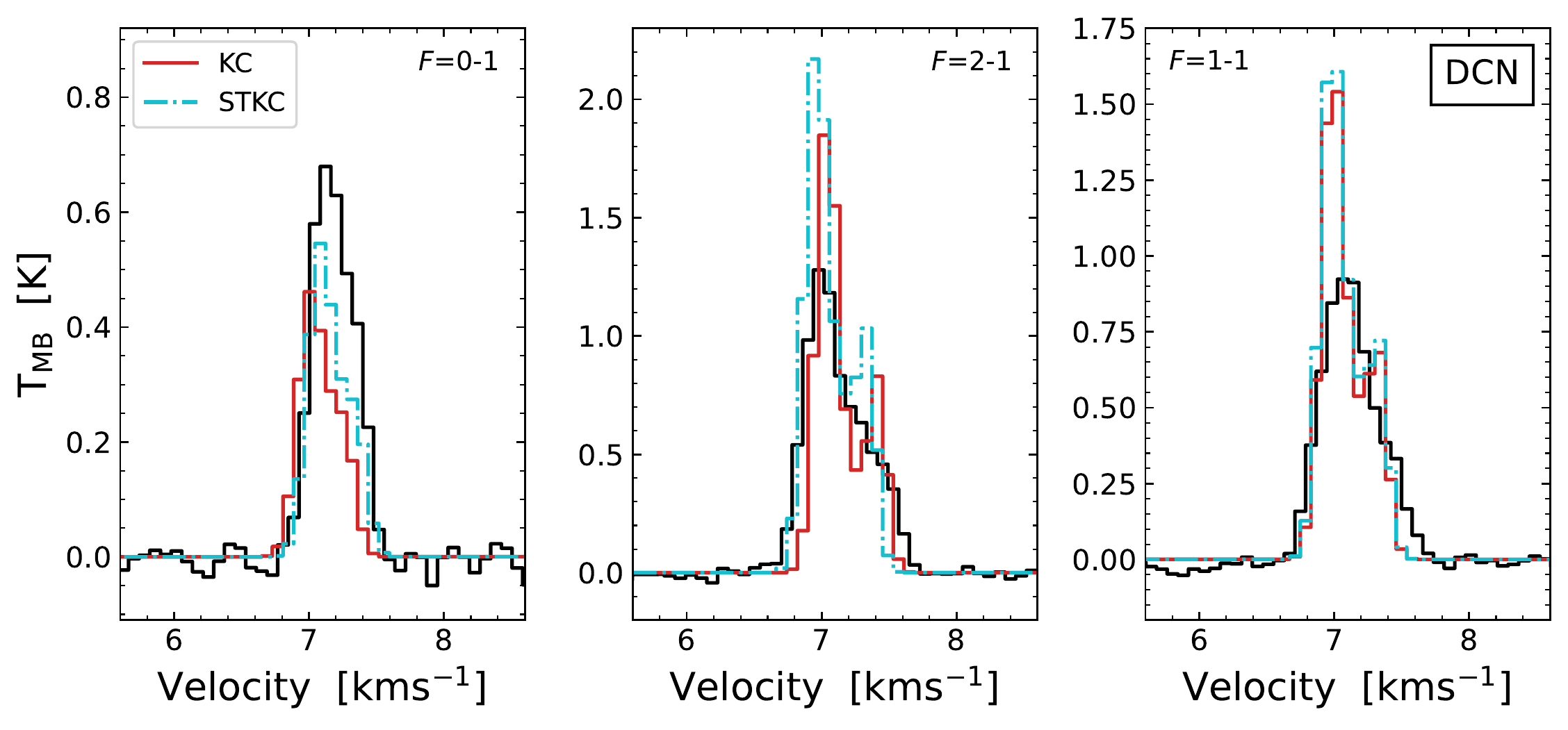}
   \includegraphics[width=0.49\textwidth]{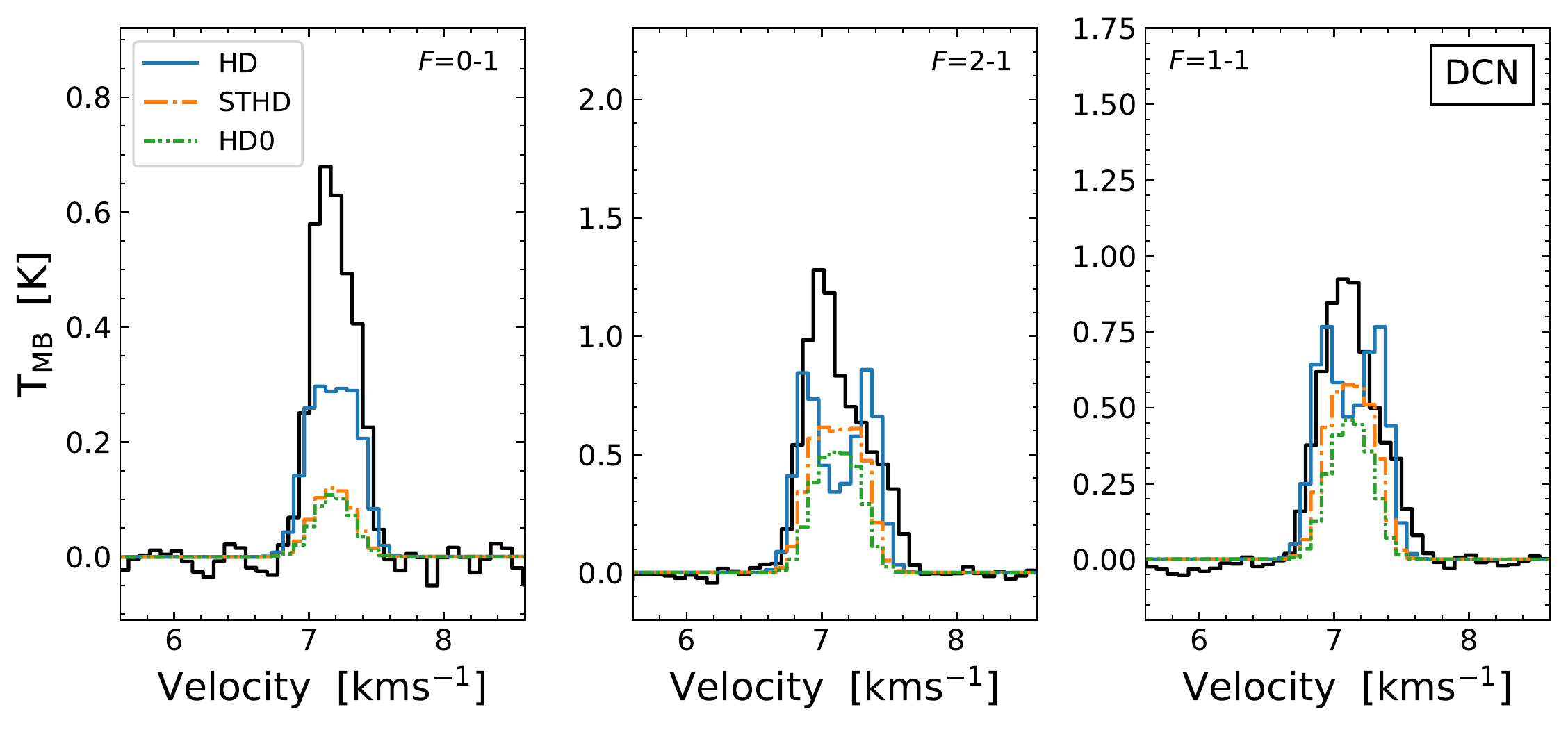}
   \includegraphics[width=0.49\textwidth]{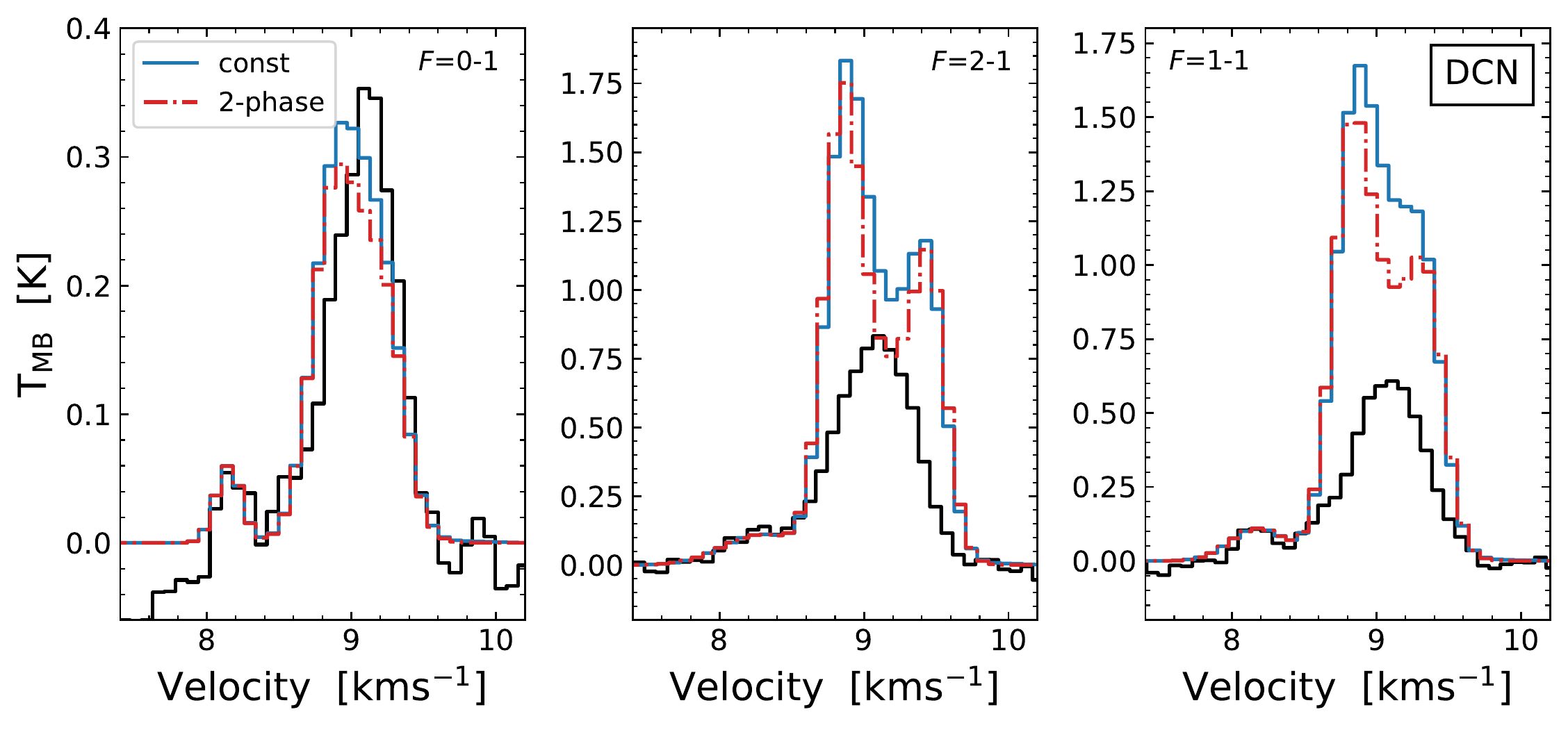}
   \caption{Synthetic spectra of DCN obtained with LOC (top rows: L1544, bottom row: HH211). Towards L1544, the best-fitting abundances and timesteps are $1\times10^{-9}$ (KC), 1.3e5\,yr (STKC), $1\times10^{-9}$ (HD), 9.8e5\,yr (STHD) for DCN. Towards HH211, the best-fitting abundances and timesteps are $4\times10^{-10}$ (const), 1e5\,yr (2-phase) for DCN.}
              \label{Fig:LOC-DCN}
\end{figure}

The synthetic spectra of HCN, \hxcn\,, and DCN are shown in Fig.~\ref{Fig:LOC-HCN} and Fig.~\ref{Fig:LOC-DCN}.
Towards HH211, we are not able to reproduce the observed spectrum of HCN. Applying the derived physical model of the protostellar core produces a self-absorbed line profile with blue asymmetry for HCN that is not observed. 
The $^{13}$C isotopologue, however, can be reproduced with the correct intensities, line shapes and line widths with both the constant  abundance and the abundance profile. 
In the case of DCN, only the weakest hyperfine component is reproduced. The more optically thick hyperfines, however, are overestimated by the model by a factor of two to three. This might indicate that DCN is more extended than what the model covers currently, resulting in more self-absorption than produced by the model. This should be addressed in future studies.

\begin{figure*}
   \centering
   \includegraphics[width=0.99\textwidth]{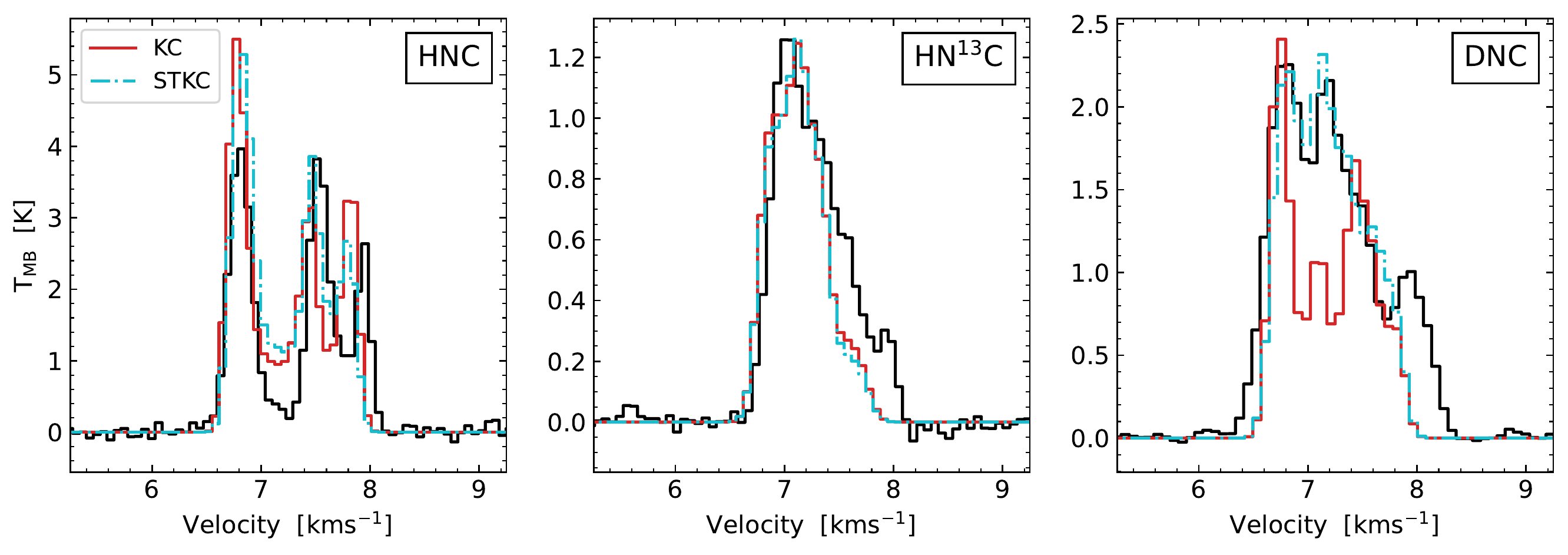}
   \includegraphics[width=0.99\textwidth]{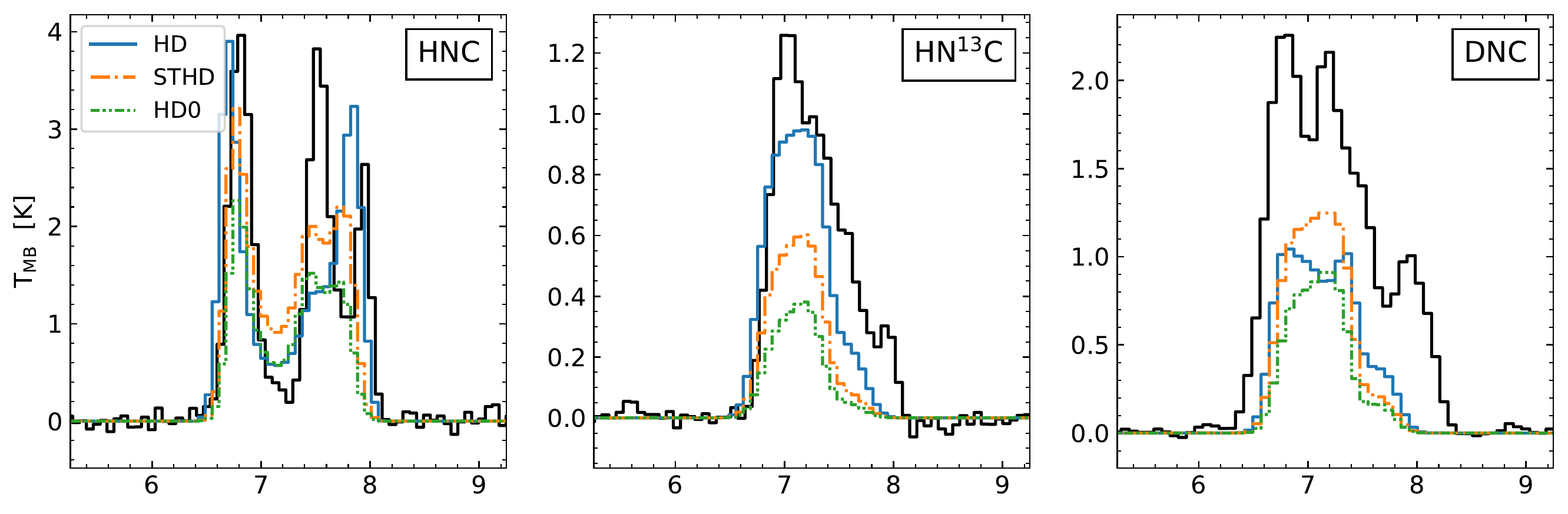}
   \includegraphics[width=0.99\textwidth]{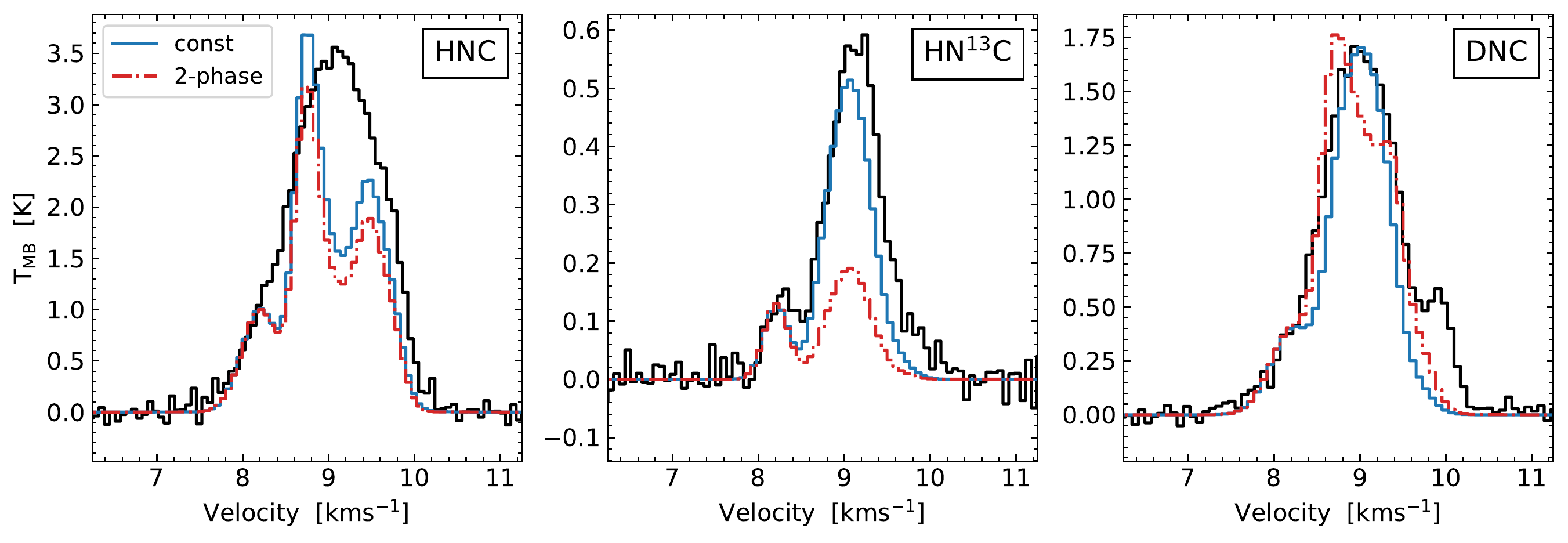}
   \caption{Synthetic spectra of HNC and isotopologues  obtained with LOC (top rows: L1544, bottom row: HH211). For L1544, the results are splitted between using the Keto-Caselli model (top) and the HDCRT model (centre) as approximation for the physical structure of the core. Towards L1544, the best-fitting abundances and timesteps are $2\times10^{-8}$ (KC), 1e5\,yr (STKC), $5\times10^{-8}$ (HD), 1e5\,yr (STHD) for HNC; $2\times10^{-10}$ (KC), 1.3e5\,yr (STKC), $2\times10^{-10}$ (HD), 9.8e5\,yr (STHD) for $\rm HN^{13}C$; $3\times10^{-9}$ (KC), 1e5\,yr (STKC), $8\times10^{-10}$ (HD), 9.8e5\,yr (STHD) for DNC. Towards HH211, the best-fitting abundances and timesteps are $5\times10^{-10}$ (const), 1e5\,yr (2-phase) for HNC $1\times10^{-11}$ (const), 1e5\,yr (2-phase) for $\rm HN^{13}C$; $7\times10^{-11}$ (const), 1e6\,yr (2-phase) for DNC.}
              \label{Fig:LOC-HNC}
\end{figure*}

The synthetic spectra of HNC, \hnxc\,, and DNC  are shown in Fig.~\ref{Fig:LOC-HNC}.
The modelled HNC shows an asymmetric self-absorption that is not observed. Also here, the linewidth towards high velocity is underestimated, similar to HNC in L1544.
Taking into account the missing hyperfine component at high velocity ($\sim$10\,km\,s$^{-1}$), \hnxc\, and DNC are well reproduced by a constant abundance of $1\times10^{-11}$ and $7\times10^{-11}$, respectively. For \hnxc\,, the synthetic spectrum derived with an abundance profile shows that also in this case, the assumption of $\rm^{12}C/^{13}C=68$ is underestimating the amount of \hnxc\, present. For DNC, the use of an abundance profile results in a self-absorbed line profile, most likely due to an overestimation of the molecular abundance in the outer regions.  

\begin{figure*}
   \centering
   \includegraphics[width=0.99\textwidth]{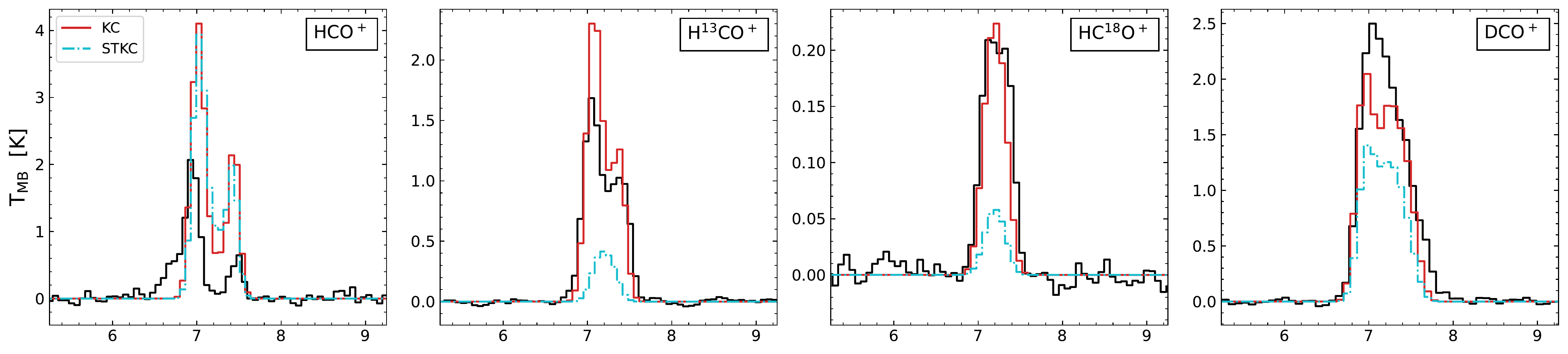}
   \includegraphics[width=0.99\textwidth]{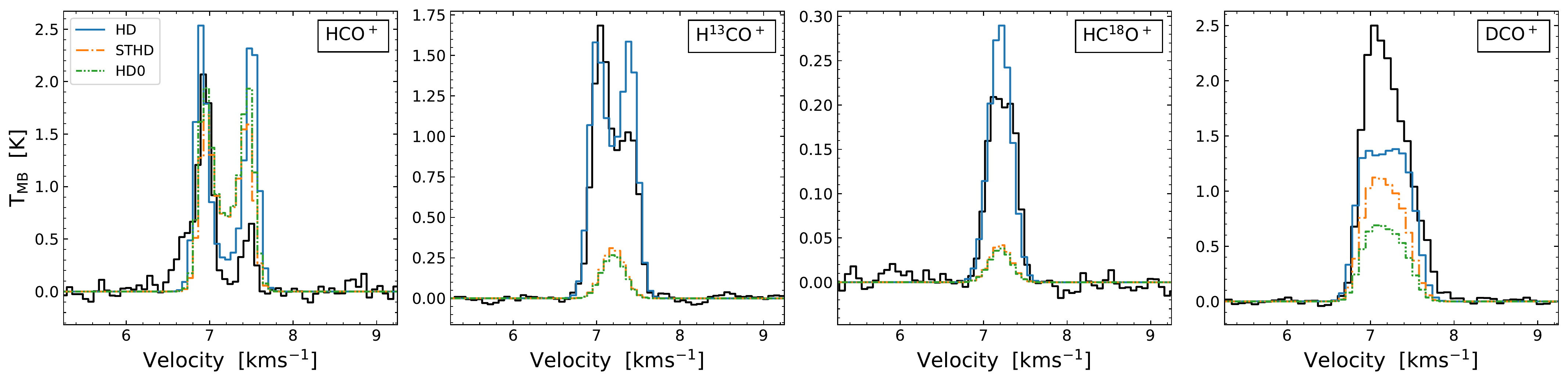}
   \includegraphics[width=0.99\textwidth]{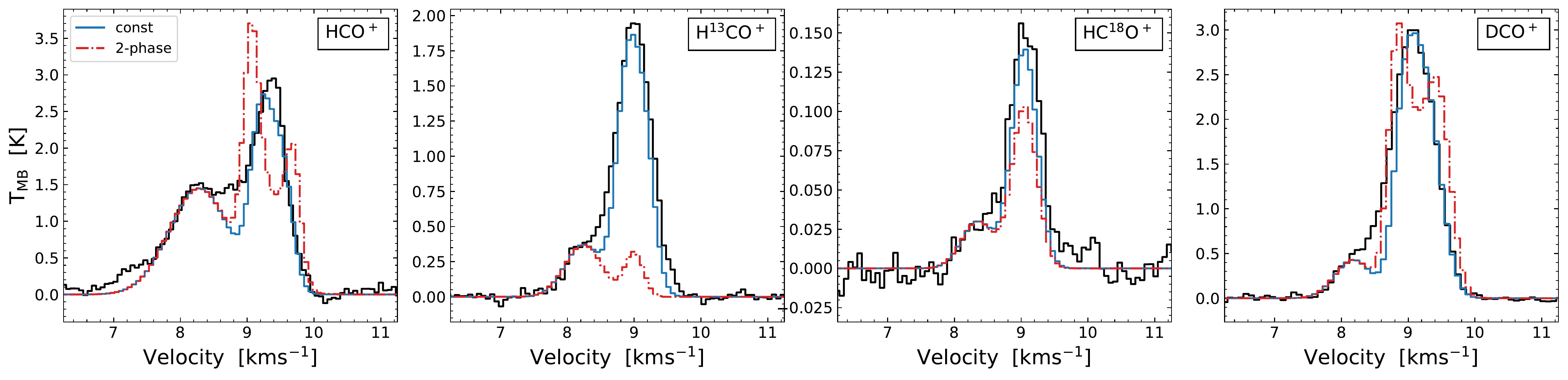}
   \caption{Synthetic spectra of \hcop\, and isotopologues obtained with LOC (top rows: L1544, bottom row: HH211). For L1544, the results are splitted between using the Keto-Caselli model (top) and the HDCRT model (centre) as approximation for the physical structure of the core. Towards L1544, the best-fitting abundances and timesteps are $7\times10^{-10}$ (KC), 9.8e5\,yr (STKC), $3\times10^{-9}$ (HD), 9.8e5\,yr (STHD) for $\rm HCO^+$; $2\times10^{-10}$ (KC), 1e5\,yr (STKC), $4\times10^{-10}$ (HD), 1e5\,yr (STHD) for $\rm H^{13}CO^+$; $6\times10^{-12}$ (KC), 1e5\,yr (STKC), $1\times10^{-11}$ (HD), 1e5\,yr (STHD) for $\rm HC^{18}O^+$; $2\times10^{-10}$ (KC), 1.7e5\,yr (STKC), $4\times10^{-10}$ (HD), 1.7e5\,yr (STHD) for $\rm DCO^+$. Towards HH211, the best-fitting abundances and timesteps are $1\times10^{-11}$ (const), 1e6\,yr (2-phase)  for $\rm HCO^+$; $7\times10^{-12}$ (const), 1e5\,yr (2-phase) for $\rm H^{13}CO^+$; $4\times10^{-13}$ (const), 1e5\,yr (2-phase) for $\rm HC^{18}O^+$; $2\times10^{-11}$ (const), 1e6\,yr (2-phase) for $\rm DCO^+$.}
              \label{Fig:LOC-HCOp}
\end{figure*}

The synthetic spectra of \hcop\, and isotopologues are shown in Fig.~\ref{Fig:LOC-HCOp}.
Towards HH211, all isotopologues of \hcop\, can be reproduced in shape and intensity, using a constant abundance profile. The line shape of the main species can be modelled, however, the constant abundance used to achieve this is of the same magnitude as for \dcop\,. This indicates that the spectrum of \hcop\, is heavily self-absorbed.
When applying an abundance profile, this self-absorption is shown in the synthetic spectrum. The same happens in the case of \dcop\,, indicating that this molecule might be slightly optically thick as well.
The $^{13}$C and $^{18}$O isotopologues cannot be reproduced with the \hcop\, abundance profile scaled down by the respective isotopic ratio. 
This highlights the necessity of new constraints on the isotopic abundance ratios in varying physical conditions.

\begin{table*}[h]
\renewcommand{\arraystretch}{1.2} 
    \centering
    \caption{Summary of the simulation setups discussed in this work.}
    \begin{tabular}{l| l }
    \hline\hline 
     Model & \multicolumn{1}{c}{Description} \\
    \hline
      \multicolumn{2}{c}{L1544}\\\hline
    KC & Constant abundance throughout the core with a drop  to zero in the central 1500\,au, \\
     & applied to the Keto-Caselli model for L1544 \citep{Keto2015} \\\hline
    STKC & Abundance profile derived with a static physical model adopting the parameters and physical structure \\
     & of the Keto-Caselli model for L1544 \\\hline
    HD & Constant abundance throughout the core with a drop to zero in the central 1500\,au, \\
     & applied to the HDCRT model for L1544 \\\hline
    STHD & Abundance profile derived with a static physical model adopting the parameters and physical structure \\
     & of the HDCRT model for L1544 \\\hline
    HD0 & Abundance profile derived from the best match between the Keto-Caselli and the fiducial \\
     & hydrodynamical model presented in \cite{Sipila2022} \\\hline
     \multicolumn{2}{c}{HH211}\\\hline
    const & Constant abundance throughout the core, applied to the physical model for HH211 derived in this work \\\hline
    2-phase & Abundance profile derived with a static physical model adopting the parameters and physical structure \\
     & of the physical model for HH211 derived in this work, applying the 2-phase model of the chemical code \\
     & (= gas phase chemistry and entire ice on the grains are active) \\
    \hline 
    \end{tabular}
    \label{tab:SummaryOfModels}
\end{table*}


\section{Discussion}\label{discussion}

\subsection{Dilution of \texorpdfstring{$^{13}$}C in CCH}
\begin{figure}
   \centering
    \includegraphics[width=\hsize]{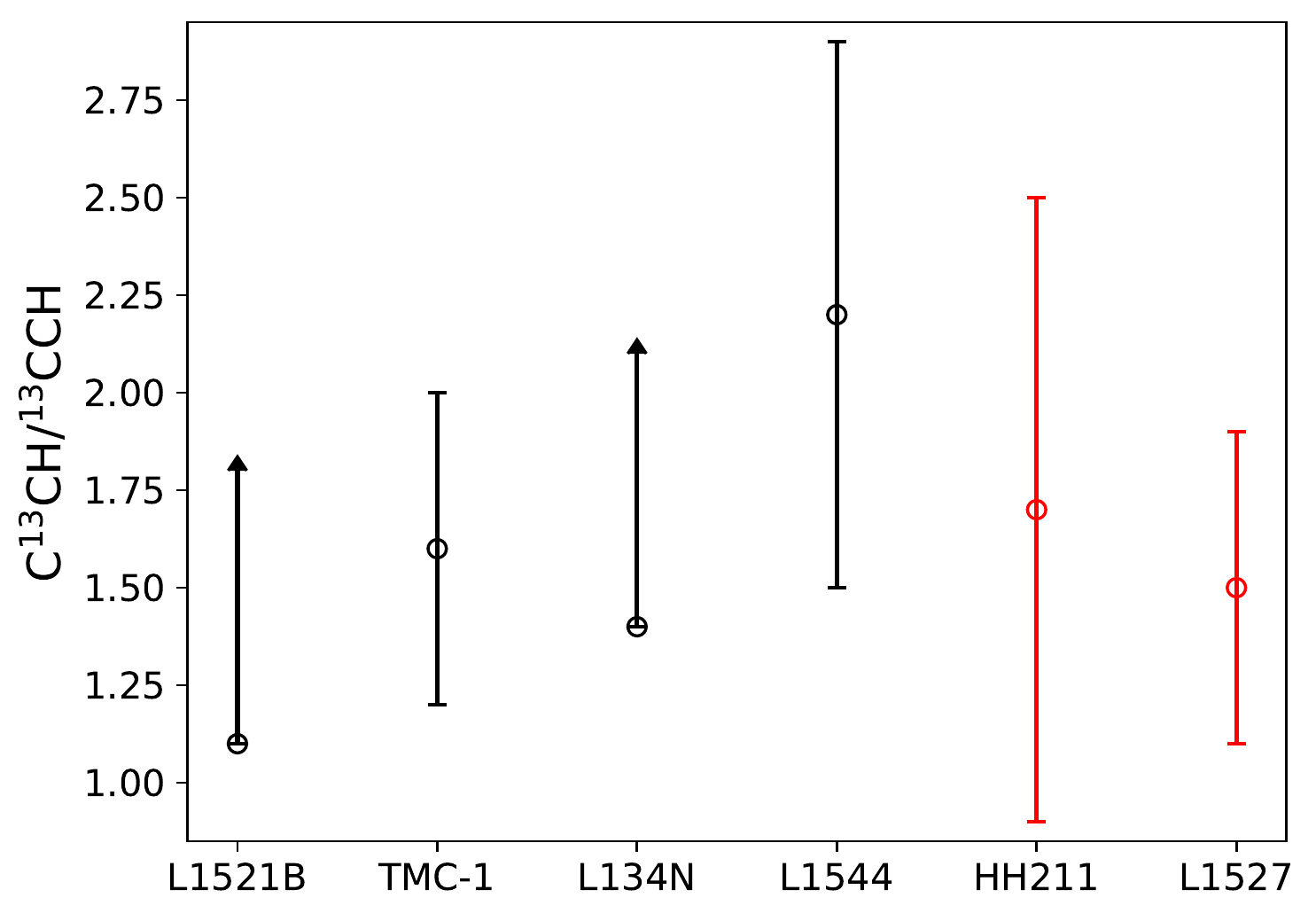}
   \caption{\cxch\,/\xcch\, ratio of starless/pre-stellar (black) and protostellar (red) cores. L1521B: \cite{Taniguchi2019}; L134N: \cite{Taniguchi2019}; TMC-1: \cite{SakaiSaruwatari2010}; L1527: \cite{Yoshida2019}.}
             \label{Fig:Evolution13C}
\end{figure}

We find that \cxch\, is more abundant than \xcch\, in L1544 and in HH211 (see Table~\ref{tab:DeuterationLevel}). This characteristic appears to be common in cold dark clouds, and was observed already in the starless cores TMC-1, L1521B and L134N \citep{Turner2001,SakaiSaruwatari2010,Taniguchi2019}, and in the protostellar cores L1527 and L483 \citep{SakaiSaruwatari2010,Agundez2019}.
\cite{Taniguchi2019} conclude that this difference in abundances is caused during the formation pathway of the molecule and the isotopomer-exchange reaction after it is formed. This reaction is exothermic:
\begin{equation}
    ^{13}\mathrm{C}\mathrm{CH} + \mathrm{H} \rightleftharpoons \mathrm{C}^{13}\mathrm{CH} + \mathrm{H} + 8\,\mathrm{K}\,,
\end{equation}
thus, at low temperatures the forward reaction is more efficient, increasing the abundance of the  $\mathrm{C}^{13}\mathrm{CH}$ isotopomer. 
Chemical models of \cite{Furuya2011} predicted an increase of the \cxch\,/\xcch\, ratio with pre-stellar evolution. 
Literature values of this ratio are collected in Fig.~\ref{Fig:Evolution13C}, which shows the ratios observed towards the starless cores L1521B, TMC-1, L134N, the very evolved pre-stellar core L1544, and the protostellar cores HH211 and L1527. 
However, due to the large uncertainties, no conclusive statement can be made. More observations are needed to confirm any trends.

The carbon isotopic ratios observed for CCH are 
$200\pm50$ and $90\pm20$ (L1544), and $160\pm60$ and $100\pm30$ (HH211) for \xcch\, and \cxch\,, respectively. This is significantly higher than the canonical value for the ISM, 68, and is caused by the dilution of $^{13}$C in carbon-chain molecules in the local ISM.
In dark clouds, $^{13}\mathrm{C}$ is mainly locked in $^{13}\mathrm{CO}$ due to the reaction
\begin{equation}
    \mathrm{CO} + ^{13}\mathrm{C}^+ \longrightarrow ^{13}\mathrm{CO} + \mathrm{C}^+ + 35\,\mathrm{K} \;.
\end{equation}
As this reaction is exothermic, the forward reaction is more efficient at low temperatures. 
Therefore, molecules such as CCH or c-C$_3$H$_2$ that are produced from C$^+$ show a significantly higher $^{12}$C/$^{13}$C ratio in molecular clouds. 
This dilution of $^{13}$C species was predicted by \cite{Langer1984}, and confirmed by previous studies \citep{SakaiSaruwatari2010,Agundez2019,Taniguchi2019,Yoshida2019}. In \cite{Colzi2020}, this behaviour is also predicted for CH.

\subsection{Deuterium fractionation}
Due to the dilution of the $^{13}$C isotopologues of CCH, the derived deuterium fractions have to be interpreted as upper limits, as the applied $^{12}$C/$^{13}$C ratio of 68 is underestimating the real value. However, as it was possible to derive the column density of CCH using emission of the main species directly, we use these results for a comparison of the deuteration in the two cores.
CCH shows a high and similar level of deuteration in both cores ($\approx$10\%) and is therefore consistent with other carbon chains such as c-C$_3$H$_2$ \citep[e.g.][]{Chantzos2018}.
Measurements towards other cores report lower deuteration levels (L183: 0.06(4), \citealt{Turner2001}; TMC-1: 0.05(2), \citealt{Turner2001}; L1527: 0.04(1), \citealt{Yoshida2019}). 
Evolutionary and environmental effects might play a role in this difference, as well as differences in the excitation temperatures used to derive the column densities. 

The deuterium fraction of HCN appears to be slightly larger in HH211 with respect to L1544. However, due to the uncertainties in the carbon isotopic ratio of HCN, these numbers have to be interpreted with caution. 
Therefore, no conclusive statement can be made based on our errorbars, and further studies are necessary to eventually confirm this trend. 
The D/H ratio for HCN of the cold dark cloud L183 \citep[0.05(2),][]{Turner2001} is consistent with our measurements towards L1544.
\cite{Roberts2002} derived slightly lower values in HH211, N(DCN)/N(HCN$)=0.038(8)$, which can be explained by different approximations used in the calculations.
The measurements towards the protostellar core L1527 are consistent with our observations \citep[0.05(1),][]{Yoshida2019}.

For HNC, we derive a deuterium fraction of 0.08(2) for both L1544 and HH211. 
Previous measurements of the deuteration in the two cores show moderate deuterium fractions (L1544: 0.03, \citealt{Hirota2003}; HH211: 0.07, \citealt{Imai2018}). However, the authors considered the emission of \hnxc\, and DNC to be optically thin and apply simple one-component Gaussian fits to obtain the line parameters.
Especially in the case of DNC this can lead to an underestimation of the column density, and subsequently of the deuterium fraction.
Observations towards other cores show moderate deuterium fractions (L183: 0.05(2), \citealt{Turner2001}; L1527: 0.045, \citealt{Yoshida2019}). 

The deuterium fractions of \hcop\, derived towards L1544 and HH211 are consistent with previous measurements in similar objects \citep[see e.g.][]{Turner2001,Jorgensen2004,Koumpia2017,Yoshida2019}.
For L1544, the D/H ratios we derived using \hxcop\, and \hcxop\, are in agreement with previous observations by \cite{Caselli2002a} and \cite{Redaelli2019}, who report a ratio of around 4\%.

Towards HH211, the result derived using the $^{18}$O isotopologue is increased by a factor of two compared to the D/H ratio derived from \hxcop\,. This is most likely caused by an overestimation of the elemental carbon isotopic ratio. Chemical models predict that the \hcop\,/\hxcop\, ratio is actually lower than the canonical $^{12}$C/$^{13}$C ratio \citep{Colzi2020}, because of isotopic exchange reactions that are important at low temperatures.
The canonical ratio of $^{13}$C/$^{18}$O is 8.2, whereas our observations give 10(1) and 20(4) for L1544 and HH211, respectively. This shows that especially towards HH211, the carbon fractionation of \hcop\, is higher than in the local ISM, resulting in HCO$^+$/H$^{13}$CO$^+<68$. Hence, by using 68, the deuterium fraction derived from \hxcop\, is underestimated.

The comparison of the D/H ratio derived from \hcxop\, in the two cores shows that the ratio towards the protostellar core is higher by a factor of almost two. 
This difference is caused by some effect that is either increasing or decreasing the observed column densities of the isotopologues. 
The latter one could be an effect of isotope-selective photodissociation which is impacting the $^{18}$O  more than the D isotopologue. 
The higher abundance of CO leads to a stronger self-shielding of the molecule compared to $\rm C^{18}O$. This results in a local enhancement of CO/C$^{18}$O, that is likely larger in HH211 than in L1544. This is reflected in an increased D/H ratio towards HH211, as \dcop\, is created directly from CO, and therefore it is more efficiently formed than \hcxop\,.


In conclusion, we do not see a general trend in the deuteration efficiency in simple molecules going from a pre-stellar to a protostellar core, as it was indicated previously by observations of c-C$_3$H$_2$ \citep{Chantzos2018}. 
This might be due to a more efficient deuteration of c-C$_3$H$_2$ happening on the surface of the dust grains in the pre-stellar phase that is not applicable for the simple molecules studied in this work.
Additionally, the fact that the deuterium fraction is similar in two cores in different environments seems to suggest that the deuterium fractionation is not sensitive to the initial conditions of its surroundings. Similar results have been found for N$_2$H$^+$ \citep{Crapsi2005,Emprechtinger2009} and c-C$_3$H$_2$ \citep{Chantzos2018}.
Nevertheless, more comprehensive surveys including multiple pre-stellar and protostellar cores and covering more molecules are necessary to provide larger statistics and confirm any trends.

\subsection{Radiative transfer modelling}

\subsubsection{L1544}
Our radiative transfer simulations show that the Keto-Caselli model works fine to explain moderately optically thin lines such as \hxcn\,, \hnxc\,, DNC, and \hxcop\, that trace regions with moderate to high density (few 10$^5$\,cm$^{-3}$).
The model shows problems when it comes to optically thin lines like \hcxop\,, where the line shape is affected by the freeze-out of CO in addition to contraction motions.
In the case of CCH, the model fails to reproduce the line shapes of the optically thin hf components, which show a red asymmetry. As the lines are optically thin, they do not trace any expansion motion like it was seen for HCN. Instead, this is likely an effect of the asymmetric distribution of the molecule across the core. In L1544, carbon chains are known to peak towards the south-east of the core \citep{Spezzano2017,Giers2022}. Due to the velocity gradient across the core \citep{Spezzano2016b}, the spectra observed towards the south are redshifted with respect to the rest velocity of the system. As CCH is most probably peaking in the south, where the gas motion is redshifted, the contribution of the redshifted lines is more prominent, resulting in an asymmetric line shape. This behaviour is also observed for cyanopolyynes by \cite{Bianchi2023}. As we assume a spherically symmetric core in the simulations, and therefore a symmetric distribution of the molecule, this line shape is not reproducible.

The HDCRT model, on the other hand, struggles to explain the blue asymmetry observed in the moderately optically thin lines.
This is most likely caused by an overestimation of the molecular abundances in the outer regions, resulting in symmetric double-peak profiles that trace more static rather than infalling layers.

In most cases, the simulations fail to reproduce the observed intensities of the rarer isotopologues when applying the abundance profiles of the main isotopologues scaled by the respective isotope ratio. 
This highlights the importance of considering the carbon and oxygen isotope chemistry in the chemical modelling, which will be addressed in future work.

In the case of the optically thick and more widespread main isotopologues, the results are better reproduced when applying the HDCRT model, while the Keto-Caselli model struggles to explain the lines. 
The only exception here is HNC, where the Keto-Caselli model is sufficient to reproduce the observed intensities. This might be an indication that HNC is actually tracing slightly different layers compared to HCN. 
In the case of CCH, the HDCRT model works fine for the optically thick components that trace a static inbetween layer.

In the case of HCN, the radiative transfer modelling in this work showed that the Keto-Caselli model is not sufficient to explain the line shapes and intensities of the molecule's hyperfine components observed towards L1544. 
Instead, it is necessary to use a physical model that shows expansion motions in the outer layers to reproduce the red asymmetry in the central hyperfine component of HCN.
Contrary to this, \cite{Redaelli2022} show that for the blue asymmetry and strong self-absorption observed in \hcop\,, it is necessary to add an extended low-density and contracting envelope to the Keto-Caselli model.
One possibility to solve this discrepancy could be that HCN is tracing an even outer layer than \hcop\,. As HCN can be efficiently formed also without CO, the layers where this process occurs could be warm enough to be associated with expansion motions.

\subsubsection{HH211}

The physical model derived in this work struggles to explain the very optically thick main isotopologues (HCN, HNC, \hcop\,). 
Both a constant abundance across the core and the abundance profiles predict high molecular abundances in the outer regions of the core. 
This leads to a strong self-absorption in the synthetic spectra that is not observed. 
To get a more realistic estimate of the molecular abundances across the core, it would be necessary to run more tests of the physical conditions and the parameter space used in the chemical modelling. However, this is beyond the scope of this paper.

Furthermore, very abundant and widespread molecules like \hcop\,, HCN, and possibly also HNC also trace the outflow of the protostar and the extended structures of the core's environment. 
These kind of large-scale structures are not considered in the simplistic and spherically symmetric physical model derived in this work and will likely be different for the sources studied here. L1544 is located in an isolated region in Taurus. In contrast, HH211 is forming in an active region within Perseus, with low density material traced by $^{13}$CO spreading over several km\,s$^{-1}$ around HH211 \citep{Sun2006}. The complex and turbulent environment towards HH211 is reflected in the line profiles for the more abundant species. For instance, our observations detect a second velocity component blueshifted by 1\,km\,s$^{-1}$, which is associated with large-scale emission extending to the north-east of HH211 \citep{Sun2006}.
To correctly model the observed line shapes of molecular tracers like \hcop\,, HCN, and HNC, these structures have to be taken into account in the density, velocity, and temperature profiles of the core.

On the other hand, we are able to reproduce all optically thin and moderately optically thick molecular lines observed towards the protostellar core. For these spectra, the simplifying assumptions of spherical symmetry and constant abundance throughout the core seem to be sufficient to reproduce the molecular emission. This is most likely due to the fact that these molecules are less abundant and trace higher-density regions, and are therefore less affected by the kinematics of the large-scale structures surrounding the core as well as the extended outflow emission.

\section{Conclusions}\label{conclusion}
We presented a survey of ground-state rotational lines of simple molecules, which allow us to compare the levels of deuteration in the very evolved pre-stellar core L1544 and the very young protostellar core HH211. 
In a non-LTE approach, we used radiative transfer simulations and molecular abundance profiles derived from chemical modelling to reproduce the observed molecular spectra.
By applying new hyperfine-resolved collisional rate coefficients, we take into account the specific spectroscopy of the respective isotopologues.

Our main results can be summarised as follows:
\begin{itemize}
    \item The similar levels of deuteration show that the deuterium fractionation seems to be equally efficient towards both cores. We do not see a general trend in the level of deuteration when going from a pre-stellar core to a protostellar core, suggesting that the protostellar envelope still retains the chemical composition of the original pre-stellar core. In addition, the deuterium fraction seems to be independent of the initial conditions present in the molecular clouds where the two cores are embedded. 
    \item Towards HH211, the D/H ratio of \hcop\, derived using the $^{18}$O isotopologue is higher by a factor of two compared to L1544. This is likely an effect of isotope-selective photodissociation, creating a local enhancement of CO/C$^{18}$O in the protostellar core.
    \item The $^{13}$C dilution of carbon-chain species in dark clouds leads to an increased $^{12}$C/$^{13}$C ratio for CCH. On the other hand, the increased $^{13}$C/$^{18}$O ratio of \hcop\, indicates a decreased $^{12}$C/$^{13}$C ratio for the molecule, caused by the isotopic exchange reaction being more effective at low temperatures. This highlights the uncertainties when dealing with $^{13}$C isotopologues and the influence of the applied carbon isotopic ratio.
    \item The central hyperfine component of HCN (1-0) observed towards L1544 shows a red asymmetry, tracing expansion motions caused by external heating in the outer layers of the core. 
    This stands in contrast to \hcop\, that has been shown to trace infall motion in the outer layers and low-density envelope of L1544. This discrepancy could be explained by HCN tracing an even outer layer than \hcop\,. 
    \item The radiative transfer modelling showed that the new collisional rate coefficients for \hnxc\, and DNC are incomplete because they only consider the splitting by $^{14}$N and therefore miss a part of the hyperfine structure. This issue is addressed by work underway that includes the effect of $^{13}$C and D.
    \item The radiative transfer modelling of HH211 using the physical structure derived in this work is successful for optically thin emission lines, but shows problems with optically thick lines that also trace more complex structures and are influenced by them.
    \item The radiative transfer modelling of L1544 applying the Keto-Caselli model works well for moderately optically thin lines that trace inner layers of the core, but shows problems with optically thin lines influenced by CO depletion (\hcxop\,) and asymmetric distribution across the core (CCH).
    \item The radiative transfer modelling of L1544 applying the HDCRT model is successful at reproducing the optically thick hyperfine components of CCH, which show symmetric double-peak profiles, and the red asymmetry observed in the central hyperfine component of HCN (1-0).
    \item The modelling results of both cores show that to correctly model emission lines, it is crucial to include the outer layers of the cores to consider the effects of extended structures.
\end{itemize}


Future projects with more comprehensive surveys, including multiple pre-stellar and protostellar cores and covering more molecules, will be able to provide larger statistics and confirm the trends reported in this work. 
New detailed chemical models including the time-dependent variations in $^{12}$C/$^{13}$C ratios (\citealt{Colzi2020}, Sipilä et al. subm.) will aid in further constraining the lines of C-containing molecules, including isotopologues.
Additionally, modelling of molecules like CS or H$_2$CO will help to further constrain the diffuse envelope surrounding the pre-stellar core L1544.

\begin{acknowledgements}
    We wish to thank the anonymous referee for their constructive comments.
    K.G. thanks Maria Jos\'{e} Maureira for useful discussions. S.S. and K.G. wish to thank the Max Planck Society for the Max Planck Research Group funding. All others authors affiliated to the MPE wish to thank the Max Planck Society for financial support. The Onsala Space Observatory national research infrastructure is funded through Swedish Research Council grant No 2017-00648. C.T.B. and F.L. acknowledge financial support from the European Research Council (Consolidator Grant COLLEXISM, Grant No. 811363). 
\end{acknowledgements}

%
\bibliographystyle{aa} 
\bibliography{mybib.bib} 

\begin{thebibliography}{119}
\expandafter\ifx\csname natexlab\endcsname\relax\def\natexlab#1{#1}\fi

\bibitem[{{Ag{\'u}ndez} {et~al.}(2019){Ag{\'u}ndez}, {Marcelino}, {Cernicharo},
  {Roueff}, \& {Tafalla}}]{Agundez2019}
{Ag{\'u}ndez}, M., {Marcelino}, N., {Cernicharo}, J., {Roueff}, E., \&
  {Tafalla}, M. 2019, \aap, 625, A147

\bibitem[{{Ahrens} {et~al.}(2002){Ahrens}, {Lewen}, {Takano}, {Winnewisser},
  {Urban}, {Negirev}, \& {Koroliev}}]{Ahrens2002}
{Ahrens}, V., {Lewen}, F., {Takano}, S., {et~al.} 2002, Zeitschrift
  Naturforschung Teil A, 57, 669

\bibitem[{Alexander \& Dagdigian(1985)}]{alexander1985collision}
Alexander, M.~H. \& Dagdigian, P.~J. 1985, The Journal of chemical physics, 83,
  2191

\bibitem[{{Altwegg} {et~al.}(2015){Altwegg}, {Balsiger}, {Bar-Nun},
  {Berthelier}, {Bieler}, {Bochsler}, {Briois}, {Calmonte}, {Combi}, {De
  Keyser}, {Eberhardt}, {Fiethe}, {Fuselier}, {Gasc}, {Gombosi}, {Hansen},
  {H{\"a}ssig}, {J{\"a}ckel}, {Kopp}, {Korth}, {LeRoy}, {Mall}, {Marty},
  {Mousis}, {Neefs}, {Owen}, {R{\`e}me}, {Rubin}, {S{\'e}mon}, {Tzou}, {Waite},
  \& {Wurz}}]{Altwegg2015}
{Altwegg}, K., {Balsiger}, H., {Bar-Nun}, A., {et~al.} 2015, Science, 347,
  1261952

\bibitem[{{Andr{\'e}} {et~al.}(2016){Andr{\'e}}, {Rev{\'e}ret}, {K{\"o}nyves},
  {Arzoumanian}, {Tig{\'e}}, {Gallais}, {Roussel}, {Le Pennec}, {Rodriguez},
  {Doumayrou}, {Dubreuil}, {Lortholary}, {Martignac}, {Talvard}, {Delisle},
  {Visticot}, {Dumaye}, {De Breuck}, {Shimajiri}, {Motte}, {Bontemps},
  {Hennemann}, {Zavagno}, {Russeil}, {Schneider}, {Palmeirim}, {Peretto},
  {Hill}, {Minier}, {Roy}, \& {Rygl}}]{Andre2016}
{Andr{\'e}}, P., {Rev{\'e}ret}, V., {K{\"o}nyves}, V., {et~al.} 2016, \aap,
  592, A54

\bibitem[{{Arce} \& {Sargent}(2004)}]{Arce2004}
{Arce}, H.~G. \& {Sargent}, A.~I. 2004, \apj, 612, 342

\bibitem[{{Arzoumanian} {et~al.}(2011){Arzoumanian}, {Andr{\'e}}, {Didelon},
  {K{\"o}nyves}, {Schneider}, {Men'shchikov}, {Sousbie}, {Zavagno}, {Bontemps},
  {di Francesco}, {Griffin}, {Hennemann}, {Hill}, {Kirk}, {Martin}, {Minier},
  {Molinari}, {Motte}, {Peretto}, {Pezzuto}, {Spinoglio}, {Ward-Thompson},
  {White}, \& {Wilson}}]{Arzoumanian2011}
{Arzoumanian}, D., {Andr{\'e}}, P., {Didelon}, P., {et~al.} 2011, \aap, 529, L6

\bibitem[{{Belitsky} {et~al.}(2015){Belitsky}, {Lapkin}, {Fredrixon}, {Sundin},
  {Helldner}, {Pettersson}, {Ferm}, {Pantaleev}, {Billade}, {Bergman},
  {Olofsson}, {Lerner}, {Strandberg}, {Whale}, {Pavolotsky}, {Flygare},
  {Olofsson}, \& {Conway}}]{OSO3mm}
{Belitsky}, V., {Lapkin}, I., {Fredrixon}, M., {et~al.} 2015, \aap, 580, A29

\bibitem[{{Bergin} \& {Tafalla}(2007)}]{BerginTafalla2007}
{Bergin}, E.~A. \& {Tafalla}, M. 2007, \araa, 45, 339

\bibitem[{{Bianchi} {et~al.}(2023){Bianchi}, {Remijan}, {Codella},
  {Ceccarelli}, {Lique}, {Spezzano}, {Balucani}, {Caselli}, {Herbst}, {Podio},
  {Vastel}, \& {McGuire}}]{Bianchi2023}
{Bianchi}, E., {Remijan}, A., {Codella}, C., {et~al.} 2023, \apj, 944, 208

\bibitem[{{Br{\"u}nken} {et~al.}(2004){Br{\"u}nken}, {Fuchs}, {Lewen}, {Urban},
  {Giesen}, \& {Winnewisser}}]{Brunken2004}
{Br{\"u}nken}, S., {Fuchs}, U., {Lewen}, F., {et~al.} 2004, Journal of
  Molecular Spectroscopy, 225, 152

\bibitem[{{Busemann} {et~al.}(2006){Busemann}, {Young}, {O'D. Alexander},
  {Hoppe}, {Mukhopadhyay}, \& {Nittler}}]{Busemann2006}
{Busemann}, H., {Young}, A.~F., {O'D. Alexander}, C.~M., {et~al.} 2006,
  Science, 312, 727

\bibitem[{{Cabezas} {et~al.}(2021){Cabezas}, {Ag{\'u}ndez}, {Marcelino},
  {Tercero}, {Cuadrado}, \& {Cernicharo}}]{Cabezas2021}
{Cabezas}, C., {Ag{\'u}ndez}, M., {Marcelino}, N., {et~al.} 2021, \aap, 654,
  A45

\bibitem[{{Caselli} {et~al.}(2012){Caselli}, {Keto}, {Bergin}, {Tafalla},
  {Aikawa}, {Douglas}, {Pagani}, {Y{\'\i}ld{\'\i}z}, {van der Tak}, {Walmsley},
  {Codella}, {Nisini}, {Kristensen}, \& {van Dishoeck}}]{Caselli2012}
{Caselli}, P., {Keto}, E., {Bergin}, E.~A., {et~al.} 2012, \apjl, 759, L37

\bibitem[{{Caselli} {et~al.}(2022){Caselli}, {Pineda}, {Sipil{\"a}}, {Zhao},
  {Redaelli}, {Spezzano}, {Maureira}, {Alves}, {Bizzocchi}, {Bourke},
  {Chac{\'o}n-Tanarro}, {Friesen}, {Galli}, {Harju}, {Jim{\'e}nez-Serra},
  {Keto}, {Li}, {Padovani}, {Schmiedeke}, {Tafalla}, \& {Vastel}}]{Caselli2022}
{Caselli}, P., {Pineda}, J.~E., {Sipil{\"a}}, O., {et~al.} 2022, \apj, 929, 13

\bibitem[{{Caselli} {et~al.}(2019){Caselli}, {Pineda}, {Zhao}, {Walmsley},
  {Keto}, {Tafalla}, {Chac{\'o}n-Tanarro}, {Bourke}, {Friesen}, {Galli}, \&
  {Padovani}}]{Caselli2019}
{Caselli}, P., {Pineda}, J.~E., {Zhao}, B., {et~al.} 2019, \apj, 874, 89

\bibitem[{{Caselli} {et~al.}(1999){Caselli}, {Walmsley}, {Tafalla}, {Dore}, \&
  {Myers}}]{Caselli1999}
{Caselli}, P., {Walmsley}, C.~M., {Tafalla}, M., {Dore}, L., \& {Myers}, P.~C.
  1999, \apjl, 523, L165

\bibitem[{{Caselli} {et~al.}(2002{\natexlab{a}}){Caselli}, {Walmsley},
  {Zucconi}, {Tafalla}, {Dore}, \& {Myers}}]{Caselli2002a}
{Caselli}, P., {Walmsley}, C.~M., {Zucconi}, A., {et~al.} 2002{\natexlab{a}},
  \apj, 565, 331

\bibitem[{{Caselli} {et~al.}(2002{\natexlab{b}}){Caselli}, {Walmsley},
  {Zucconi}, {Tafalla}, {Dore}, \& {Myers}}]{Caselli2002b}
{Caselli}, P., {Walmsley}, C.~M., {Zucconi}, A., {et~al.} 2002{\natexlab{b}},
  \apj, 565, 344

\bibitem[{{Ceccarelli} {et~al.}(2014){Ceccarelli}, {Caselli},
  {Bockel{\'e}e-Morvan}, {Mousis}, {Pizzarello}, {Robert}, \&
  {Semenov}}]{Ceccarelli2014}
{Ceccarelli}, C., {Caselli}, P., {Bockel{\'e}e-Morvan}, D., {et~al.} 2014, in
  Protostars and Planets VI, ed. H.~{Beuther}, R.~S. {Klessen}, C.~P.
  {Dullemond}, \& T.~{Henning}, 859

\bibitem[{{Chac{\'o}n-Tanarro} {et~al.}(2019){Chac{\'o}n-Tanarro}, {Caselli},
  {Bizzocchi}, {Pineda}, {Sipil{\"a}}, {Vasyunin}, {Spezzano}, {Punanova},
  {Giuliano}, \& {Lattanzi}}]{ChaconTanarro2019}
{Chac{\'o}n-Tanarro}, A., {Caselli}, P., {Bizzocchi}, L., {et~al.} 2019, \aap,
  622, A141

\bibitem[{{Chantzos} {et~al.}(2018){Chantzos}, {Spezzano}, {Caselli},
  {Chac{\'o}n-Tanarro}, {Bizzocchi}, {Sipil{\"a}}, \&
  {Giuliano}}]{Chantzos2018}
{Chantzos}, J., {Spezzano}, S., {Caselli}, P., {et~al.} 2018, \apj, 863, 126

\bibitem[{{Cleeves} {et~al.}(2014){Cleeves}, {Bergin}, {Alexander}, {Du},
  {Graninger}, {{\"O}berg}, \& {Harries}}]{Cleeves2014}
{Cleeves}, L.~I., {Bergin}, E.~A., {Alexander}, C. M.~O.~D., {et~al.} 2014,
  Science, 345, 1590

\bibitem[{{Colzi} {et~al.}(2020){Colzi}, {Sipil{\"a}}, {Roueff}, {Caselli}, \&
  {Fontani}}]{Colzi2020}
{Colzi}, L., {Sipil{\"a}}, O., {Roueff}, E., {Caselli}, P., \& {Fontani}, F.
  2020, \aap, 640, A51

\bibitem[{{Crapsi} {et~al.}(2005){Crapsi}, {Caselli}, {Walmsley}, {Myers},
  {Tafalla}, {Lee}, \& {Bourke}}]{Crapsi2005}
{Crapsi}, A., {Caselli}, P., {Walmsley}, C.~M., {et~al.} 2005, \apj, 619, 379

\bibitem[{{Crapsi} {et~al.}(2007){Crapsi}, {Caselli}, {Walmsley}, \&
  {Tafalla}}]{Crapsi2007}
{Crapsi}, A., {Caselli}, P., {Walmsley}, M.~C., \& {Tafalla}, M. 2007, \aap,
  470, 221

\bibitem[{{Crimier} {et~al.}(2010){Crimier}, {Ceccarelli}, {Maret},
  {Bottinelli}, {Caux}, {Kahane}, {Lis}, \& {Olofsson}}]{Crimier2010}
{Crimier}, N., {Ceccarelli}, C., {Maret}, S., {et~al.} 2010, \aap, 519, A65

\bibitem[{{Dagdigian}(2018)}]{Dagdigian2018}
{Dagdigian}, P.~J. 2018, \mnras, 479, 3227

\bibitem[{{Dalgarno} \& {Lepp}(1984)}]{DalgarnoLepp1984}
{Dalgarno}, A. \& {Lepp}, S. 1984, \apjl, 287, L47

\bibitem[{Denis-Alpizar {et~al.}(2013)Denis-Alpizar, Kalugina, Stoecklin, Vera,
  \& Lique}]{denis2013new}
Denis-Alpizar, O., Kalugina, Y., Stoecklin, T., Vera, M.~H., \& Lique, F. 2013,
  The Journal of chemical physics, 139, 224301

\bibitem[{{Denis-Alpizar} {et~al.}(2020){Denis-Alpizar}, {Stoecklin}, {Dutrey},
  \& {Guilloteau}}]{DenisAlpizar2020}
{Denis-Alpizar}, O., {Stoecklin}, T., {Dutrey}, A., \& {Guilloteau}, S. 2020,
  \mnras, 497, 4276

\bibitem[{{Dubernet} {et~al.}(2013){Dubernet}, {Alexander}, {Ba},
  {Balakrishnan}, {Balan{\c{c}}a}, {Ceccarelli}, {Cernicharo}, {Daniel},
  {Dayou}, {Doronin}, {Dumouchel}, {Faure}, {Feautrier}, {Flower}, {Grosjean},
  {Halvick}, {K{\l}os}, {Lique}, {McBane}, {Marinakis}, {Moreau}, {Moszynski},
  {Neufeld}, {Roueff}, {Schilke}, {Spielfiedel}, {Stancil}, {Stoecklin},
  {Tennyson}, {Yang}, {Vasserot}, \& {Wiesenfeld}}]{Dubernet2013}
{Dubernet}, M.~L., {Alexander}, M.~H., {Ba}, Y.~A., {et~al.} 2013, \aap, 553,
  A50

\bibitem[{Dumouchel {et~al.}(2011)Dumouchel, K{\l}os, \&
  Lique}]{dumouchel2011rotational}
Dumouchel, F., K{\l}os, J., \& Lique, F. 2011, Physical Chemistry Chemical
  Physics, 13, 8204

\bibitem[{{Dunham} {et~al.}(2015){Dunham}, {Allen}, {Evans},
  {Broekhoven-Fiene}, {Cieza}, {Di Francesco}, {Gutermuth}, {Harvey},
  {Hatchell}, {Heiderman}, {Huard}, {Johnstone}, {Kirk}, {Matthews}, {Miller},
  {Peterson}, \& {Young}}]{Dunham2015}
{Dunham}, M.~M., {Allen}, L.~E., {Evans}, Neal~J., I., {et~al.} 2015, \apjs,
  220, 11

\bibitem[{{Dunham} {et~al.}(2008){Dunham}, {Crapsi}, {Evans}, {Bourke},
  {Huard}, {Myers}, \& {Kauffmann}}]{Dunham2008}
{Dunham}, M.~M., {Crapsi}, A., {Evans}, Neal~J., I., {et~al.} 2008, \apjs, 179,
  249

\bibitem[{{Emprechtinger} {et~al.}(2009){Emprechtinger}, {Caselli}, {Volgenau},
  {Stutzki}, \& {Wiedner}}]{Emprechtinger2009}
{Emprechtinger}, M., {Caselli}, P., {Volgenau}, N.~H., {Stutzki}, J., \&
  {Wiedner}, M.~C. 2009, \aap, 493, 89

\bibitem[{{Enoch} {et~al.}(2006){Enoch}, {Young}, {Glenn}, {Evans}, {Golwala},
  {Sargent}, {Harvey}, {Aguirre}, {Goldin}, {Haig}, {Huard}, {Lange},
  {Laurent}, {Maloney}, {Mauskopf}, {Rossinot}, \& {Sayers}}]{Enoch2006}
{Enoch}, M.~L., {Young}, K.~E., {Glenn}, J., {et~al.} 2006, \apj, 638, 293

\bibitem[{{Evans}(1999)}]{Evans1999}
{Evans}, Neal~J., I. 1999, \araa, 37, 311

\bibitem[{{Ferrer Asensio} {et~al.}(2022){Ferrer Asensio}, {Spezzano},
  {Caselli}, {Alves}, {Sipil{\"a}}, {Redaelli}, {Bizzocchi}, {Lique}, \&
  {Mullins}}]{FerrerAsensio2022}
{Ferrer Asensio}, J., {Spezzano}, S., {Caselli}, P., {et~al.} 2022, \aap, 667,
  A119

\bibitem[{{Fuchs} {et~al.}(2004){Fuchs}, {Bruenken}, {Fuchs}, {Thorwirth},
  {Ahrens}, {Lewen}, {Urban}, {Giesen}, \& {Winnewisser}}]{Fuchs2004}
{Fuchs}, U., {Bruenken}, S., {Fuchs}, G.~W., {et~al.} 2004, Zeitschrift
  Naturforschung Teil A, 59, 861

\bibitem[{{Furuya} {et~al.}(2011){Furuya}, {Aikawa}, {Sakai}, \&
  {Yamamoto}}]{Furuya2011}
{Furuya}, K., {Aikawa}, Y., {Sakai}, N., \& {Yamamoto}, S. 2011, \apj, 731, 38

\bibitem[{{Galli} {et~al.}(2019){Galli}, {Loinard}, {Bouy}, {Sarro},
  {Ortiz-Le{\'o}n}, {Dzib}, {Olivares}, {Heyer}, {Hernandez},
  {Rom{\'a}n-Z{\'u}{\~n}iga}, {Kounkel}, \& {Covey}}]{Galli2019}
{Galli}, P.~A.~B., {Loinard}, L., {Bouy}, H., {et~al.} 2019, \aap, 630, A137

\bibitem[{{Giers} {et~al.}(2022){Giers}, {Spezzano}, {Alves}, {Caselli},
  {Redaelli}, {Sipil{\"a}}, {Ben Khalifa}, {Wiesenfeld}, {Br{\"u}nken}, \&
  {Bizzocchi}}]{Giers2022}
{Giers}, K., {Spezzano}, S., {Alves}, F., {et~al.} 2022, \aap, 664, A119

\bibitem[{{Ginsburg} \& {Mirocha}(2011)}]{pyspeckit}
{Ginsburg}, A. \& {Mirocha}, J. 2011, {PySpecKit: Python Spectroscopic
  Toolkit}, Astrophysics Source Code Library, record ascl:1109.001

\bibitem[{{Ginsburg} {et~al.}(2022){Ginsburg}, {Sokolov}, {de Val-Borro},
  {Rosolowsky}, {Pineda}, {Sip{\H{o}}cz}, \& {Henshaw}}]{pyspeckit2}
{Ginsburg}, A., {Sokolov}, V., {de Val-Borro}, M., {et~al.} 2022, \aj, 163, 291

\bibitem[{{Goicoechea} {et~al.}(2022){Goicoechea}, {Lique}, \&
  {Santa-Maria}}]{Goicoechea:22}
{Goicoechea}, J.~R., {Lique}, F., \& {Santa-Maria}, M.~G. 2022, \aap, 658, A28

\bibitem[{{Goldsmith}(2001)}]{Goldsmith2001}
{Goldsmith}, P.~F. 2001, \apj, 557, 736

\bibitem[{{Goldsmith} \& {Langer}(1999)}]{GoldsmithLanger1999}
{Goldsmith}, P.~F. \& {Langer}, W.~D. 1999, \apj, 517, 209

\bibitem[{Green(1975)}]{green1975rotational}
Green, S. 1975, The Journal of Chemical Physics, 62, 2271

\bibitem[{{Gueth} \& {Guilloteau}(1999)}]{GuethGuilloteau1999}
{Gueth}, F. \& {Guilloteau}, S. 1999, \aap, 343, 571

\bibitem[{{Harju} {et~al.}(2017){Harju}, {Sipil{\"a}}, {Br{\"u}nken},
  {Schlemmer}, {Caselli}, {Juvela}, {Menten}, {Stutzki}, {Asvany},
  {Kami{\'n}ski}, {Okada}, \& {Higgins}}]{Harju2017}
{Harju}, J., {Sipil{\"a}}, O., {Br{\"u}nken}, S., {et~al.} 2017, \apj, 840, 63

\bibitem[{{Hatchell} {et~al.}(1998){Hatchell}, {Millar}, \&
  {Rodgers}}]{Hatchell1998}
{Hatchell}, J., {Millar}, T.~J., \& {Rodgers}, S.~D. 1998, \aap, 332, 695

\bibitem[{Hern{\'a}ndez~Vera {et~al.}(2017)Hern{\'a}ndez~Vera, Lique,
  Dumouchel, Hily-Blant, \& Faure}]{hernandez2017rotational}
Hern{\'a}ndez~Vera, M., Lique, F., Dumouchel, F., Hily-Blant, P., \& Faure, A.
  2017, Monthly Notices of the Royal Astronomical Society, 468, 1084

\bibitem[{{Hildebrand}(1983)}]{Hildebrand1983}
{Hildebrand}, R.~H. 1983, \qjras, 24, 267

\bibitem[{{Hily-Blant} {et~al.}(2013){Hily-Blant}, {Bonal}, {Faure}, \&
  {Quirico}}]{HilyBlant2013}
{Hily-Blant}, P., {Bonal}, L., {Faure}, A., \& {Quirico}, E. 2013, \icarus,
  223, 582

\bibitem[{{Hily-Blant} {et~al.}(2010){Hily-Blant}, {Walmsley}, {Pineau Des
  For{\^e}ts}, \& {Flower}}]{HilyBlant2010}
{Hily-Blant}, P., {Walmsley}, M., {Pineau Des For{\^e}ts}, G., \& {Flower}, D.
  2010, \aap, 513, A41

\bibitem[{{Hirota} {et~al.}(2003){Hirota}, {Ikeda}, \& {Yamamoto}}]{Hirota2003}
{Hirota}, T., {Ikeda}, M., \& {Yamamoto}, S. 2003, \apj, 594, 859

\bibitem[{{Imai} {et~al.}(2018){Imai}, {Sakai}, {L{\'o}pez-Sepulcre},
  {Higuchi}, {Zhang}, {Oya}, {Watanabe}, {Sakai}, {Ceccarelli}, {Lefloch}, \&
  {Yamamoto}}]{Imai2018}
{Imai}, M., {Sakai}, N., {L{\'o}pez-Sepulcre}, A., {et~al.} 2018, \apj, 869, 51

\bibitem[{{Jensen} {et~al.}(2023){Jensen}, {Spezzano}, {Caselli}, {Grassi}, \&
  {Haugb{\o}lle}}]{Jensen2023arXiv}
{Jensen}, S.~S., {Spezzano}, S., {Caselli}, P., {Grassi}, T., \&
  {Haugb{\o}lle}, T. 2023, arXiv e-prints, arXiv:2305.05932

\bibitem[{{J{\o}rgensen} {et~al.}(2004){J{\o}rgensen}, {Sch{\"o}ier}, \& {van
  Dishoeck}}]{Jorgensen2004}
{J{\o}rgensen}, J.~K., {Sch{\"o}ier}, F.~L., \& {van Dishoeck}, E.~F. 2004,
  \aap, 416, 603

\bibitem[{{Juvela}(2020)}]{Juvela2020}
{Juvela}, M. 2020, \aap, 644, A151

\bibitem[{{Keto} \& {Caselli}(2010)}]{Keto2010}
{Keto}, E. \& {Caselli}, P. 2010, \mnras, 402, 1625

\bibitem[{{Keto} {et~al.}(2015){Keto}, {Caselli}, \& {Rawlings}}]{Keto2015}
{Keto}, E., {Caselli}, P., \& {Rawlings}, J. 2015, \mnras, 446, 3731

\bibitem[{{Koumpia} {et~al.}(2017){Koumpia}, {Semenov}, {van der Tak},
  {Boogert}, \& {Caux}}]{Koumpia2017}
{Koumpia}, E., {Semenov}, D.~A., {van der Tak}, F.~F.~S., {Boogert}, A.~C.~A.,
  \& {Caux}, E. 2017, \aap, 603, A88

\bibitem[{{Langer} {et~al.}(1984){Langer}, {Graedel}, {Frerking}, \&
  {Armentrout}}]{Langer1984}
{Langer}, W.~D., {Graedel}, T.~E., {Frerking}, M.~A., \& {Armentrout}, P.~B.
  1984, \apj, 277, 581

\bibitem[{Lanza \& Lique(2014)}]{lanza2014hyperfine}
Lanza, M. \& Lique, F. 2014, The Journal of chemical physics, 141, 164321

\bibitem[{{Launhardt} {et~al.}(2013){Launhardt}, {Stutz}, {Schmiedeke},
  {Henning}, {Krause}, {Balog}, {Beuther}, {Birkmann}, {Hennemann},
  {Kainulainen}, {Khanzadyan}, {Linz}, {Lippok}, {Nielbock}, {Pitann}, {Ragan},
  {Risacher}, {Schmalzl}, {Shirley}, {Stecklum}, {Steinacker}, \&
  {Tackenberg}}]{Launhardt2013}
{Launhardt}, R., {Stutz}, A.~M., {Schmiedeke}, A., {et~al.} 2013, \aap, 551,
  A98

\bibitem[{{Lee} {et~al.}(2018){Lee}, {Li}, {Hirano}, {Shang}, {Ho}, \&
  {Zhang}}]{LeeLi2018}
{Lee}, C.-F., {Li}, Z.-Y., {Hirano}, N., {et~al.} 2018, \apj, 863, 94

\bibitem[{{Magalh{\~a}es} {et~al.}(2018){Magalh{\~a}es}, {Hily-Blant}, {Faure},
  {Hernandez-Vera}, \& {Lique}}]{Magalhaes2018}
{Magalh{\~a}es}, V.~S., {Hily-Blant}, P., {Faure}, A., {Hernandez-Vera}, M., \&
  {Lique}, F. 2018, \aap, 615, A52

\bibitem[{{Mangum} \& {Shirley}(2015)}]{Mangum2015}
{Mangum}, J.~G. \& {Shirley}, Y.~L. 2015, \pasp, 127, 266

\bibitem[{{Mathews} {et~al.}(2013){Mathews}, {Klaassen}, {Juh{\'a}sz},
  {Harsono}, {Chapillon}, {van Dishoeck}, {Espada}, {de Gregorio-Monsalvo},
  {Hales}, {Hogerheijde}, {Mottram}, {Rawlings}, {Takahashi}, \&
  {Testi}}]{Mathews2013}
{Mathews}, G.~S., {Klaassen}, P.~D., {Juh{\'a}sz}, A., {et~al.} 2013, \aap,
  557, A132

\bibitem[{{Mccarthy} {et~al.}(1995){Mccarthy}, {Gottlieb}, \&
  {Thaddeus}}]{McCarthy1995}
{Mccarthy}, M.~C., {Gottlieb}, C.~A., \& {Thaddeus}, P. 1995, Journal of
  Molecular Spectroscopy, 173, 303

\bibitem[{{McCaughrean} {et~al.}(1994){McCaughrean}, {Rayner}, \&
  {Zinnecker}}]{McCaughrean1994}
{McCaughrean}, M.~J., {Rayner}, J.~T., \& {Zinnecker}, H. 1994, \apjl, 436,
  L189

\bibitem[{{Milam} {et~al.}(2005){Milam}, {Savage}, {Brewster}, {Ziurys}, \&
  {Wyckoff}}]{Milam2005}
{Milam}, S.~N., {Savage}, C., {Brewster}, M.~A., {Ziurys}, L.~M., \& {Wyckoff},
  S. 2005, \apj, 634, 1126

\bibitem[{{Motte} \& {Andr{\'e}}(2001)}]{MotteAndre2001}
{Motte}, F. \& {Andr{\'e}}, P. 2001, \aap, 365, 440

\bibitem[{{M{\"u}ller} {et~al.}(2001){M{\"u}ller}, {Thorwirth}, {Roth}, \&
  {Winnewisser}}]{Mueller2001}
{M{\"u}ller}, H.~S.~P., {Thorwirth}, S., {Roth}, D.~A., \& {Winnewisser}, G.
  2001, \aap, 370, L49

\bibitem[{Navarro-Almaida {et~al.}(2022)Navarro-Almaida, Bop, Lique, Esplugues,
  Rodr{\'\i}guez-Baras, Kramer, Romero, Caselli, Rivi{\'e}re-Marichalar, Kirk,
  {et~al.}}]{navarro2022linking}
Navarro-Almaida, D., Bop, C., Lique, F., {et~al.} 2022, arXiv preprint
  arXiv:2212.07675

\bibitem[{{Ortiz-Le{\'o}n} {et~al.}(2018){Ortiz-Le{\'o}n}, {Loinard}, {Dzib},
  {Galli}, {Kounkel}, {Mioduszewski}, {Rodr{\'\i}guez}, {Torres}, {Hartmann},
  {Boden}, {Evans}, {Brice{\~n}o}, \& {Tobin}}]{OrtizLeon2018}
{Ortiz-Le{\'o}n}, G.~N., {Loinard}, L., {Dzib}, S.~A., {et~al.} 2018, \apj,
  865, 73

\bibitem[{{Padovani} {et~al.}(2009){Padovani}, {Walmsley}, {Tafalla}, {Galli},
  \& {M{\"u}ller}}]{Padovani2009}
{Padovani}, M., {Walmsley}, C.~M., {Tafalla}, M., {Galli}, D., \& {M{\"u}ller},
  H.~S.~P. 2009, \aap, 505, 1199

\bibitem[{{Padovani} {et~al.}(2011){Padovani}, {Walmsley}, {Tafalla},
  {Hily-Blant}, \& {Pineau Des For{\^e}ts}}]{Padovani2011}
{Padovani}, M., {Walmsley}, C.~M., {Tafalla}, M., {Hily-Blant}, P., \& {Pineau
  Des For{\^e}ts}, G. 2011, \aap, 534, A77

\bibitem[{{Pagani} {et~al.}(2012){Pagani}, {Bourgoin}, \& {Lique}}]{Pagani2012}
{Pagani}, L., {Bourgoin}, A., \& {Lique}, F. 2012, \aap, 548, L4

\bibitem[{{Pagani} {et~al.}(1992){Pagani}, {Salez}, \& {Wannier}}]{Pagani1992}
{Pagani}, L., {Salez}, M., \& {Wannier}, P.~G. 1992, \aap, 258, 479

\bibitem[{{Parise} {et~al.}(2004){Parise}, {Castets}, {Herbst}, {Caux},
  {Ceccarelli}, {Mukhopadhyay}, \& {Tielens}}]{Parise2004}
{Parise}, B., {Castets}, A., {Herbst}, E., {et~al.} 2004, \aap, 416, 159

\bibitem[{{Persson} {et~al.}(2018){Persson}, {J{\o}rgensen}, {M{\"u}ller},
  {Coutens}, {van Dishoeck}, {Taquet}, {Calcutt}, {van der Wiel}, {Bourke}, \&
  {Wampfler}}]{Persson2018}
{Persson}, M.~V., {J{\o}rgensen}, J.~K., {M{\"u}ller}, H.~S.~P., {et~al.} 2018,
  \aap, 610, A54

\bibitem[{{Pety}(2005)}]{Pety2005}
{Pety}, J. 2005, in SF2A-2005: Semaine de l'Astrophysique Francaise, ed.
  F.~{Casoli}, T.~{Contini}, J.~M. {Hameury}, \& L.~{Pagani}, 721

\bibitem[{{Pineda} {et~al.}(2022){Pineda}, {Arzoumanian}, {Andr{\'e}},
  {Friesen}, {Zavagno}, {Clarke}, {Inoue}, {Chen}, {Lee}, {Soler}, \&
  {Kuffmeier}}]{Pineda2022arXiv}
{Pineda}, J.~E., {Arzoumanian}, D., {Andr{\'e}}, P., {et~al.} 2022, arXiv
  e-prints, arXiv:2205.03935

\bibitem[{{Pineda} {et~al.}(2019){Pineda}, {Zhao}, {Schmiedeke}, {Segura-Cox},
  {Caselli}, {Myers}, {Tobin}, \& {Dunham}}]{Pineda2019}
{Pineda}, J.~E., {Zhao}, B., {Schmiedeke}, A., {et~al.} 2019, \apj, 882, 103

\bibitem[{{Plummer}(1911)}]{Plummer1911}
{Plummer}, H.~C. 1911, \mnras, 71, 460

\bibitem[{{Qu{\'e}nard} {et~al.}(2018){Qu{\'e}nard}, {Bottinelli}, {Caux}, \&
  {Wakelam}}]{Quenard2018}
{Qu{\'e}nard}, D., {Bottinelli}, S., {Caux}, E., \& {Wakelam}, V. 2018, \mnras,
  477, 5312

\bibitem[{{Qu{\'e}nard} {et~al.}(2017){Qu{\'e}nard}, {Vastel}, {Ceccarelli},
  {Hily-Blant}, {Lefloch}, \& {Bachiller}}]{Quenard2017}
{Qu{\'e}nard}, D., {Vastel}, C., {Ceccarelli}, C., {et~al.} 2017, \mnras, 470,
  3194

\bibitem[{{Redaelli} {et~al.}(2019){Redaelli}, {Bizzocchi}, {Caselli},
  {Sipil{\"a}}, {Lattanzi}, {Giuliano}, \& {Spezzano}}]{Redaelli2019}
{Redaelli}, E., {Bizzocchi}, L., {Caselli}, P., {et~al.} 2019, \aap, 629, A15

\bibitem[{{Redaelli} {et~al.}(2022){Redaelli}, {Chac{\'o}n-Tanarro}, {Caselli},
  {Tafalla}, {Pineda}, {Spezzano}, \& {Sipil{\"a}}}]{Redaelli2022}
{Redaelli}, E., {Chac{\'o}n-Tanarro}, A., {Caselli}, P., {et~al.} 2022, \apj,
  941, 168

\bibitem[{{Robert}(2003)}]{Robert2003}
{Robert}, F. 2003, \ssr, 106, 87

\bibitem[{{Roberts} {et~al.}(2002){Roberts}, {Fuller}, {Millar}, {Hatchell}, \&
  {Buckle}}]{Roberts2002}
{Roberts}, H., {Fuller}, G.~A., {Millar}, T.~J., {Hatchell}, J., \& {Buckle},
  J.~V. 2002, \aap, 381, 1026

\bibitem[{{Sakai} {et~al.}(2008){Sakai}, {Sakai}, {Hirota}, \&
  {Yamamoto}}]{Sakai2008}
{Sakai}, N., {Sakai}, T., {Hirota}, T., \& {Yamamoto}, S. 2008, \apj, 672, 371

\bibitem[{{Sakai} {et~al.}(2010){Sakai}, {Saruwatari}, {Sakai}, {Takano}, \&
  {Yamamoto}}]{SakaiSaruwatari2010}
{Sakai}, N., {Saruwatari}, O., {Sakai}, T., {Takano}, S., \& {Yamamoto}, S.
  2010, \aap, 512, A31

\bibitem[{{Saykally} {et~al.}(1976){Saykally}, {Szanto}, {Anderson}, \&
  {Woods}}]{Saykally1976}
{Saykally}, R.~J., {Szanto}, P.~G., {Anderson}, T.~G., \& {Woods}, R.~C. 1976,
  \apjl, 204, L143

\bibitem[{{Schmid-Burgk} {et~al.}(2004){Schmid-Burgk}, {Muders}, {M{\"u}ller},
  \& {Brupbacher-Gatehouse}}]{SchmidBurgk2004}
{Schmid-Burgk}, J., {Muders}, D., {M{\"u}ller}, H.~S.~P., \&
  {Brupbacher-Gatehouse}, B. 2004, \aap, 419, 949

\bibitem[{{Sch{\"o}ier} {et~al.}(2005){Sch{\"o}ier}, {van der Tak}, {van
  Dishoeck}, \& {Black}}]{Schoier2005}
{Sch{\"o}ier}, F.~L., {van der Tak}, F.~F.~S., {van Dishoeck}, E.~F., \&
  {Black}, J.~H. 2005, \aap, 432, 369

\bibitem[{{Segura-Cox} {et~al.}(2016){Segura-Cox}, {Harris}, {Tobin}, {Looney},
  {Li}, {Chandler}, {Kratter}, {Dunham}, {Sadavoy}, {Perez}, \&
  {Melis}}]{SeguraCox2016}
{Segura-Cox}, D.~M., {Harris}, R.~J., {Tobin}, J.~J., {et~al.} 2016, \apjl,
  817, L14

\bibitem[{{Sipil{\"a}} {et~al.}(2019){Sipil{\"a}}, {Caselli}, {Redaelli},
  {Juvela}, \& {Bizzocchi}}]{Sipila2019}
{Sipil{\"a}}, O., {Caselli}, P., {Redaelli}, E., {Juvela}, M., \& {Bizzocchi},
  L. 2019, \mnras, 487, 1269

\bibitem[{{Sipil{\"a}} {et~al.}(2022){Sipil{\"a}}, {Caselli}, {Redaelli}, \&
  {Spezzano}}]{Sipila2022}
{Sipil{\"a}}, O., {Caselli}, P., {Redaelli}, E., \& {Spezzano}, S. 2022, \aap,
  668, A131

\bibitem[{{Spezzano} {et~al.}(2016){Spezzano}, {Bizzocchi}, {Caselli}, {Harju},
  \& {Br{\"u}nken}}]{Spezzano2016b}
{Spezzano}, S., {Bizzocchi}, L., {Caselli}, P., {Harju}, J., \& {Br{\"u}nken},
  S. 2016, \aap, 592, L11

\bibitem[{{Spezzano} {et~al.}(2017){Spezzano}, {Caselli}, {Bizzocchi},
  {Giuliano}, \& {Lattanzi}}]{Spezzano2017}
{Spezzano}, S., {Caselli}, P., {Bizzocchi}, L., {Giuliano}, B.~M., \&
  {Lattanzi}, V. 2017, \aap, 606, A82

\bibitem[{{Spezzano} {et~al.}(2022){Spezzano}, {Caselli}, {Sipil{\"a}}, \&
  {Bizzocchi}}]{Spezzano2022}
{Spezzano}, S., {Caselli}, P., {Sipil{\"a}}, O., \& {Bizzocchi}, L. 2022, \aap,
  664, L2

\bibitem[{{Sun} {et~al.}(2006){Sun}, {Kramer}, {Ossenkopf}, {Bensch},
  {Stutzki}, \& {Miller}}]{Sun2006}
{Sun}, K., {Kramer}, C., {Ossenkopf}, V., {et~al.} 2006, \aap, 451, 539

\bibitem[{{Tafalla} {et~al.}(1998){Tafalla}, {Mardones}, {Myers}, {Caselli},
  {Bachiller}, \& {Benson}}]{Tafalla1998}
{Tafalla}, M., {Mardones}, D., {Myers}, P.~C., {et~al.} 1998, \apj, 504, 900

\bibitem[{{Taniguchi} {et~al.}(2019){Taniguchi}, {Herbst}, {Ozeki}, \&
  {Saito}}]{Taniguchi2019}
{Taniguchi}, K., {Herbst}, E., {Ozeki}, H., \& {Saito}, M. 2019, \apj, 884, 167

\bibitem[{{Tinti} {et~al.}(2007){Tinti}, {Bizzocchi}, {Degli Esposti}, \&
  {Dore}}]{Tinti2007}
{Tinti}, F., {Bizzocchi}, L., {Degli Esposti}, C., \& {Dore}, L. 2007, \apjl,
  669, L113

\bibitem[{{Turner}(2001)}]{Turner2001}
{Turner}, B.~E. 2001, \apjs, 136, 579

\bibitem[{{van der Tak} {et~al.}(2020){van der Tak}, {Lique}, {Faure}, {Black},
  \& {van Dishoeck}}]{VanDerTak2020}
{van der Tak}, F. F.~S., {Lique}, F., {Faure}, A., {Black}, J.~H., \& {van
  Dishoeck}, E.~F. 2020, Atoms, 8, 15

\bibitem[{{van der Tak} {et~al.}(2009){van der Tak}, {M{\"u}ller}, {Harding},
  \& {Gauss}}]{VanDerTak2009}
{van der Tak}, F.~F.~S., {M{\"u}ller}, H.~S.~P., {Harding}, M.~E., \& {Gauss},
  J. 2009, \aap, 507, 347

\bibitem[{{van Dishoeck} {et~al.}(1995){van Dishoeck}, {Blake}, {Jansen}, \&
  {Groesbeck}}]{VanDishoeck1995}
{van Dishoeck}, E.~F., {Blake}, G.~A., {Jansen}, D.~J., \& {Groesbeck}, T.~D.
  1995, \apj, 447, 760

\bibitem[{{Vastel} {et~al.}(2006){Vastel}, {Caselli}, {Ceccarelli}, {Phillips},
  {Wiedner}, {Peng}, {Houde}, \& {Dominik}}]{Vastel2006}
{Vastel}, C., {Caselli}, P., {Ceccarelli}, C., {et~al.} 2006, \apj, 645, 1198

\bibitem[{{Walker} {et~al.}(2016){Walker}, {Kalinauskaite}, {McCarthy},
  {Trappe}, {Murphy}, {Helldner}, {Pantaleev}, \& {Flygare}}]{OSO4mm}
{Walker}, G.~W., {Kalinauskaite}, E., {McCarthy}, D.~N., {et~al.} 2016, in
  Society of Photo-Optical Instrumentation Engineers (SPIE) Conference Series,
  Vol. 9914, Millimeter, Submillimeter, and Far-Infrared Detectors and
  Instrumentation for Astronomy VIII, ed. W.~S. {Holland} \& J.~{Zmuidzinas},
  99142V

\bibitem[{{Whitworth} \& {Ward-Thompson}(2001)}]{WhitworthWardThompson2001}
{Whitworth}, A.~P. \& {Ward-Thompson}, D. 2001, \apj, 547, 317

\bibitem[{{Wilson}(1999)}]{Wilson1999}
{Wilson}, T.~L. 1999, Reports on Progress in Physics, 62, 143

\bibitem[{{Yoshida} {et~al.}(2019){Yoshida}, {Sakai}, {Nishimura}, {Tokudome},
  {Watanabe}, {Sakai}, {Takano}, \& {Yamamoto}}]{Yoshida2019}
{Yoshida}, K., {Sakai}, N., {Nishimura}, Y., {et~al.} 2019, \pasj, 71, S18

\bibitem[{{Zhang} {et~al.}(2021){Zhang}, {Wu}, {Liu}, {Tang}, {Li}, {Esimbek},
  \& {Qin}}]{ZhangWuLiu2021}
{Zhang}, C., {Wu}, Y., {Liu}, X.~C., {et~al.} 2021, \aap, 648, A83

\end{thebibliography}
%

\FloatBarrier

\begin{appendix}
\section{Observed lines}
\begin{figure*}[!b]
   \centering
    \includegraphics[width=0.95\textwidth]{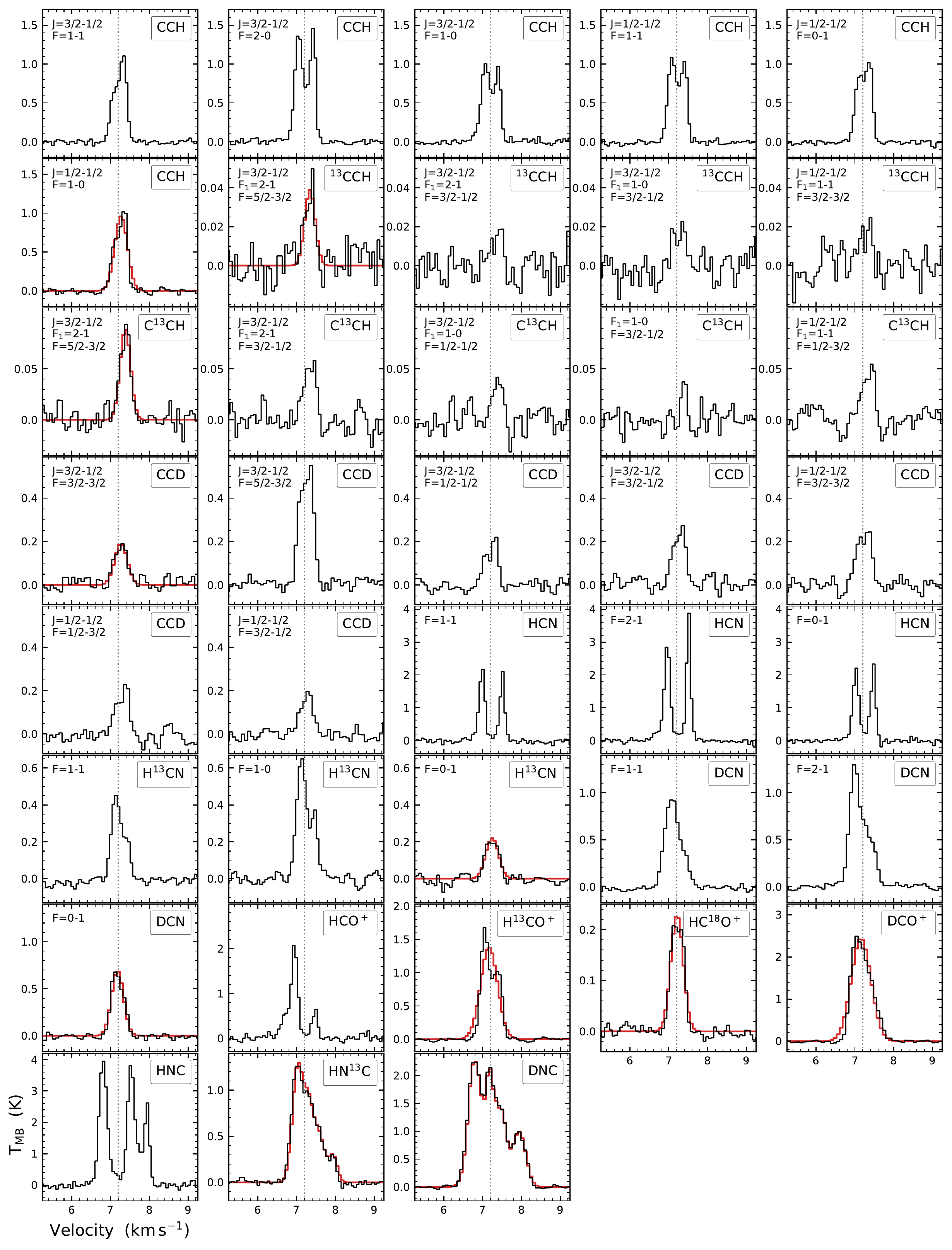}
   \caption{Ground-state rotational lines observed towards the pre-stellar core L1544. The respective hyperfine transition is given in the upper left corner of each plot. The dotted line indicates the rest velocity of the system (7.2\,km\,s$^{-1}$). The red curves show the fits used in the LTE analysis.}
   \label{Fig:L1544all}
\end{figure*}

\begin{figure*}[!b]
   \centering
    \includegraphics[width=0.95\textwidth]{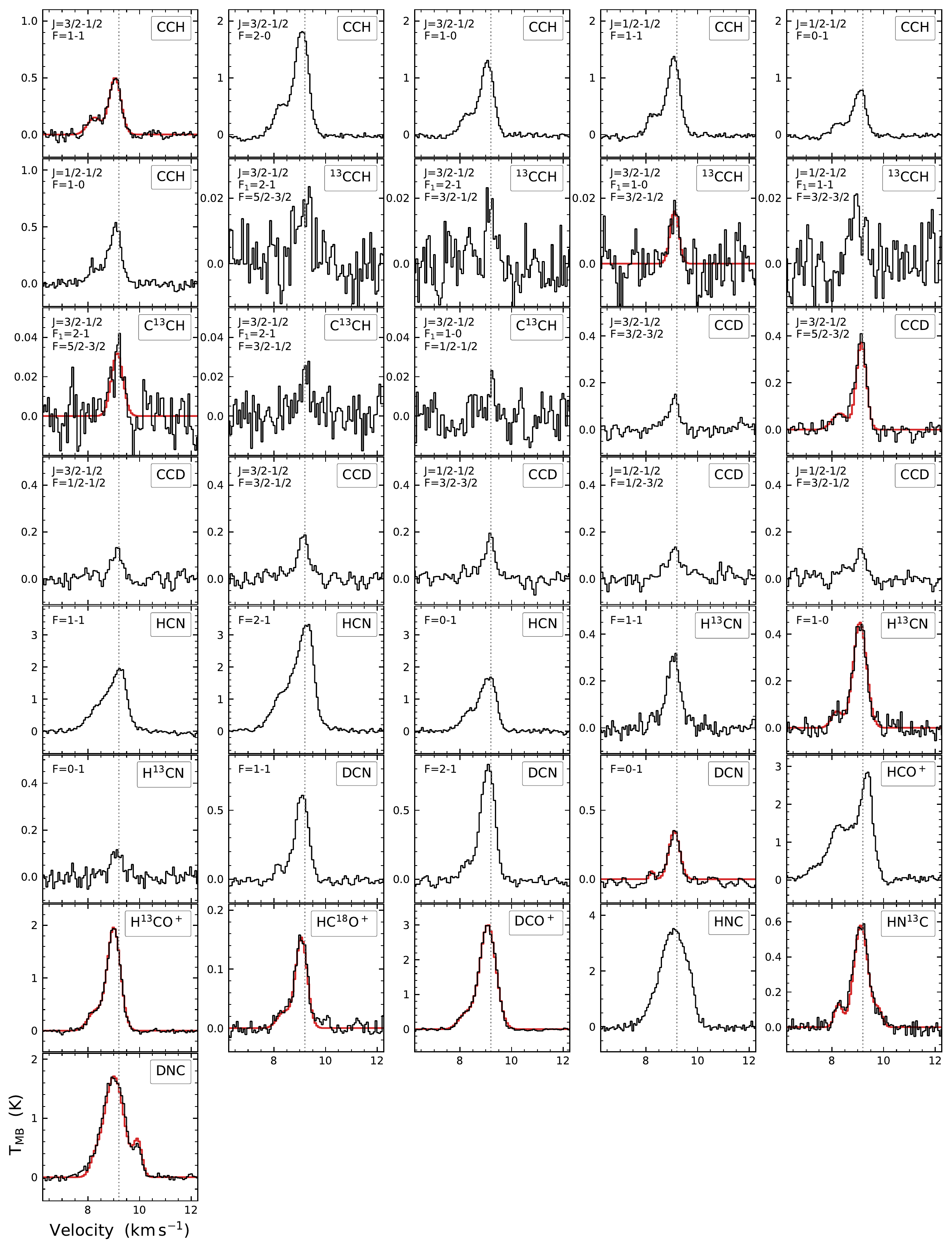}
   \caption{Ground-state rotational lines observed towards the protostellar core HH211. The respective hyperfine transition is given in the upper left corner of each plot. The dotted line indicates the rest velocity of the system (9.2\,km\,s$^{-1}$). The red curves show the fits used in the LTE analysis.}
    \label{Fig:HH211all}
\end{figure*}

Table~\ref{Tab:AllObervedLines} lists all emission lines observed towards L1544 and HH211, along with the respective transition, frequency and telescope beam size. Fig.~\ref{Fig:L1544all} and Fig.~\ref{Fig:HH211all} show the corresponding spectra for L1544 and HH211, respectively, with the hyperfine transition given in the top left corner. 
\begin{table*}[h]
    \centering
    \caption{Properties of the observed lines.}
    \label{Tab:AllObervedLines}
    \begin{tabular}{l l c l c c }
    \hline\hline 
    \noalign{\smallskip}
     Molecule & Transition & Relative & Frequency & HPBW & Ref. \\
      & & Intensity & (MHz) &  & \\
      \noalign{\smallskip}
      \hline
      \noalign{\smallskip}
      CCH  & $N=1-0$, $J=3/2-1/2$, $F=1-1$ & 0.043 & 87284.11(1) & 43.5 & 1 \\
      & $N=1-0$, $J=3/2-1/2$, $F=2-0$ & 0.417 & 87316.90(1) & 43.5 & 1 \\
    & $N=1-0$, $J=3/2-1/2$, $F=1-0$ & 0.208 & 87328.59(1) & 43.5 &  1\\
    & $N=1-0$, $J=1/2-1/2$, $F=1-1$ & 0.208 & 87401.99(1) & 43.5 &  1\\
    & $N=1-0$, $J=1/2-1/2$, $F=0-1$ & 0.083 & 87407.17(1) & 43.5 & 1 \\
    & $N=1-0$, $J=1/2-1/2$, $F=1-0$ & 0.043 & 87446.47(1) & 43.5 &  1\\
      \xcch\,  & $N=1-0$, $J=3/2-1/2$, $F_1=2-1$, $F=5/2-3/2$ & 0.250 & 84119.33(2) & 45.9 & 2\\
        & $N=1-0$, $J=3/2-1/2$, $F_1=2-1$, $F=3/2-1/2$ & 0.153 & 84124.14(2) & 45.9 & 2 \\
       & $N=1-0$, $J=3/2-1/2$, $F_1=1-0$, $F=3/2-1/2$ & 0.167 & 84153.31(2) & 45.9 & 2 \\
       & $N=1-0$, $J=1/2-1/2$, $F_1=1-1$, $F=3/2-3/2$ & 0.152 & 84206.87(2) & 45.8 & 2 \\
      \cxch\,  & $N=1-0$, $J=3/2-1/2$, $F_1=2-1$, $F=5/2-3/2$ & 0.263 & 85229.335(4) & 44.8 & 2 \\
       & $N=1-0$, $J=3/2-1/2$, $F_1=2-1$, $F=3/2-1/2$ & 0.165 & 85232.805(4) & 44.8  & 2 \\
       & $N=1-0$, $J=3/2-1/2$, $F_1=1-0$, $F=1/2-1/2$ & 0.085 & 85247.728(4) & 44.8 & 2 \\
       & $N=1-0$, $J=3/2-1/2$, $F_1=1-0$, $F=3/2-1/2$ & 0.168 & 85256.988(4) & 44.8 & 2 \\
       & $N=1-0$, $J=1/2-1/2$, $F_1=1-1$, $F=1/2-3/2$ & 0.081 & 85303.990(4) & 44.7 & 2 \\
       & $N=1-0$, $J=1/2-1/2$, $F_1=1-1$, $F=3/2-3/2$ & 0.158 & 85307.459(4) & 44.7 & 2 \\
       \ccd\,  & $N=1-0$, $J=3/2-1/2$, $F=3/2-3/2$ & 0.085 & 72101.811(5) & 58.2 & 3 \\
       & $N=1-0$, $J=3/2-1/2$, $F=5/2-3/2$ & 0.333 & 72107.721(3) & 58.2 & 3 \\
        & $N=1-0$, $J=3/2-1/2$, $F=1/2-1/2$ & 0.101 & 72109.050(5) & 58.2 & 3 \\
        & $N=1-0$, $J=3/2-1/2$, $F=3/2-1/2$ & 0.137 & 72112.295(5) & 58.2 & 3 \\
       & $N=1-0$, $J=1/2-1/2$, $F=3/2-3/2$ & 0.137 & 72187.708(5) & 58.1 & 3 \\
        & $N=1-0$, $J=1/2-1/2$, $F=1/2-3/2$ & 0.101 & 72189.726(6) & 58.1 & 3 \\
       & $N=1-0$, $J=1/2-1/2$, $F=3/2-1/2$ & 0.085 & 72198.193(6) & 58.1 & 3 \\
      HCN   & $J=1-0$, $F=1-1$ & 0.333 & 88630.4156(2) & 43.0 &  4 \\
            & $J=1-0$, $F=2-1$ & 0.556 & 88631.8475(3) & 43.0 & 4 \\
            & $J=1-0$, $F=0-1$ & 0.111 & 88633.9357(3) & 43.0 & 4 \\
      \hxcn\,   & $J=1-0$, $F=1-1$ & 0.333 & 86338.7352(1) & 43.9 & 5 \\
                & $J=1-0$, $F=1-0$ & 0.556 & 86340.1666(1) & 43.9 & 5 \\
                & $J=1-0$, $F=0-1$ & 0.111 & 86342.2543(3) & 43.9 & 5 \\
      DCN   & $J=1-0$, $F=1-1$ & 0.333 & 72413.50(1) & 57.9 & 6 \\
            & $J=1-0$, $F=2-1$ & 0.556 & 72414.93(1) & 57.9 & 6 \\ 
            & $J=1-0$, $F=0-1$ & 0.111 & 72417.03(1) & 57.9 & 6 \\
      \hcop\, & $J=1-0$ & - & 89188.525(4) & 42.8 & 7 \\
      \hxcop\, & $J=1-0$ & - & 86754.288(5) & 43.7 & 8 \\
      \hcxop\, & $J=1-0$ & - & 85162.223(5) & 44.9 & 8 \\
      \dcop\, & $J=1-0$ & - & 72039.3124(8) & 58.3 & 9 \\
      HNC  & $J=1-0$ & - & 90663.568(4) & 42.2 & 10 \\
      \hnxc\,  & $J=1-0$ & - & 87090.825(4) & 43.6 & 9 \\
      DNC  & $J=1-0$ & - & 76305.700(1) & 53.9 & 9 \\
     \hline 
     \noalign{\smallskip}
    \end{tabular}
    \tablebib{
    (1) \cite{Padovani2009}; (2) \cite{McCarthy1995}; (3) \cite{Cabezas2021}; (4) \cite{Ahrens2002}; (5) \cite{Fuchs2004}; (6) \cite{Brunken2004}; (7) \cite{Tinti2007};  (8) \cite{SchmidBurgk2004}; (9) \cite{VanDerTak2009}; (10) \cite{Saykally1976}.
    }
\end{table*}

\section{Parameters of the additional velocity component towards HH211}
Table~\ref{tab:HH2112Gaussians} presents the best-fit parameters of the additional velocity component observed towards HH211, along with the corresponding column density. The parameters were obtained by applying a two-component Gaussian fit to the spectral lines. In the case of \hnxc\, and DNC, we applied a three-component Gaussian to account for their more separate hyperfine component at high velocity.
The additional velocity component is located at approximately 8.22\,km\,s$^{-1}$. 

Table~\ref{tab:HH2112GaussiansDeuteration} compares the level of deuteration between the additional component (add.) and the main component. Within errorbars, the deuterium fractionation is similar for CCH, HCN, and HNC, indicating that main and additional component trace the same gas along the line-of-sight. 
For \hcop\,, the D/H ratios differ by a factor of two between the two components. 
However, with this little amount of data on the additional velocity component, we cannot make any conclusive statement about this. Most likely, the additional velocity components of \hcop\, and isotopologues are tracing the same gas as the other molecules in this study, plus some additional gas or structure.

\begin{table*}[h]
    \centering
    \caption{Best-fit parameters and column densities of the additional velocity component towards HH211.}
    \begin{tabular}{l l l l l l}
    \hline\hline 
    \noalign{\smallskip}
     Molecule & $T_\mathrm{mb,peak}$ & $V_\mathrm{LSR}$\tablefootmark{a} & $\Delta v$ & $\int T_\mathrm{mb}\mathrm{d}v$ & $N$   \\
      & (K) & (km\,s$^{-1}$) & (km\,s$^{-1}$) & (K\,km\,s$^{-1}$) & ($\times10^{12}$\,cm$^{-2}$) \\
    \noalign{\smallskip}
    \hline
    \noalign{\smallskip}
    \noalign{\smallskip}
    CCH   & 0.15(1) & 8.22 & 0.62(9) & 0.10(2) & 70(20) \\
    \ccd\,   & 0.070(3) & 8.27 & 0.7(1) & 0.053(9) & 6(1) \\
    \hxcn\,   & 0.07(2) & 8.16 & 0.5(2) & 0.04(2) & 0.17(8) \\
    DCN   & 0.06(7) & 8.20 & 0.2(3) & 0.01(2) & 0.5(1.0) \\
    \hxcop\,   & 0.36(1) & 8.24 & 0.66(5) & 0.25(2) & 0.30(5) \\
    \hcxop\,  & 0.027(4) & 8.35 & 0.6(2) & 0.017(5) & 0.020(7) \\
    \dcop\,  & 0.43(2) & 8.18 & 0.65(4) & 0.30(2) & 0.46(7) \\
    \hnxc\,  & 0.13(2) & 8.22 & 0.36(6) & 0.05(1) & 0.15(4) \\
    DNC   & 0.40(7) & 8.22 & 0.6(2) & 0.25(8) & 1.0(4) \\
    \noalign{\smallskip}
    \hline 
    \noalign{\smallskip}
    \end{tabular}
    \tablefoot{
    \tablefoottext{a}{The uncertainties of the fits are smaller than the velocity resolution. Therefore, the error on $V_\mathrm{LSR}$ is given by the observed spectral resolution, $0.07$\,km\,s$^{-1}$. }}
    \label{tab:HH2112Gaussians}
\end{table*}

\begin{table*}[h]
    \centering
    \caption{Comparison of column density ratios in the main component and the additional velocity component observed towards HH211.}
    \begin{tabular}{l l l}
    \hline\hline 
    \noalign{\smallskip}
       & HH211 (main) & HH211 (add.)  \\
    \noalign{\smallskip}
    \hline
    \noalign{\smallskip}
    N(\ccd\,)/N(CCH)           & 0.10(3)  & 0.09(3) \\
    N(DCN)/N(\hxcn\,$\times$68)       & 0.07(2)  & 0.04(8) \\
    N(\dcop\,)/N(\hxcop\,$\times$68)  & 0.037(5) & 0.022(5) \\
    N(\dcop\,)/N(\hcxop\,$\times$557) & 0.09(2) & 0.04(2) \\
    N(DNC)/N(\hnxc\,$\times$68)       & 0.08(3) & 0.10(4) \\    
    \noalign{\smallskip}
    \hline 
    \noalign{\smallskip}
    \end{tabular}
    \label{tab:HH2112GaussiansDeuteration}
\end{table*}

\section{Collisional rate coefficients}\label{collisionrates}

Within the Born-Oppenheimer approximation, the interactionpotential is the same for HCN-H$_2$ and DCN-H$_2$ (HNC-H$_2$, DNC-H$_2$, HN$^{13}$C-H$_2$) and only depends on the mutual distances of the atoms. 
Hence, the only difference between the interaction potentials of two different isotopologues with H$_2$ is the position of the centre of mass taken for the origin of the Jacobi coordinates used to describe the geometries of the system.
Then, we employed the HCN-H$_2$ \citep{denis2013new} and HNC-H$_2$ \citep{dumouchel2011rotational} potential energy surfaces (PESs) corrected to consider the effect of isotopic substitution.

In the following, the rotational level of the targets (DCN, HNC, HN$^{13}$C, and DNC) will be denoted $j$. Only the coupling of the molecular rotation with the nuclear spin ($(I = 1)$ of the nitrogen atom will be considered. The coupling results in a weak splitting of each rotational level $j$, into three hyperfine levels (except for the $j = 0$ level which is split into a single level). Each hyperfine level is designated by a quantum number $F$ ($F = I + j$) varying between $|I - j|$ and $I + j$. Inly collisional excitation by para-H$_2$ in its ground rotational state has been considered.

Rotational (i.e. nuclear-spin free) scattering matrix, cross sections and rate coefficients for the excitation of DCN, HNC, HN$^{13}$C and DNC (the isotopologues of HCN and HNC) induced by collisions with H$_2$ were computed by \cite{navarro2022linking} for temperatures up to 30 K. These data were calculated using the quantum mechanical close-coupling approach \citep{green1975rotational}. For more computational details, we refer the readers to the work of \cite{navarro2022linking}.

To account for the hyperfine structure due to the nitrogen nuclear spin of the isotopologues mentioned above, we used the scattering matrix computed by \cite{navarro2022linking} and applied a recoupling method \citep{alexander1985collision,lanza2014hyperfine}. Therefore, we derived hyperfine resolved rate coefficients for the 25 low-lying energy levels, that is $(j,F)\leq(8,9)$, for temperatures up to 30 K.

\begin{figure*}
    \centering
    \includegraphics[width = \linewidth, trim = 6 10 13 10, clip = true]{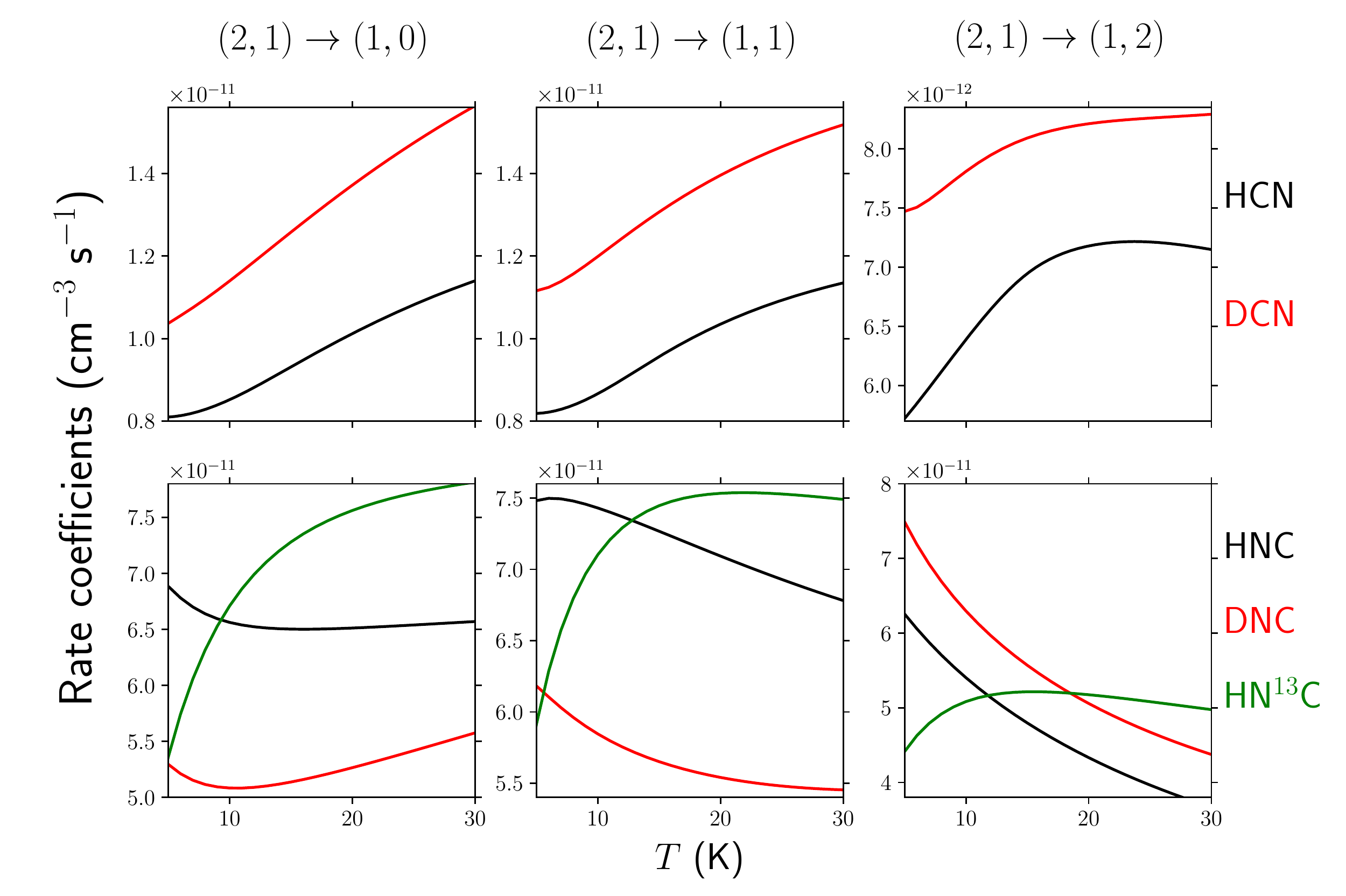}
   \caption{Temperature dependence of hyperfine resolved rate coefficients of HCN and DCN (upper panels) and HNC, DNC and HN$^{13}$C (lower panels) for the $(j,F)=(2,1)\to(1,F')$ transitions. The data of HCN are from \cite{Magalhaes2018}.}
    \label{fig:HCN-rates}
\end{figure*}

Fig.~\ref{fig:HCN-rates} displays the temperature dependence of the hyperfine resolved excitation rate coefficients of HCN, DCN, HNC, DNC and HN$^{13}$C  induced by collision with H$_2$  for the $(j,F)=(2,1)\to(1,F')$ transitions, where $F'={0,1,2}$. 
The upper panels show that the rate coefficients for DCN-H$_2$ collisions exhibit systematically a higher magnitude than those for HCN-H$_2$ collisions for all transitions, the differences being of the order of roughly 25\% -- 35\%. 
At the opposite, the collisional data for DNC-H$_2$ collisions can either overestimate or underestimate the ones for HNC-H$_2$ collisions depending on the transitions (see lower panels). It is also interesting to note that the HN$^{13}$C-H$_2$ collisional data increase with increasing temperature while those for HNC-H$_2$ (DNC-H$_2$) collisions decrease.
All this comparison clearly demonstrates the need of using isotopologues specific collisional data for modelling HCN/HNC isotopologues' observations.

\section{ Derivation of the physical structure of HH211}\label{AppPhysModelHH211}
\begin{figure}
   \centering
    \includegraphics[width=\hsize]{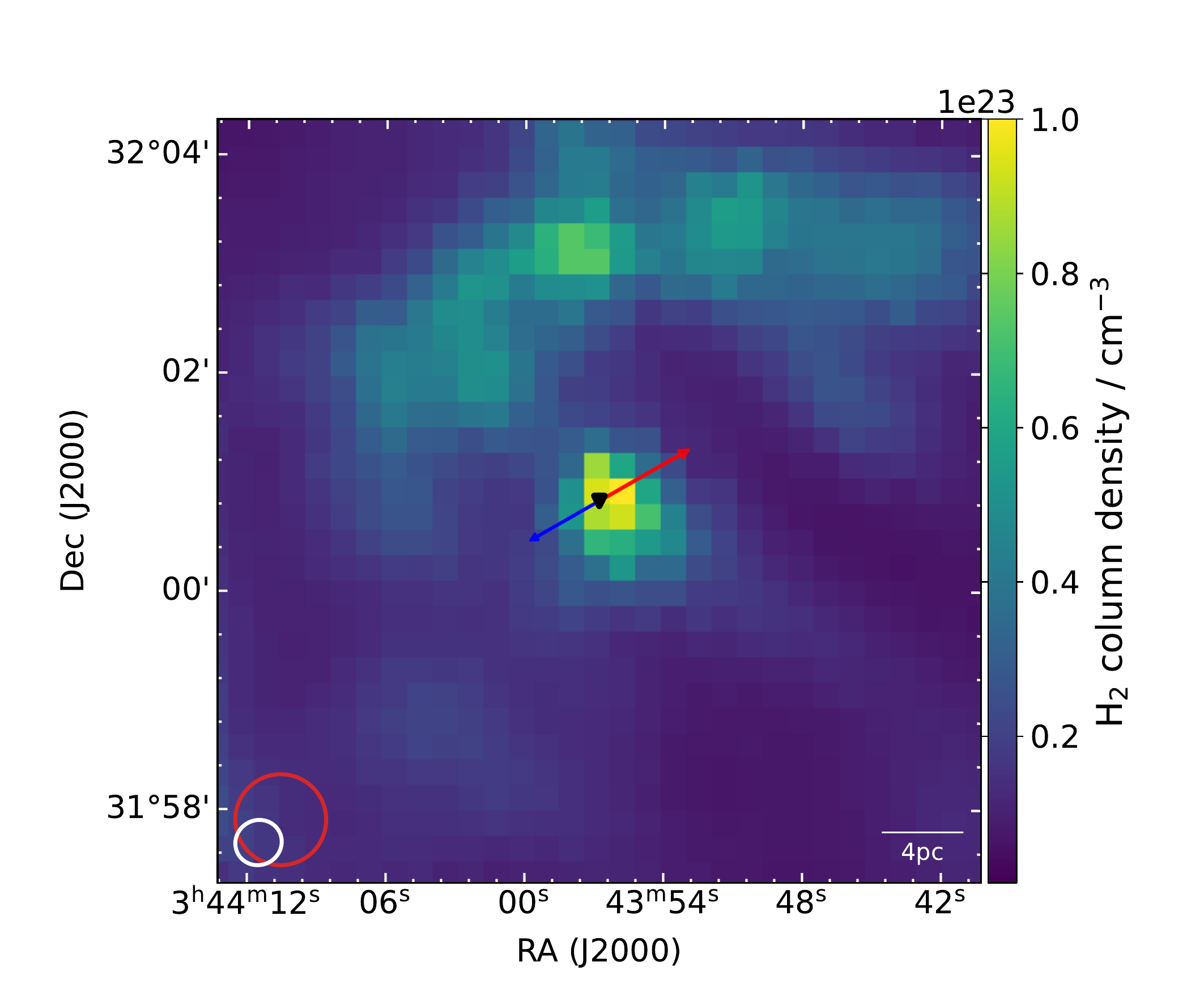}
    \includegraphics[width=\hsize]{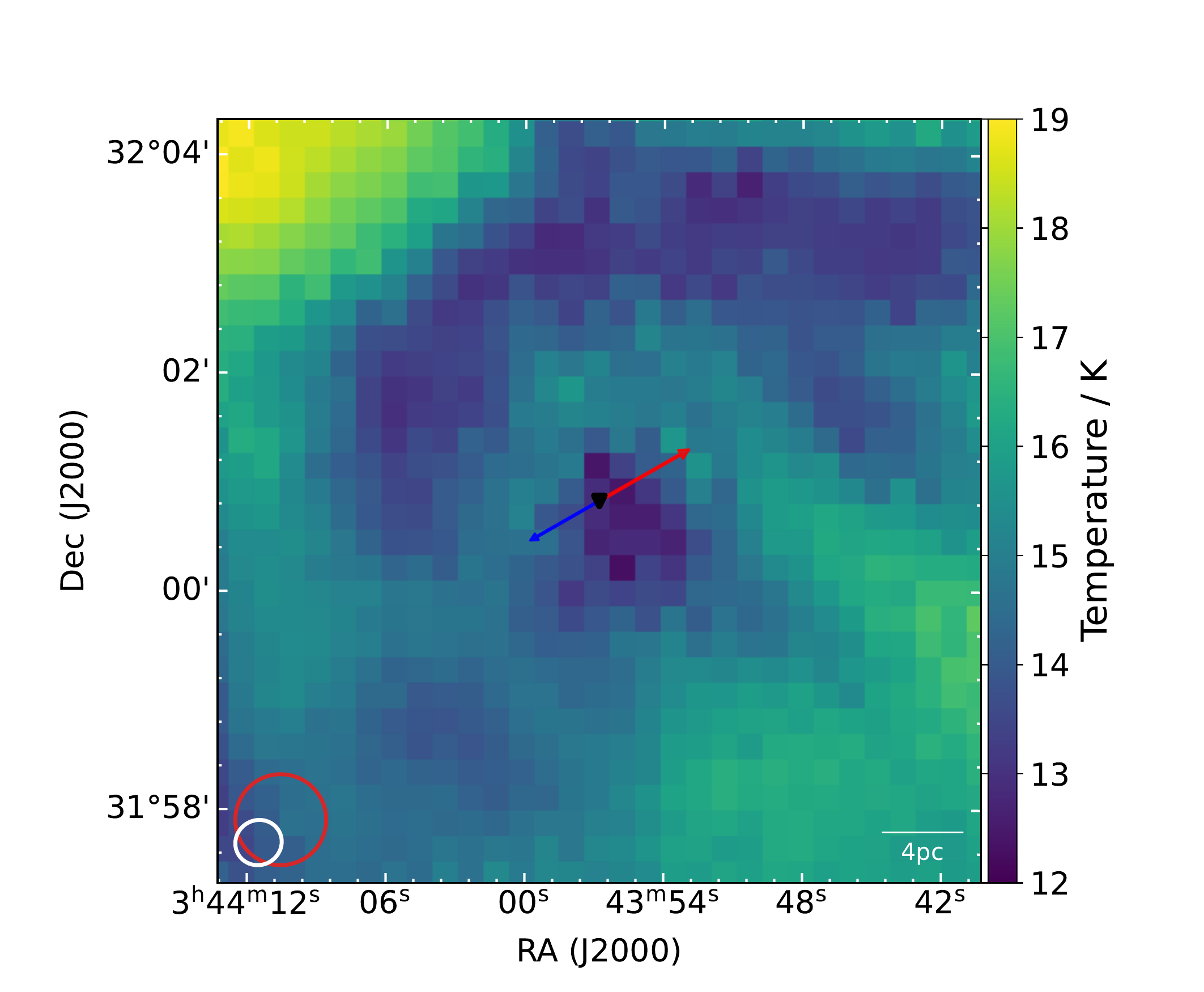}
   \caption{Analysis of \textit{Herschel} SPIRE data towards HH211. \textit{Top:} H$_2$ column density map of HH211, derived using the \textit{Herschel} SPIRE images at 250, 350 and 500\,$\mu$m. The red and blue arrows indicate the direction and size of the molecular outflow/jet from the protostellar core (marked by a black triangle). The \textit{Herschel} beam size is shown as a white circle, the average beam size of our observations is shown as red circle. \textit{Bottom:} Dust temperature map of HH211, derived using the \textit{Herschel} SPIRE images at 250, 350 and 500\,$\mu$m.}
              \label{Fig:HH211Herschel}
\end{figure}

To derive the physical structure of HH211, we used the \textit{Herschel} SPIRE dust continuum emission maps of the core from the \textit{Herschel}Science Archive (HSA). 
Using the three SPIRE bands at 250, 350 and 500\,$\mu$m, we derived the H$_2$ column density and the dust temperature map, following \cite{Harju2017}.
Therefore, we smoothed the 250\,$\mu$m and 350\,$\mu$m images to the resolution of the 500\,$\mu$m image ($35''$), and then fitted a modified blackbody to each pixel. We applied a dust emissivity spectral index of $\beta=2$, and a dust emissivity coefficient per unit mass of gas of $\kappa_{250\mu m}$= 0.1 cm$^2$ g$^{-1}$ \citep{Hildebrand1983}.
The two resulting maps are shown in Fig.~\ref{Fig:HH211Herschel}, with the direction and size of the jet indicated by red and blue arrows. 

From the H$_2$ column density map we derived a radial column density profile. 
For this, we masked out the structures north and south-east of the core, and the filamentary structure trailing the core in the south-west (see Fig.~\ref{Fig:HH211Herschel}).
To characterise the radial profile, we fitted a Plummer-like profile \citep{Plummer1911,WhitworthWardThompson2001}, that is modified by a constant term to account for the background column density \citep{Launhardt2013}:
\begin{equation}
    N_{H_2}(r)=\frac{N_{H_2,0}}{\left[1+\left(r/R_\mathrm{flat}\right)^2\right]^{\frac{p-1}{2}}} + N_\mathrm{out},
\end{equation}
where $N_{H_2,0}$ is the central column density, $R_\mathrm{flat}$ the characteristic radius of the flat inner portion of the density profile, and power-law index $p$.
The three fit parameters were then used to derive the radial volume density profile \citep{Andre2016}:
\begin{equation}\label{equ:nouter}
    n_\mathrm{outer}(r) = \frac{n_{H_2,0}}{\left[1+\left(r/R_\mathrm{flat}\right)^2\right]^{p/2}}.
\end{equation}
The central volume density $n_{H_2,0}$ is related to the central column density by $n_{H_2,0}=N_{H_2,0}/\left(A_PR_\mathrm{flat}\right)$, where $A_P$ is a constant factor related to the inclination angle of the core to the plane of sky, $A_P=\frac{1}{\cos i}\times B\left(\frac{1}{2},\frac{p-1}{2}\right)$, and $B$ is the Euler Beta function. 
Assuming $i=0^\circ$, we derived $N_0 = 9.7(6)\times10^{22}$\,cm$^{-2}$, $r_\mathrm{flat} = 8(1)\times10^{-2}$\,pc, $p = 3.0(4)$, $N_\mathrm{out} = 6(1)\times10^{21}$\,cm$^{-2}$, and $n_0 = 1.9(3)\times10^5$\,cm$^{-3}$.

For the inner regions of the core, we adapted the frequently used physical model for IRAS 16293-2422 A/B \citep{Crimier2010}, where the radial density distribution is described by a power law, $n(H_2)=r^{-1.8}$. After normalising the profile with the central pixel of the \textit{Herschel} column density map, we arrived at a volume density profile of
\begin{equation}\label{equ:ninner}
    n_\mathrm{inner}(r) =  n_1\cdot \left(\frac{r}{r_1}\right)^{-1.5},
\end{equation}
where $n_1=1.89\times10^5$\,cm$^{-3}$ and $r_1=0.0128$\,pc are the starting values of $n_\mathrm{outer}$. This Shu-like density distribution is for example also used in \cite{Quenard2018}.



To parameterise the radial temperature profile, we applied an empirical fit following \cite{Arzoumanian2011}:
\begin{equation}
    T(r) = T_\mathrm{out}-\frac{\Delta T}{\left(1+\left(\frac{r}{r_\mathrm{flat}}\right)^2\right)^{q/2}}.
\end{equation}
We fixed the values for $T_\mathrm{out}$ and $\Delta T$ to 21\,K and 8\,K, respectively, resulting in best-fit parameters of $r_\mathrm{flat}=0.17(1)$\,pc and $q=0.38(3)$.
The profile is decreasing towards the centre, with a central temperature of around 13\,K, and does not show an internal heating source. This is caused by the low spectral resolution of the \textit{Herschel} temperature map, where one pixel covers around 6000\,au. For example, \cite{Launhardt2013} show that embedded protostars only influence the temperature in the inner 5000\,au of a globule.

To account for the central heating, we adopted the radial dust temperature profile for protostellar envelopes introduced by \cite{MotteAndre2001}:
\begin{equation}
    T_\mathrm{dust}(r,L_*) \approx 38\,\mathrm{K} \times \left(\frac{r}{100\,\mathrm{AU}}\right)^{-0.4}\left(\frac{L_*}{1\,L_\odot}\right)^{0.2}.
\end{equation}
Assuming that internal heating by the accreting protostar dominates the thermal balance of the envelope, and the envelope is optically thin to the bulk of the infrared radiation.
To derive the internal luminosity of the protostar, we followed \cite{Dunham2008}, using the 70\,$\mu$m flux as an estimate:
\begin{equation}
    L_\mathrm{int}=3.3\times10^8 F_{70}^{0.94}\,L_\odot.
\end{equation}
Using the $70\,\mu$m flux of HH211 published by \cite{Dunham2015}, we derived a protostellar luminosity of $L_*=1.2(2)$\,L$_\odot$.
For the purpose of our radiative transfer simulations, we assumed the gas temperature in the core to be equal to the dust temperature. This is reasonable, as gas and dust are closely coupled at densities above $10^4$\,cm$^{-3}$ \citep{Goldsmith2001}.

For the infall velocity profile, we followed \cite{Crimier2010} and assumed a free fall velocity gradient,
\begin{equation}
    v(r) = \sqrt{\frac{2gM}{r}},
\end{equation}
where the mass $M$ is given by the protostellar mass, 0.05\,M$_\odot$ \citep{LeeLi2018}.
\cite{Pineda2019} show that HH211 is in fact not in free fall, but is rotating. However, as we are working with a 1D model of the source, the freefall velocity profile is somewhat mimicing the rotation by the steep increase of velocity towards the centre (see Pineda et al. in prep).

\begin{figure}
   \centering
    \includegraphics[width=\hsize]{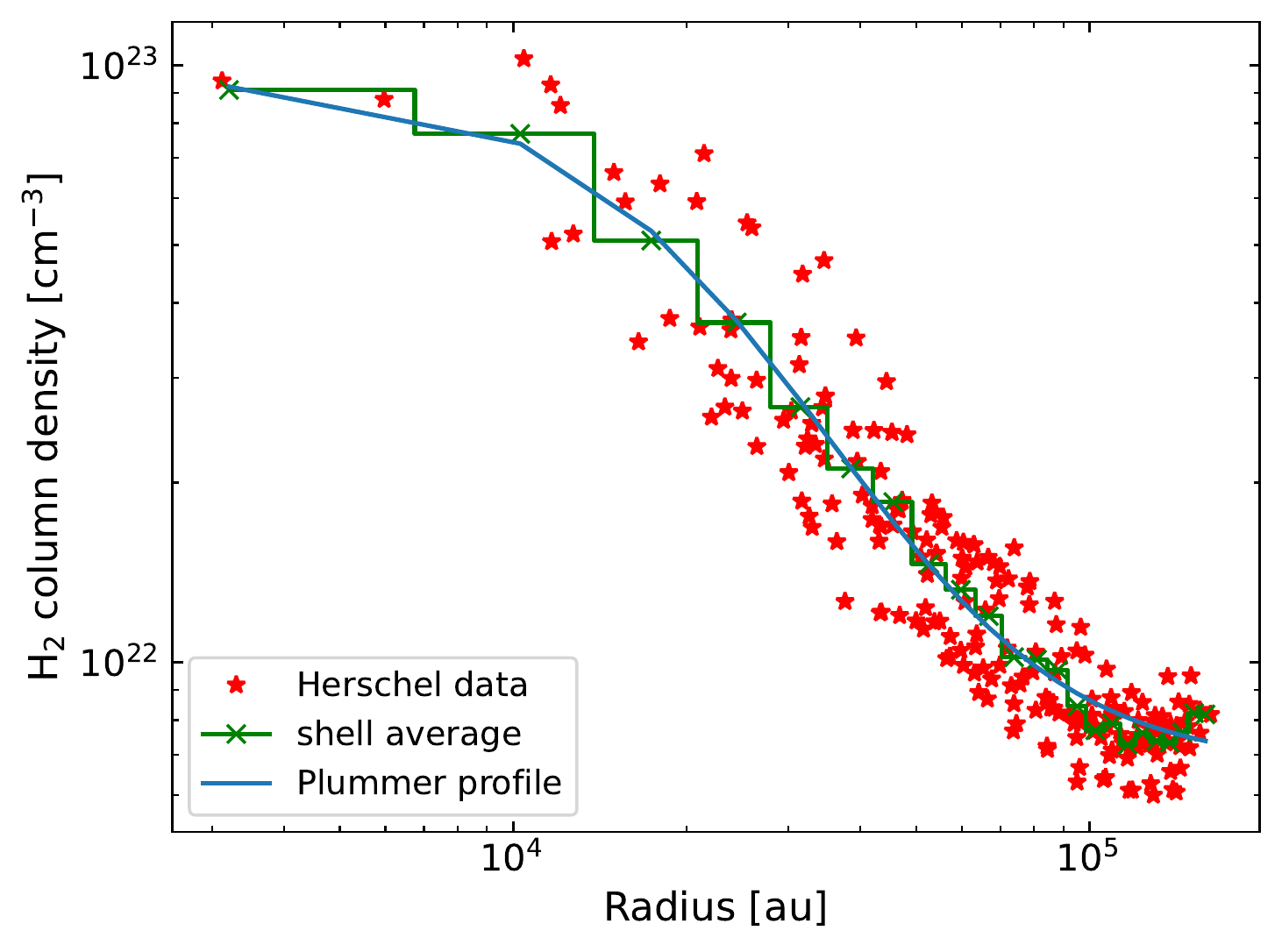}
    \includegraphics[width=\hsize]{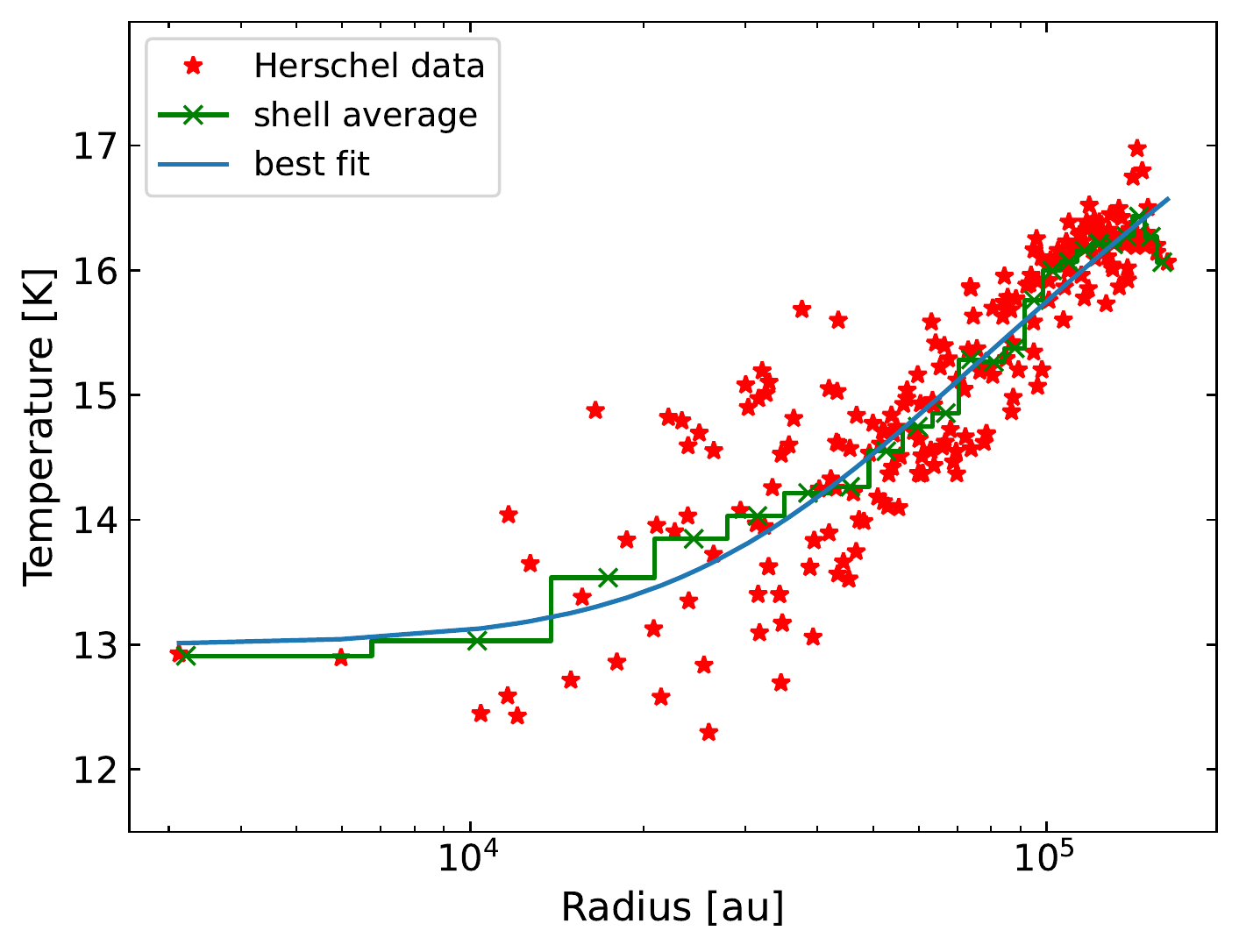}
   \caption{Analysis of the H$_2$ column density map and the temperature map derived from \textit{Herschel} data. \textit{Top:} Radial H$_2$ column density profile (red), overlaid with the fitted Plummer-like profile (blue). Given in green are the data points averaged over shells of 11\,arcsec, the size corresponding to 1/3 of the \textit{Herschel} beam size. \textit{Bottom:} Radial temperature profile (red), overlaid with the best fit (blue) and the averaged shells (green).}
              \label{Fig:HH211HerschelDensity}
\end{figure}

\section{Molecular abundance profiles of the best-fit modelling results}\label{abundanceprofiles}

\begin{figure*}
   \centering
   \includegraphics[width=0.33\textwidth]{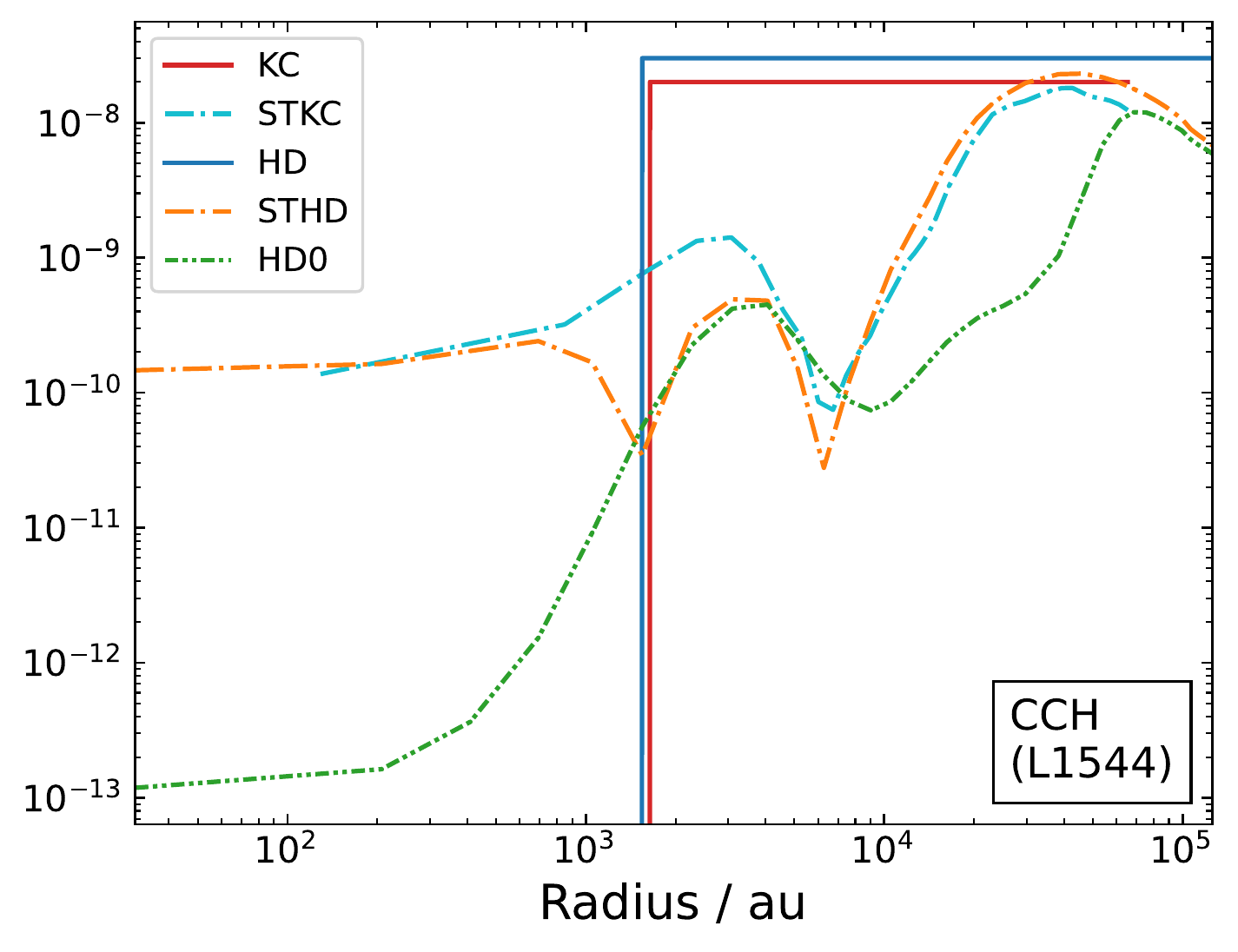}
   \includegraphics[width=0.33\textwidth]{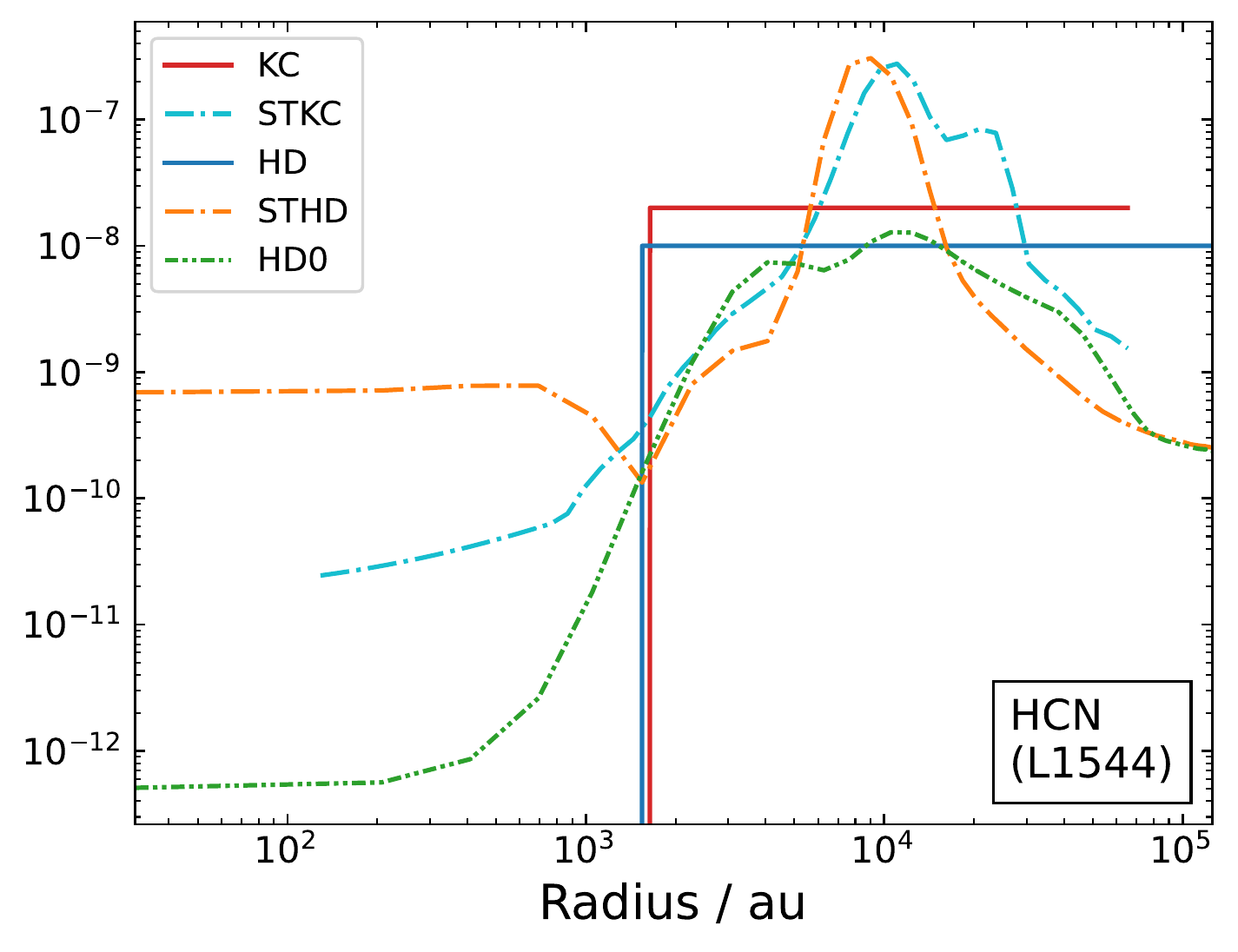}
   \includegraphics[width=0.33\textwidth]{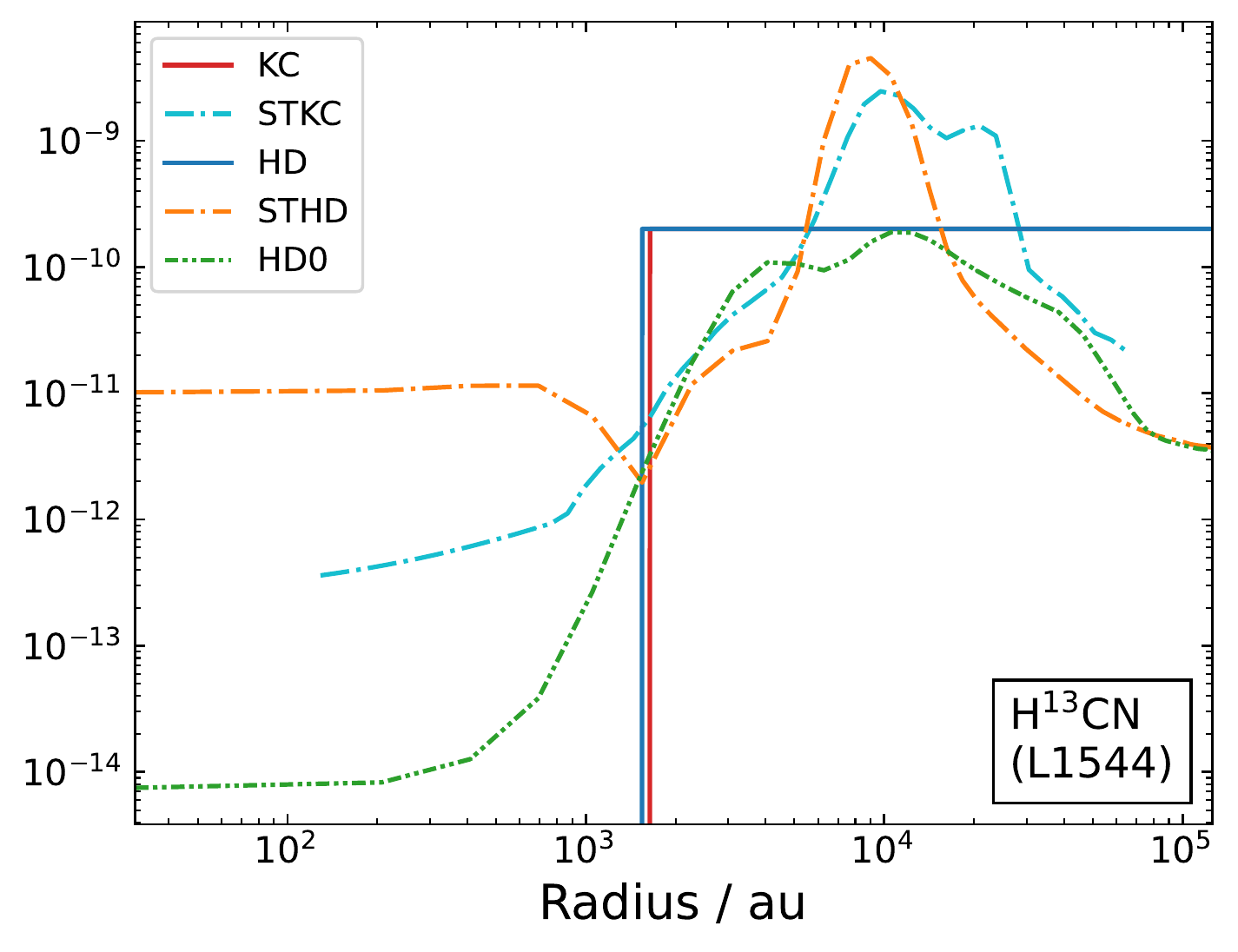}
   \includegraphics[width=0.33\textwidth]{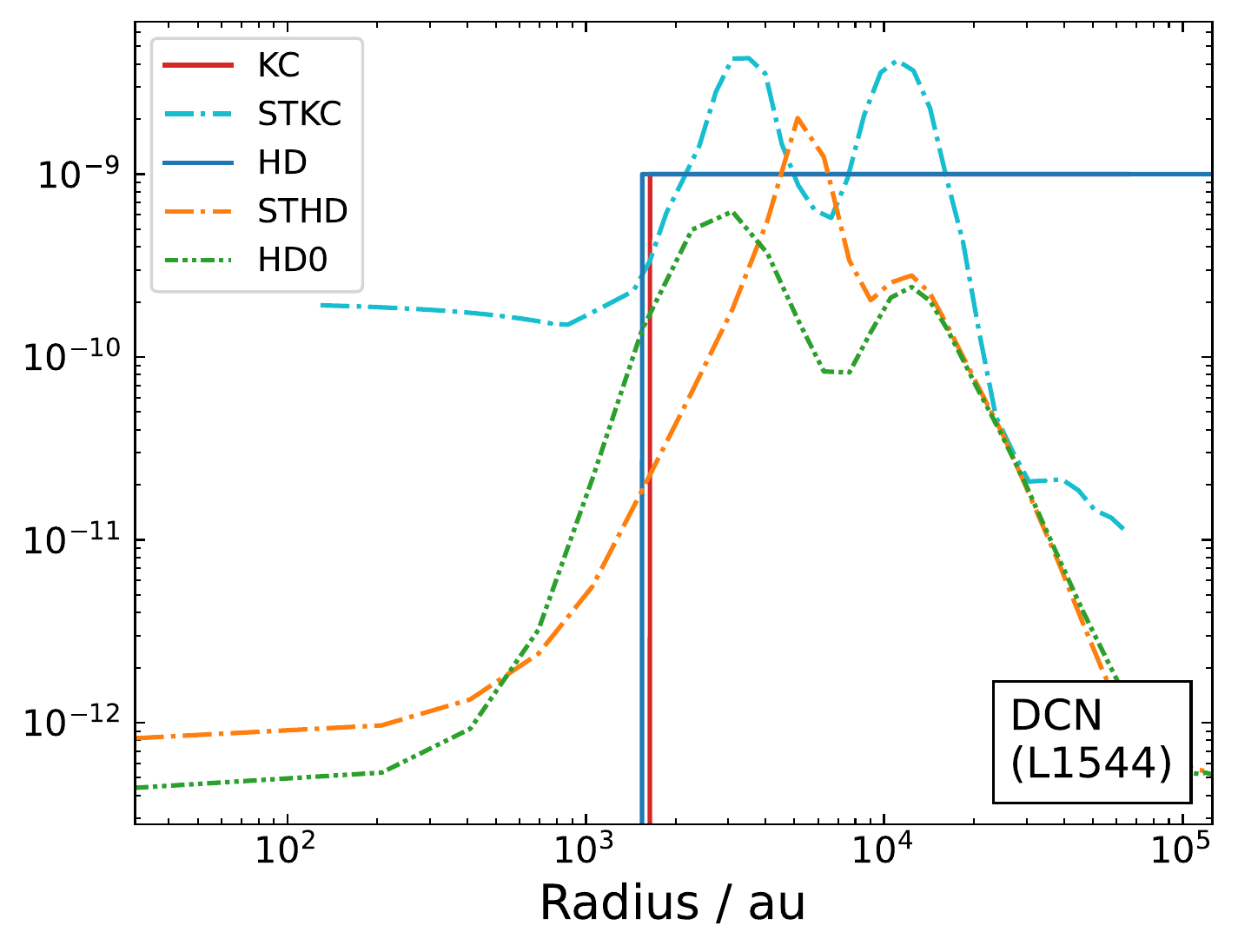}
   \includegraphics[width=0.33\textwidth]{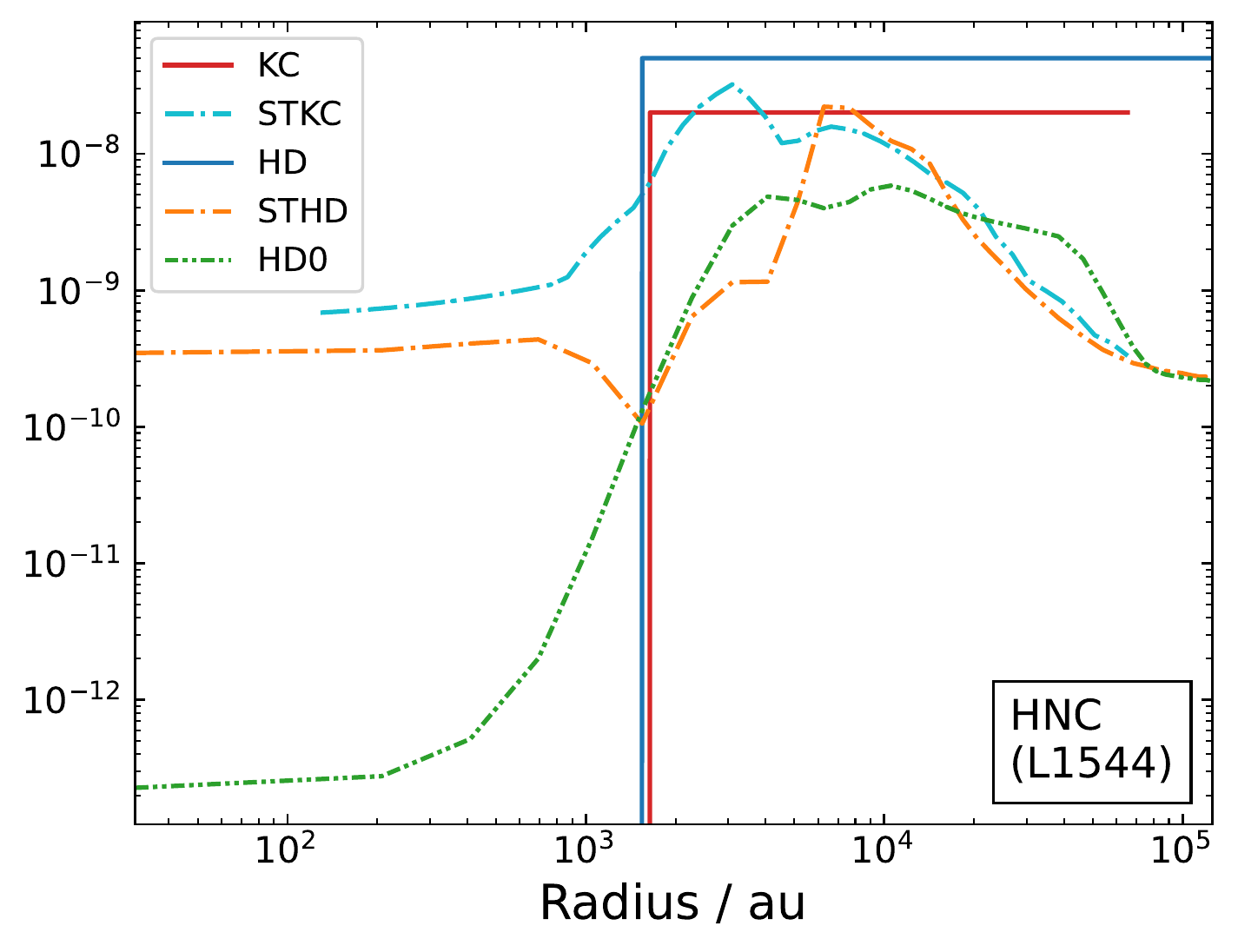}
   \includegraphics[width=0.33\textwidth]{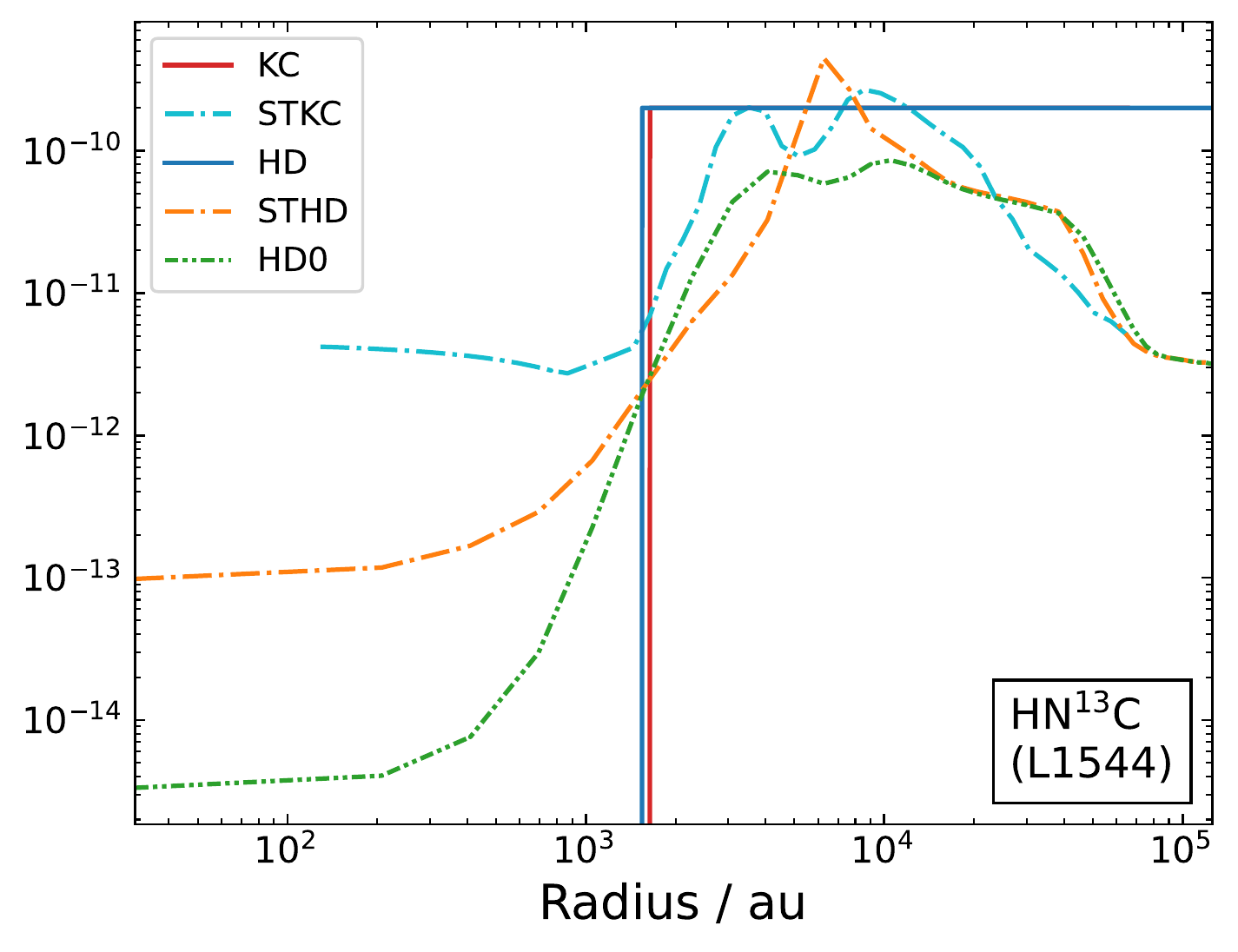}
   \includegraphics[width=0.33\textwidth]{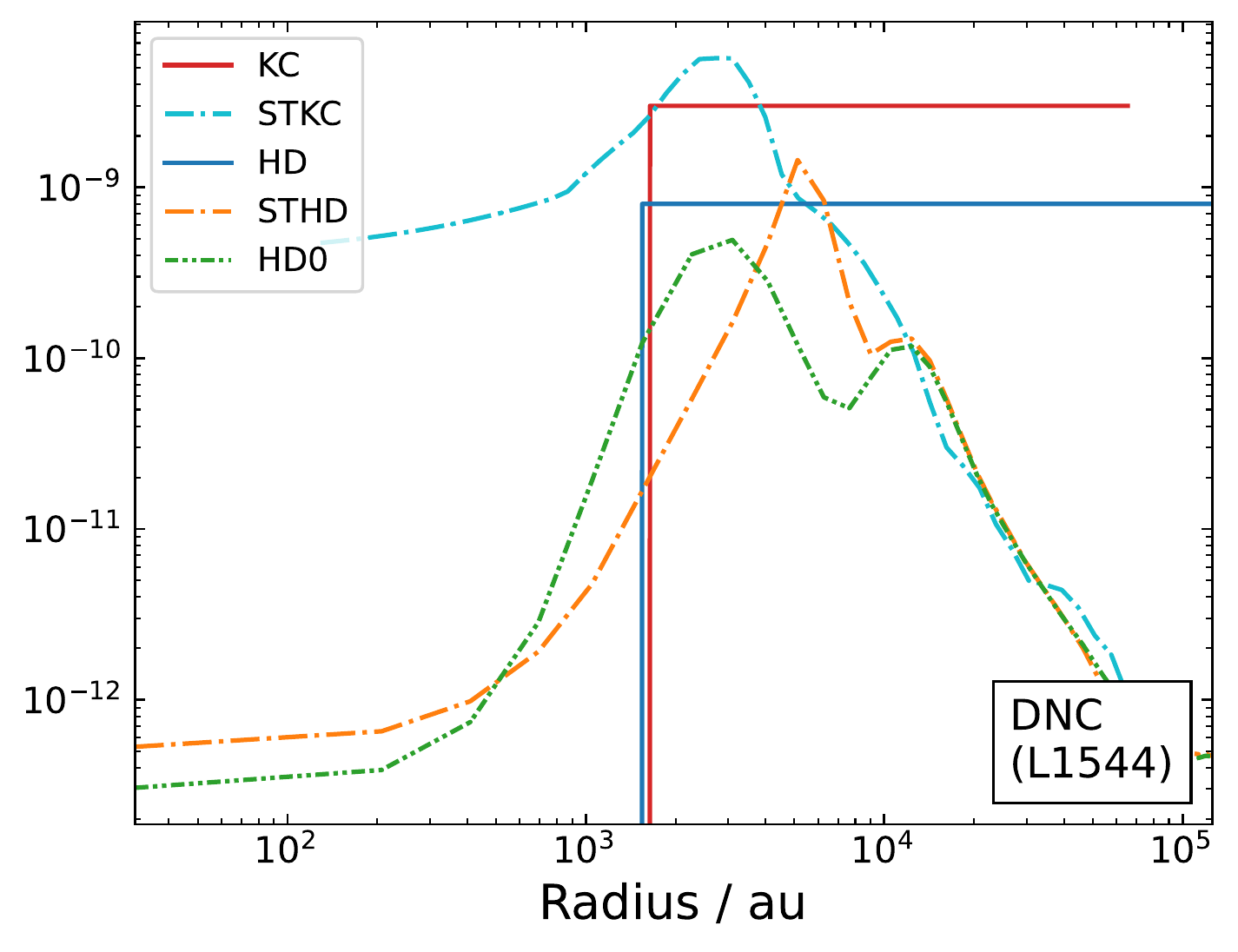}
   \includegraphics[width=0.33\textwidth]{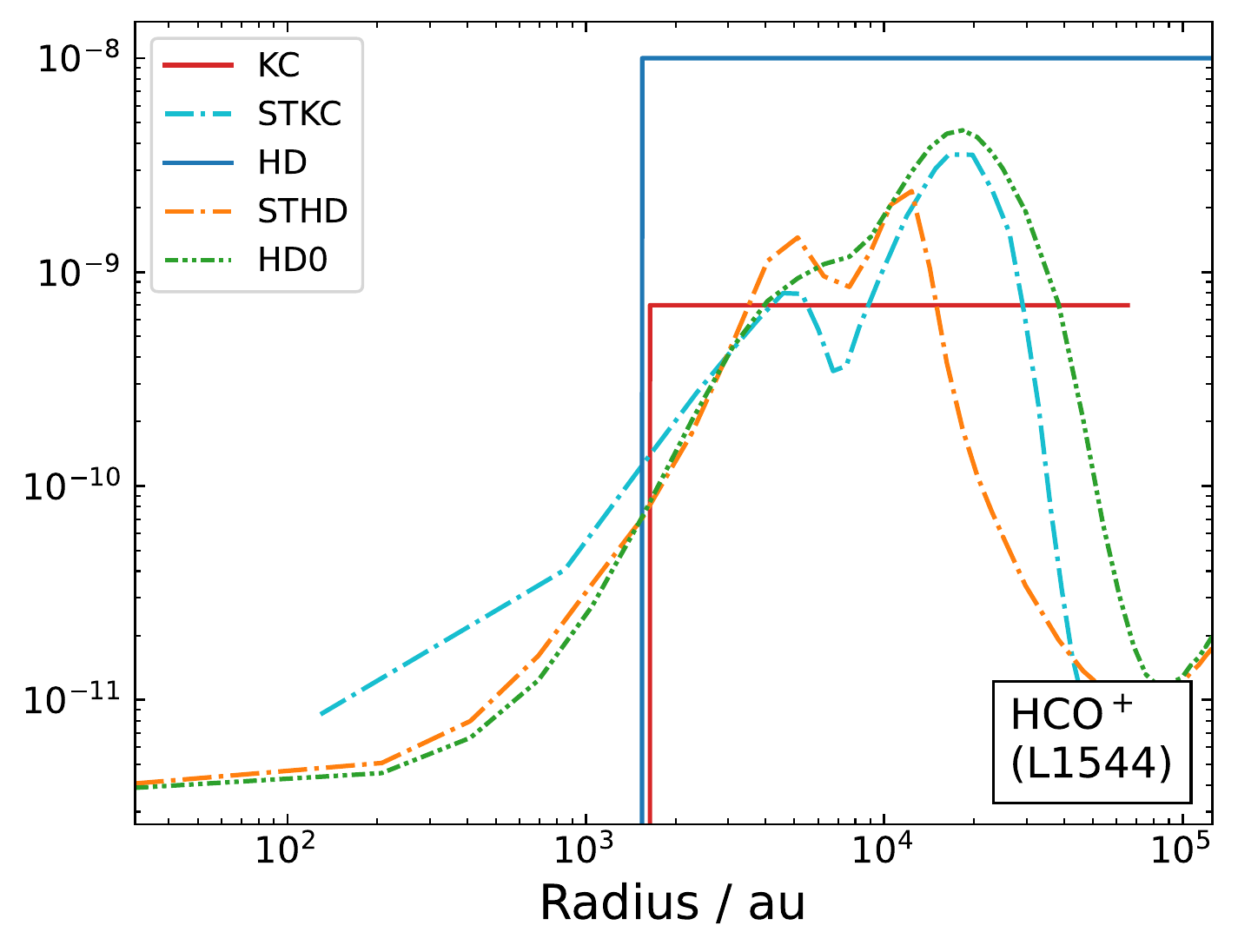}
   \includegraphics[width=0.33\textwidth]{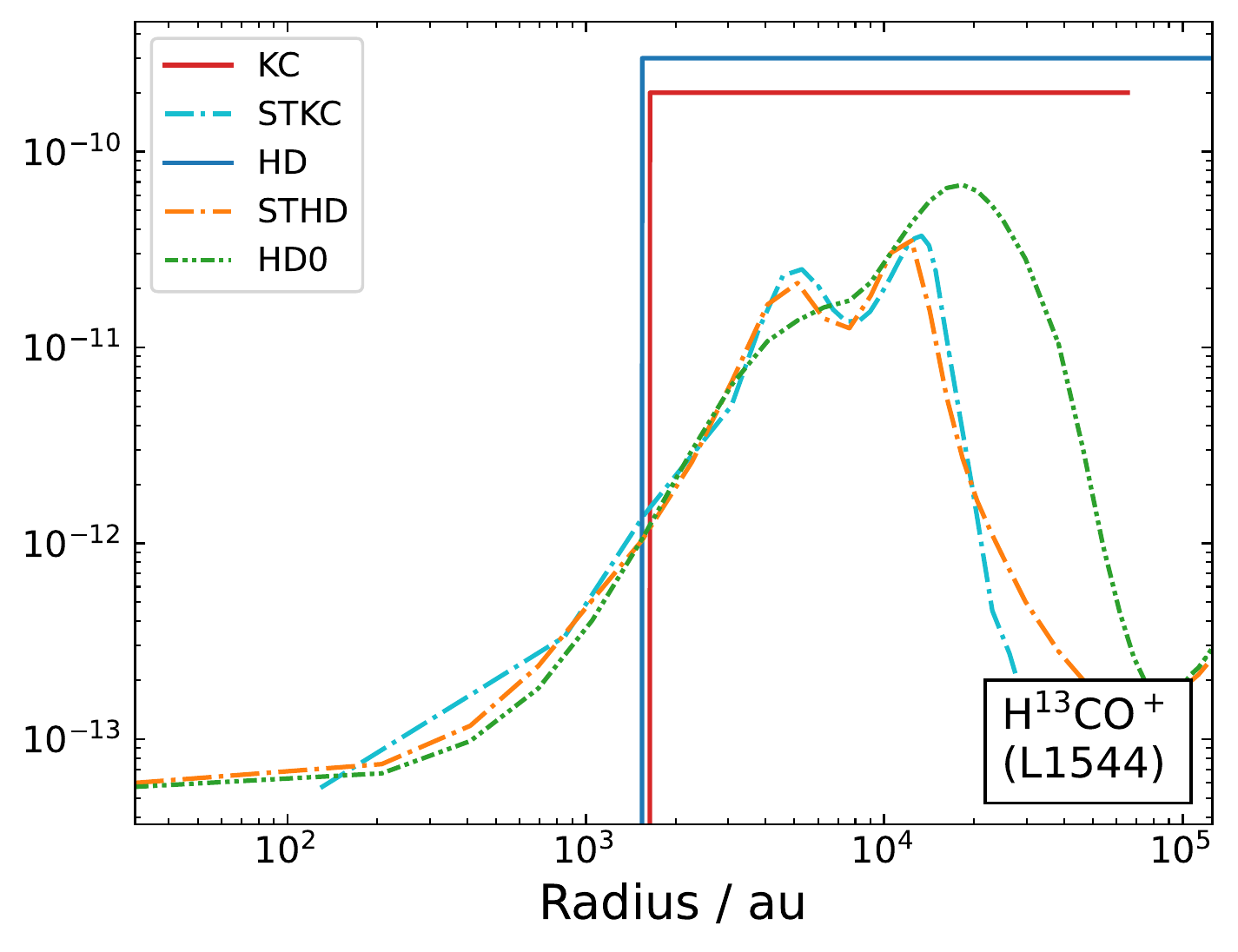}
   \includegraphics[width=0.33\textwidth]{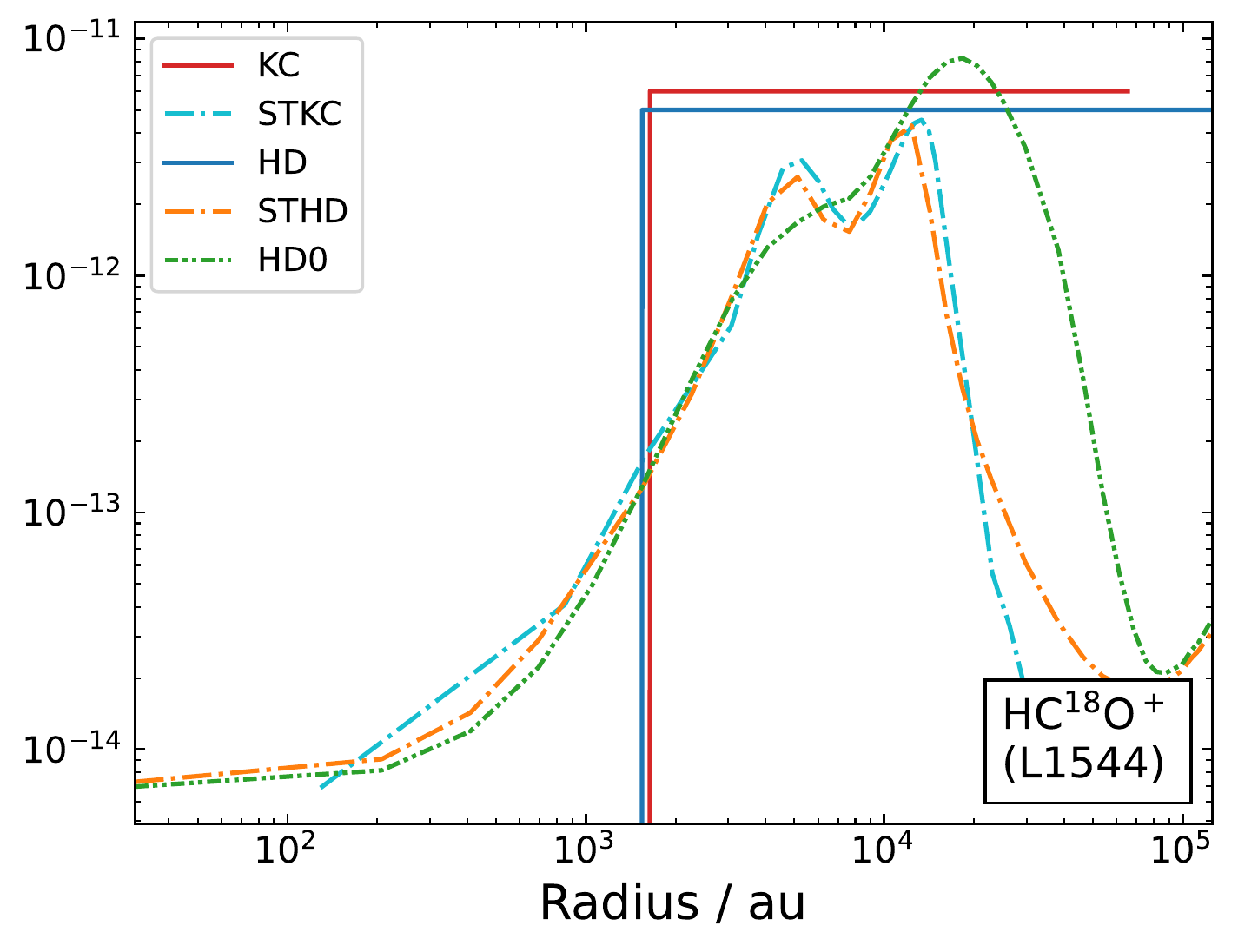}
   \includegraphics[width=0.33\textwidth]{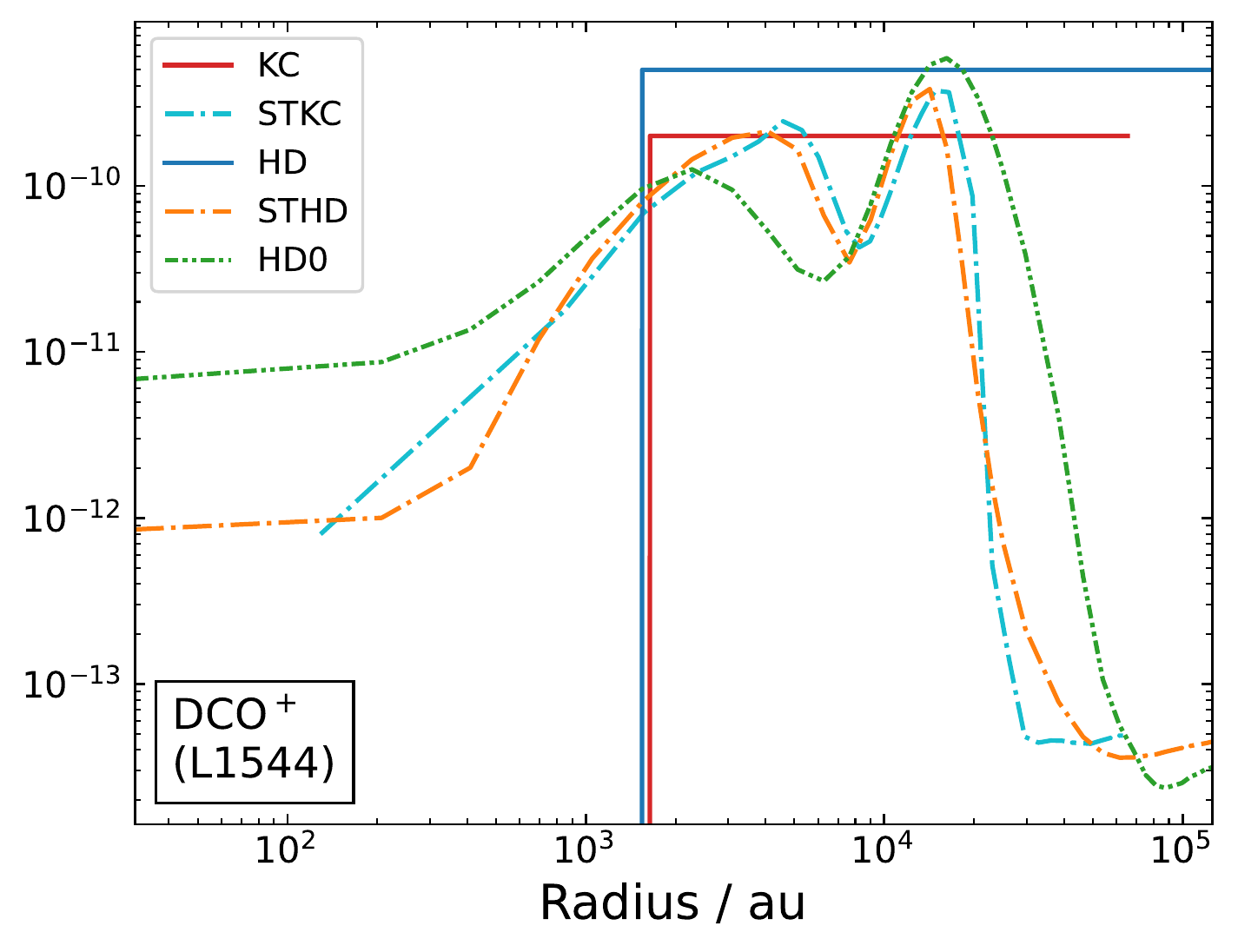}
   \caption{Fractional molecular abundance (with respect to H$_2$) profiles of the best-fit results produced with LOC for the spectra observed towards L1544. In the case of the $^{13}$C and $^{18}$O isotopologues, the abundance profiles correspond to the profiles of the main species, scaled down by the isotopic ratio (68 and 557, respectively).}
              \label{Fig:LOC-abuprofilesL1544}
\end{figure*}

\begin{figure*}
   \centering
   \includegraphics[width=0.33\textwidth]{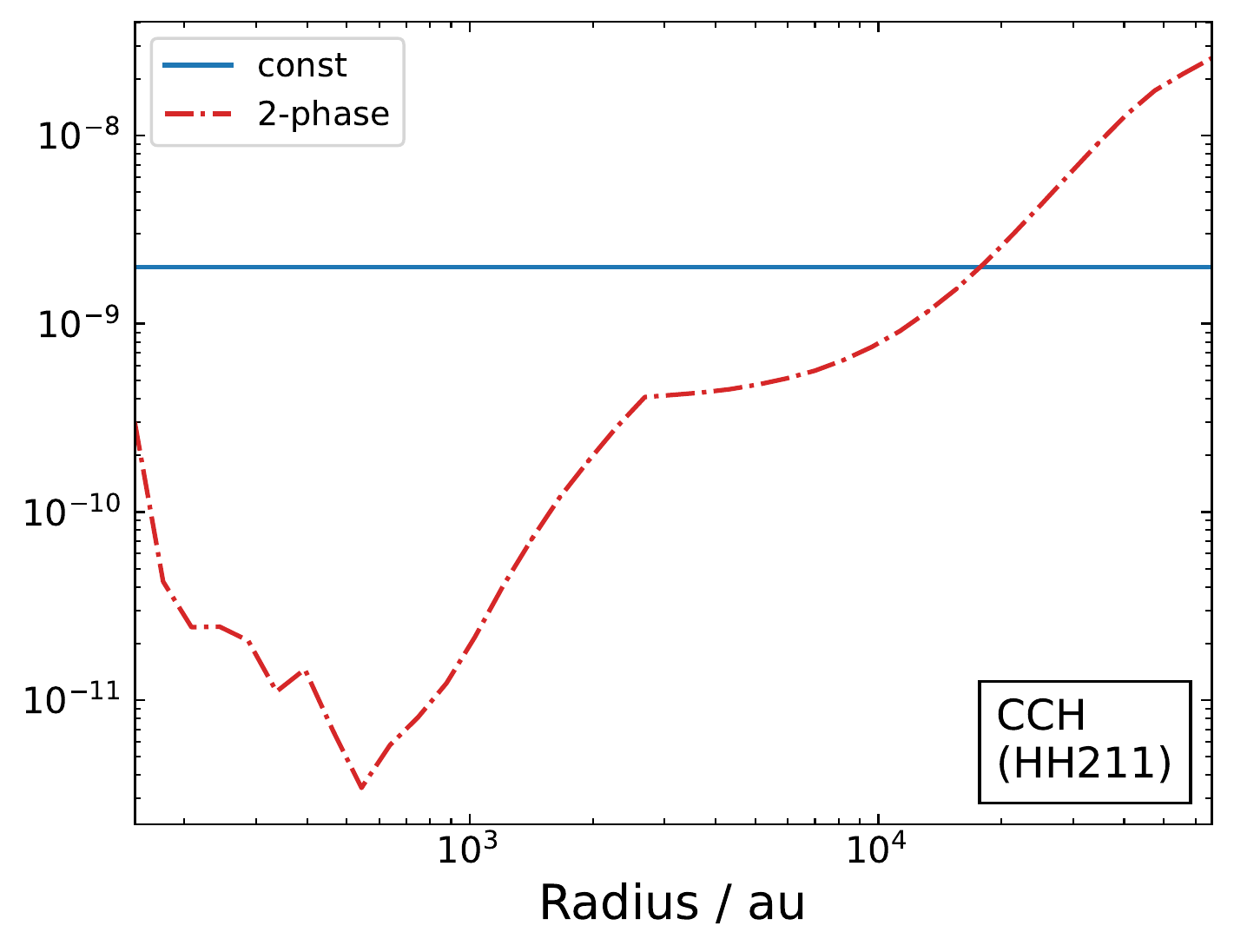}
   \includegraphics[width=0.33\textwidth]{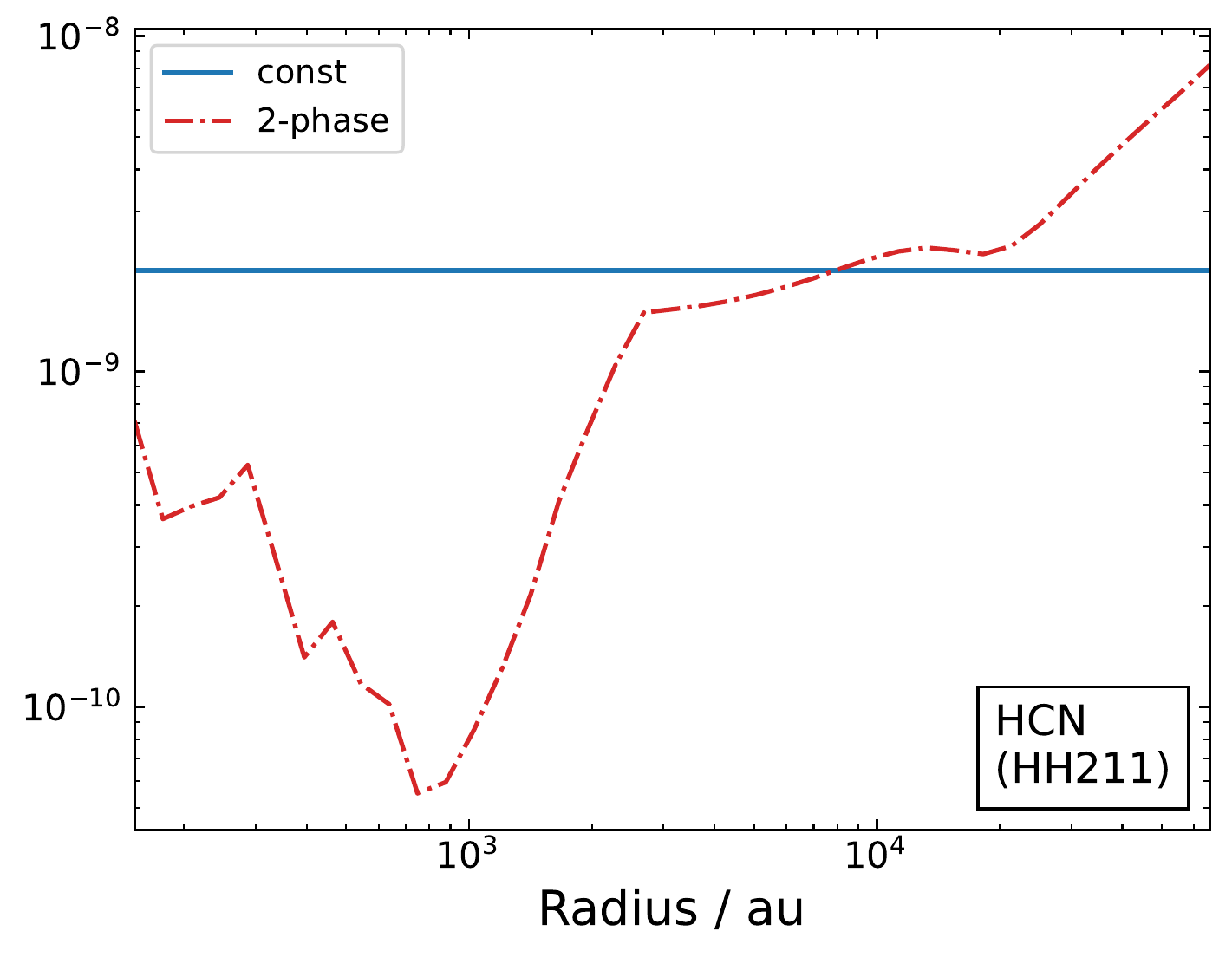}
   \includegraphics[width=0.33\textwidth]{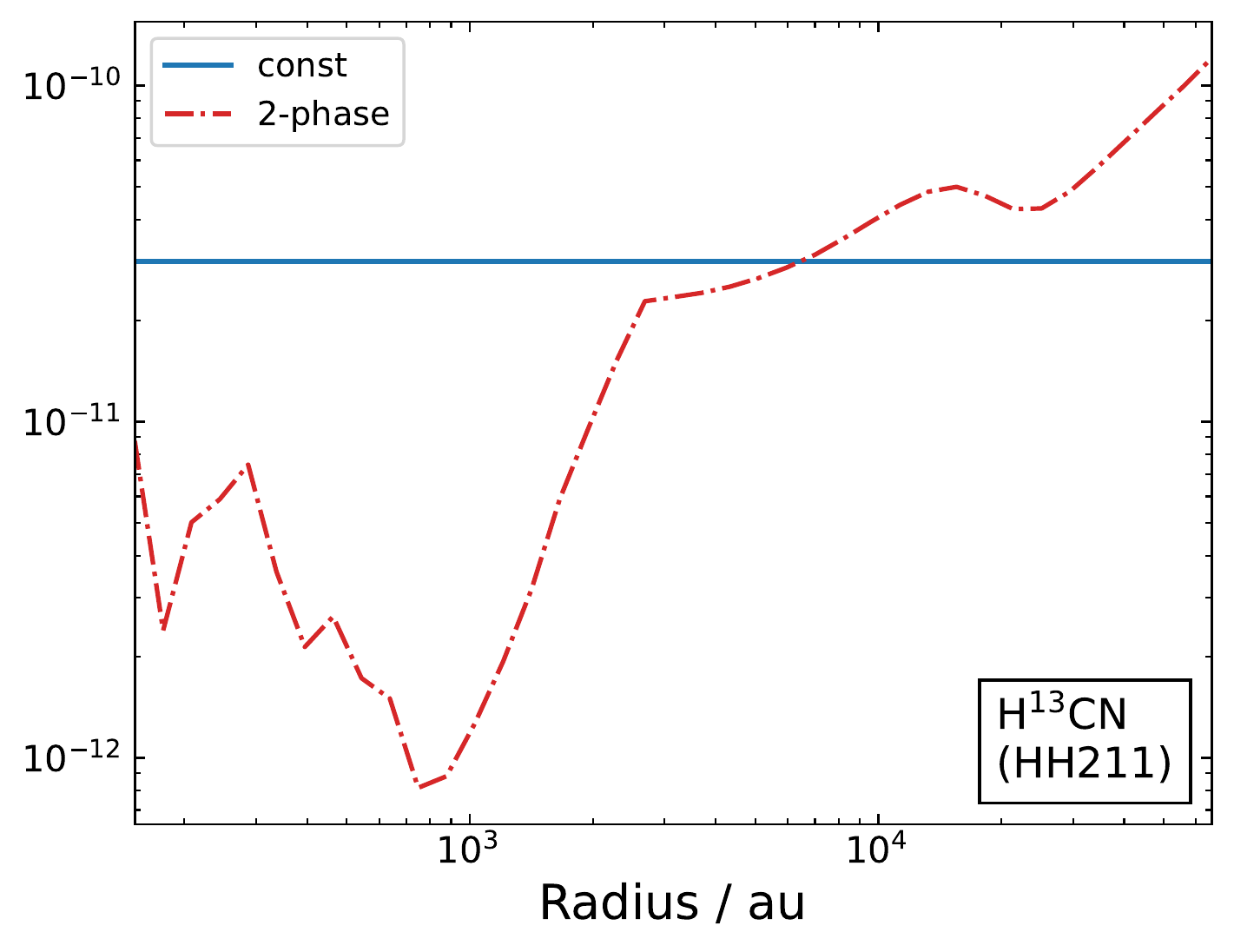}
   \includegraphics[width=0.33\textwidth]{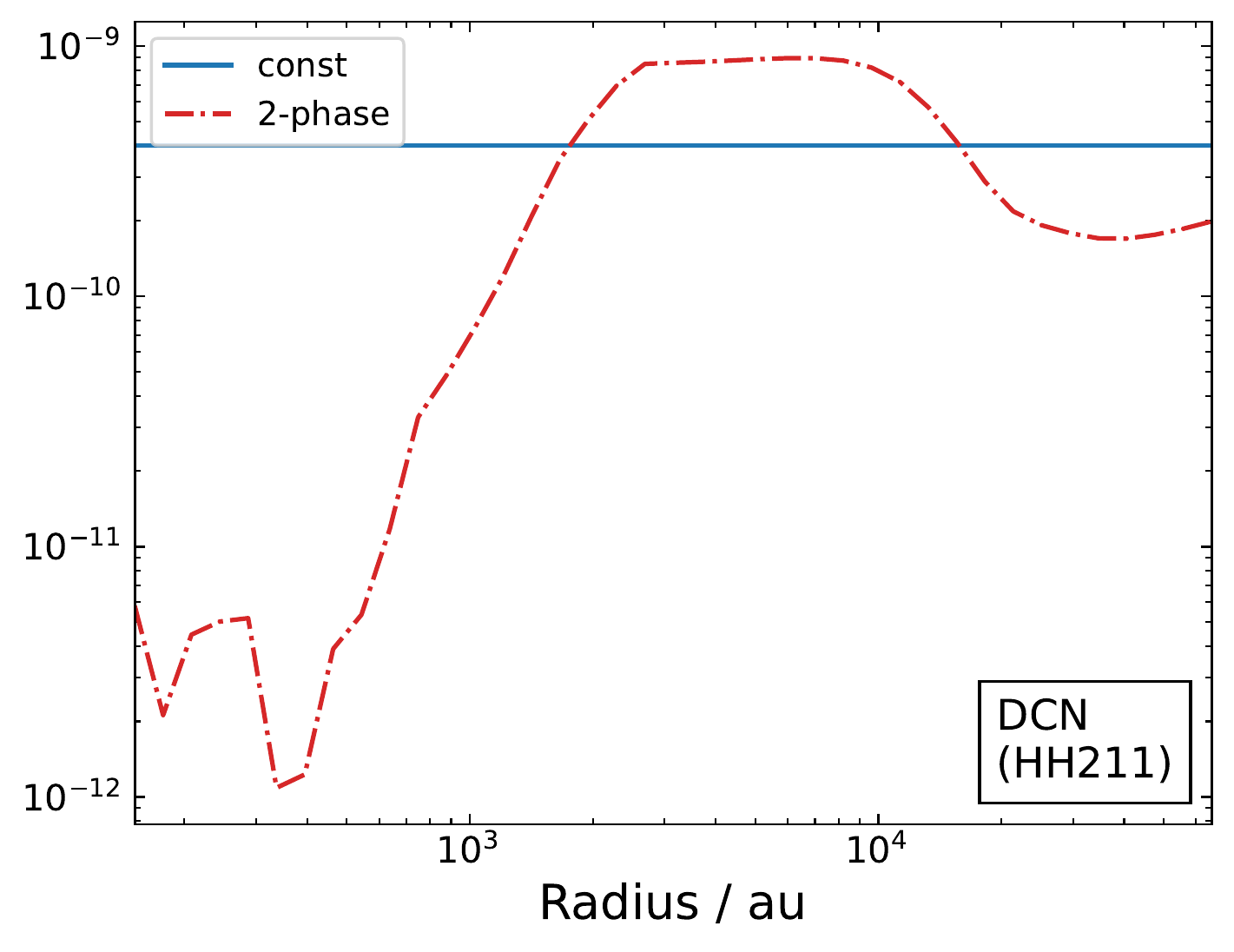}
   \includegraphics[width=0.33\textwidth]{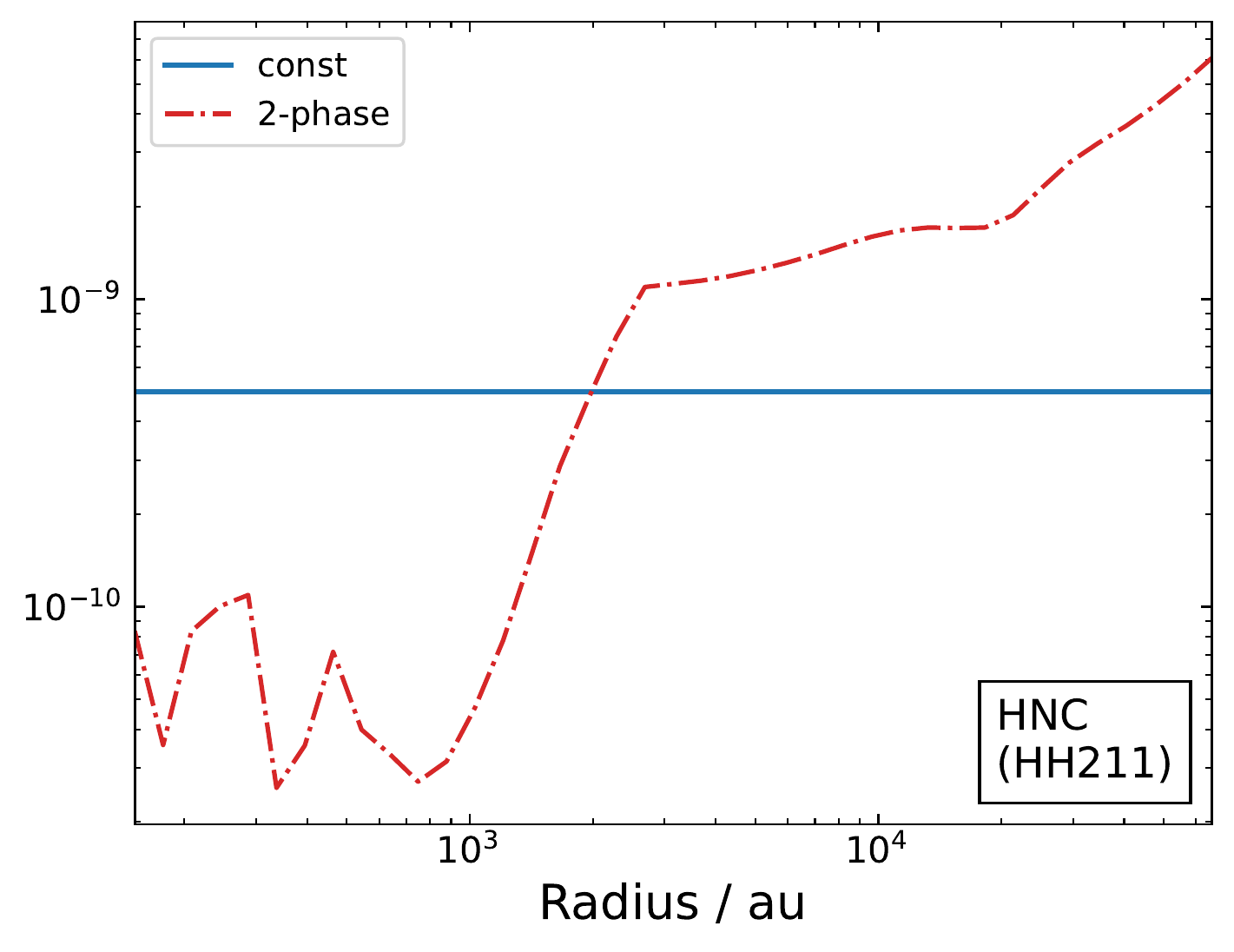}
   \includegraphics[width=0.33\textwidth]{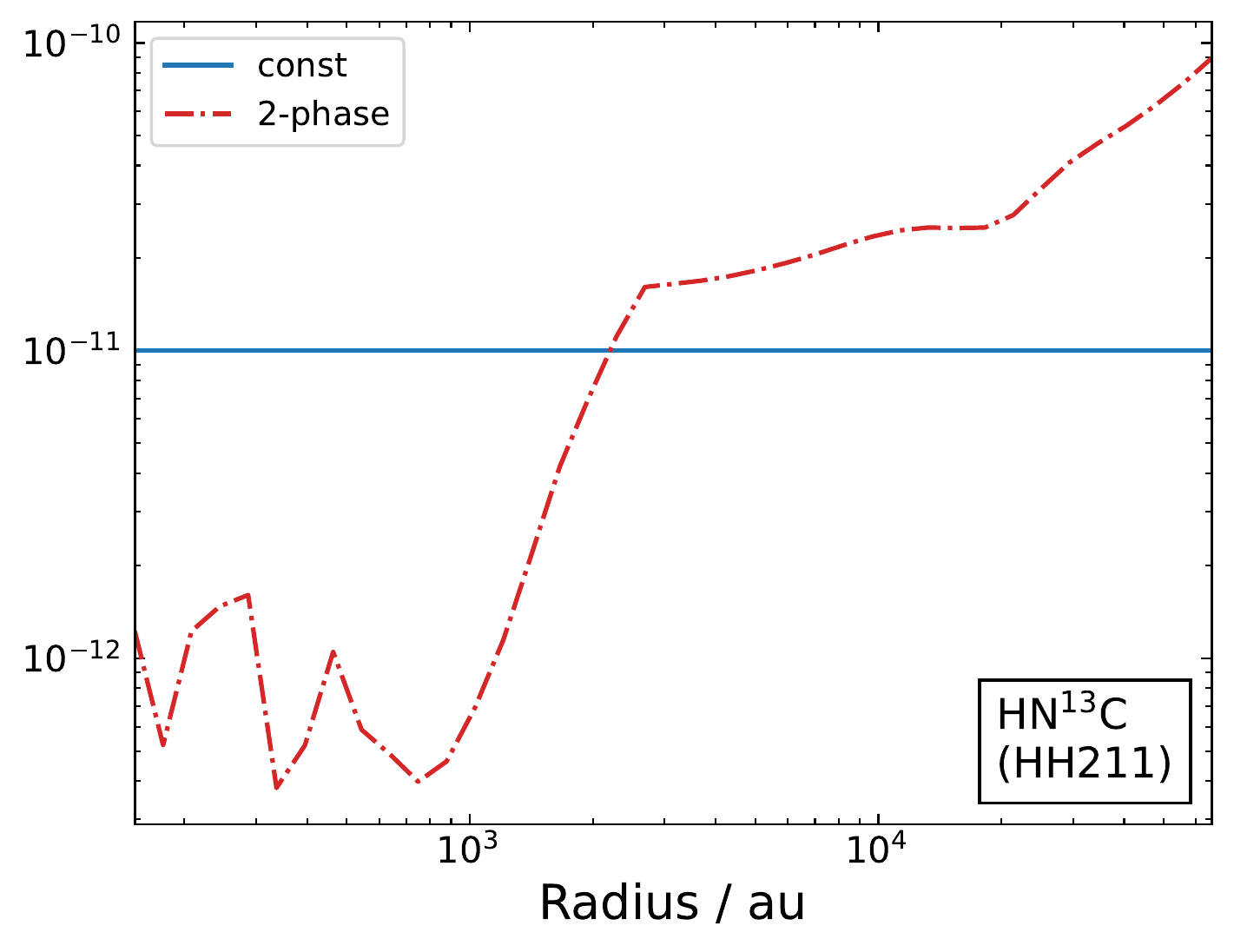}
   \includegraphics[width=0.33\textwidth]{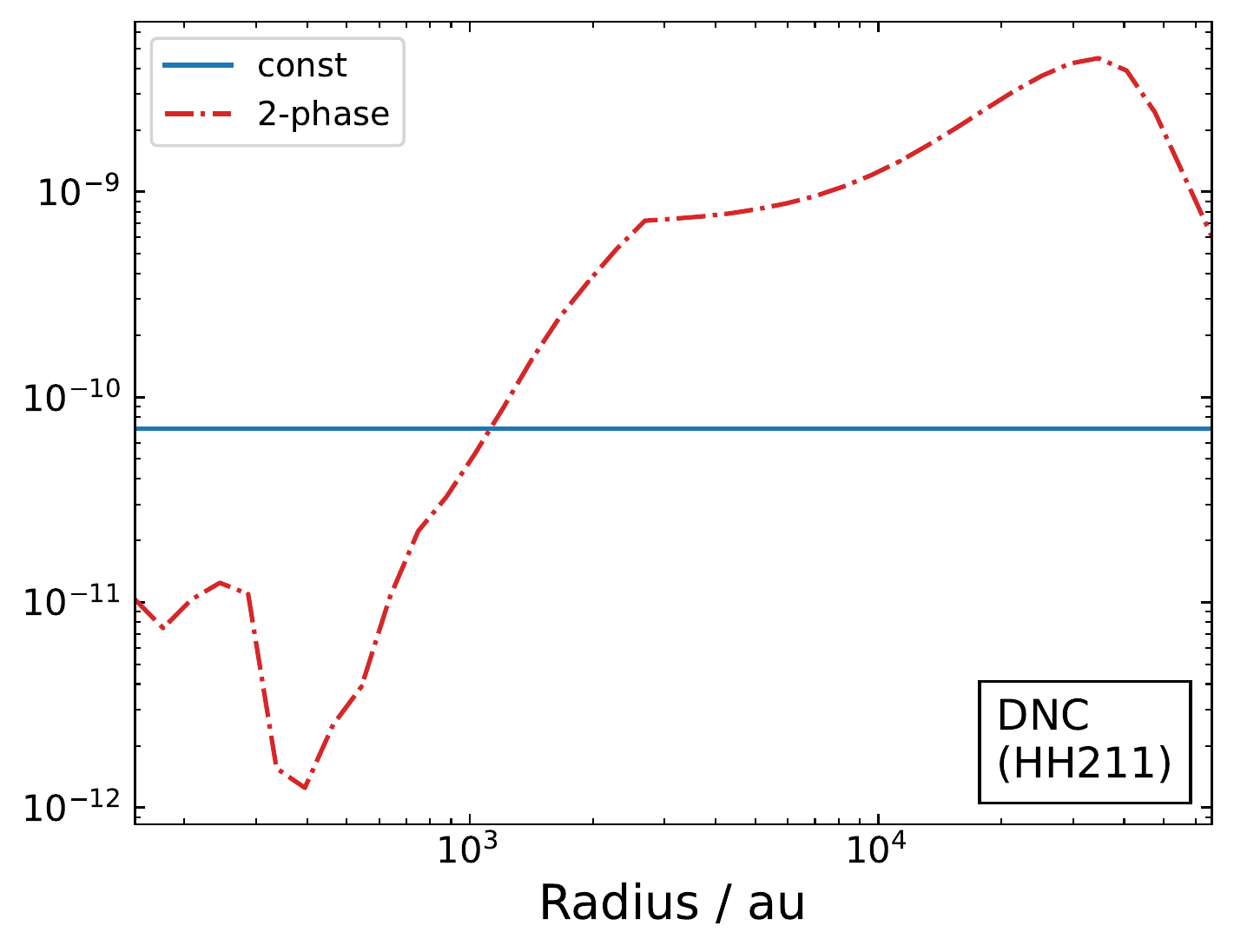}
   \includegraphics[width=0.33\textwidth]{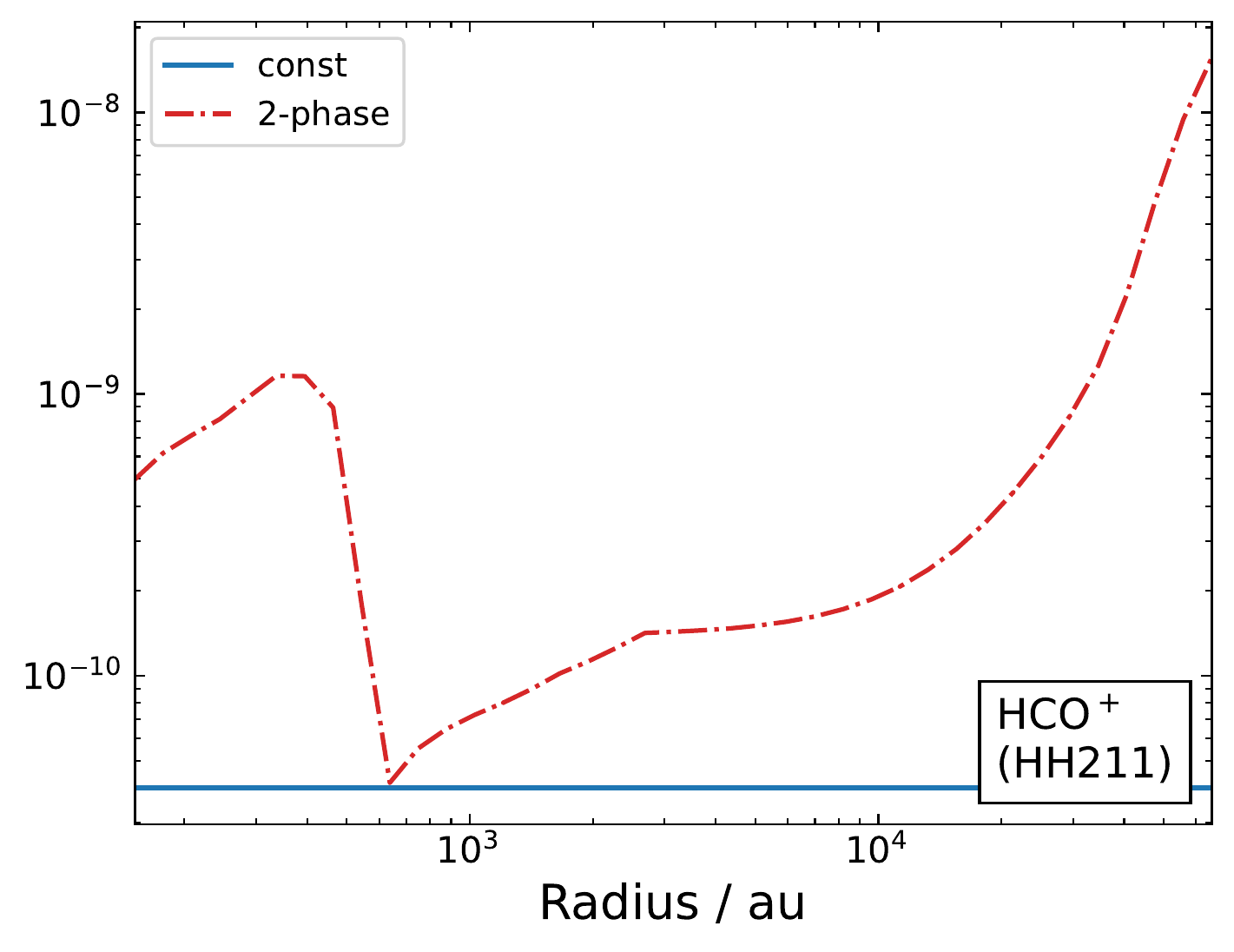}
   \includegraphics[width=0.33\textwidth]{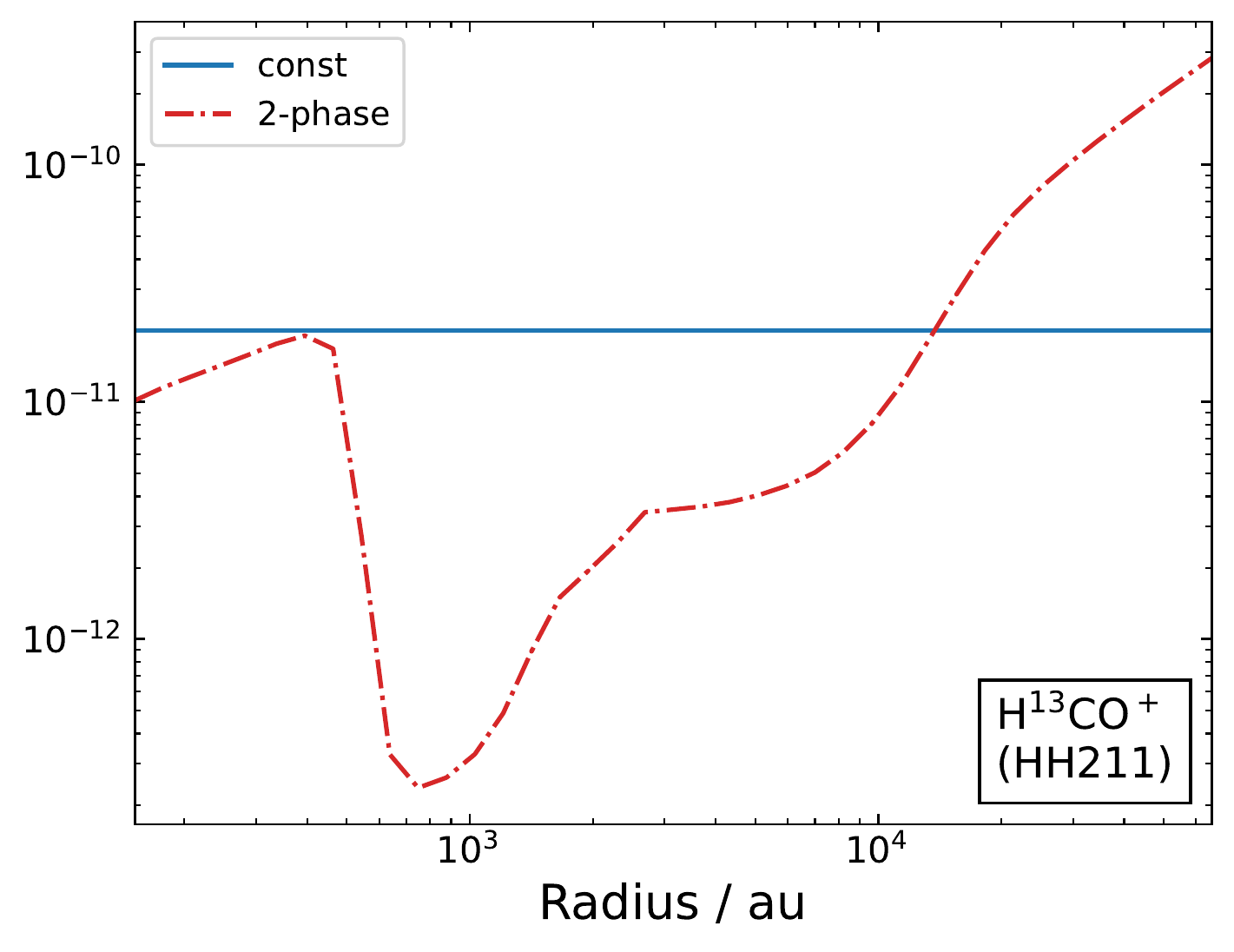}
   \includegraphics[width=0.33\textwidth]{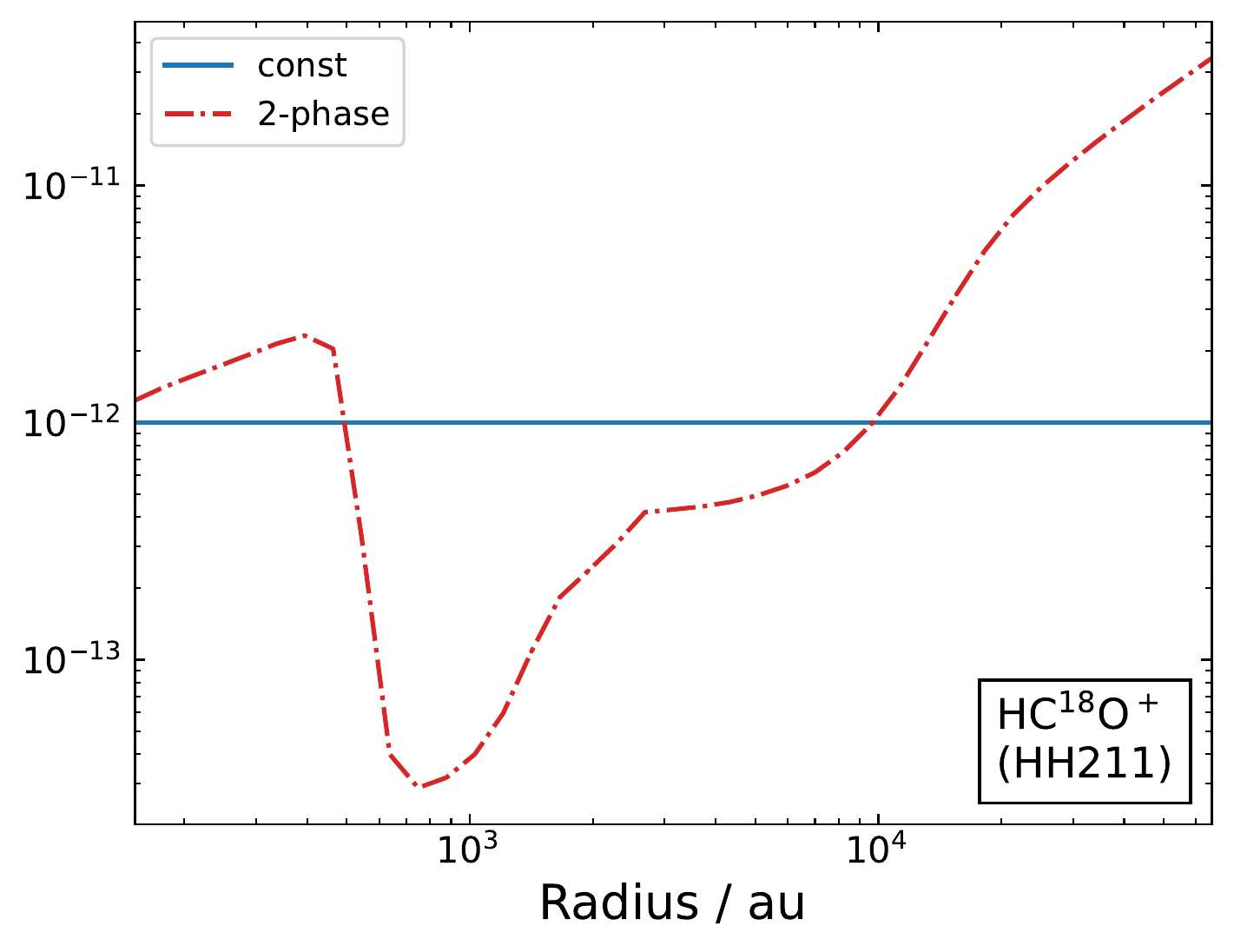}
   \includegraphics[width=0.33\textwidth]{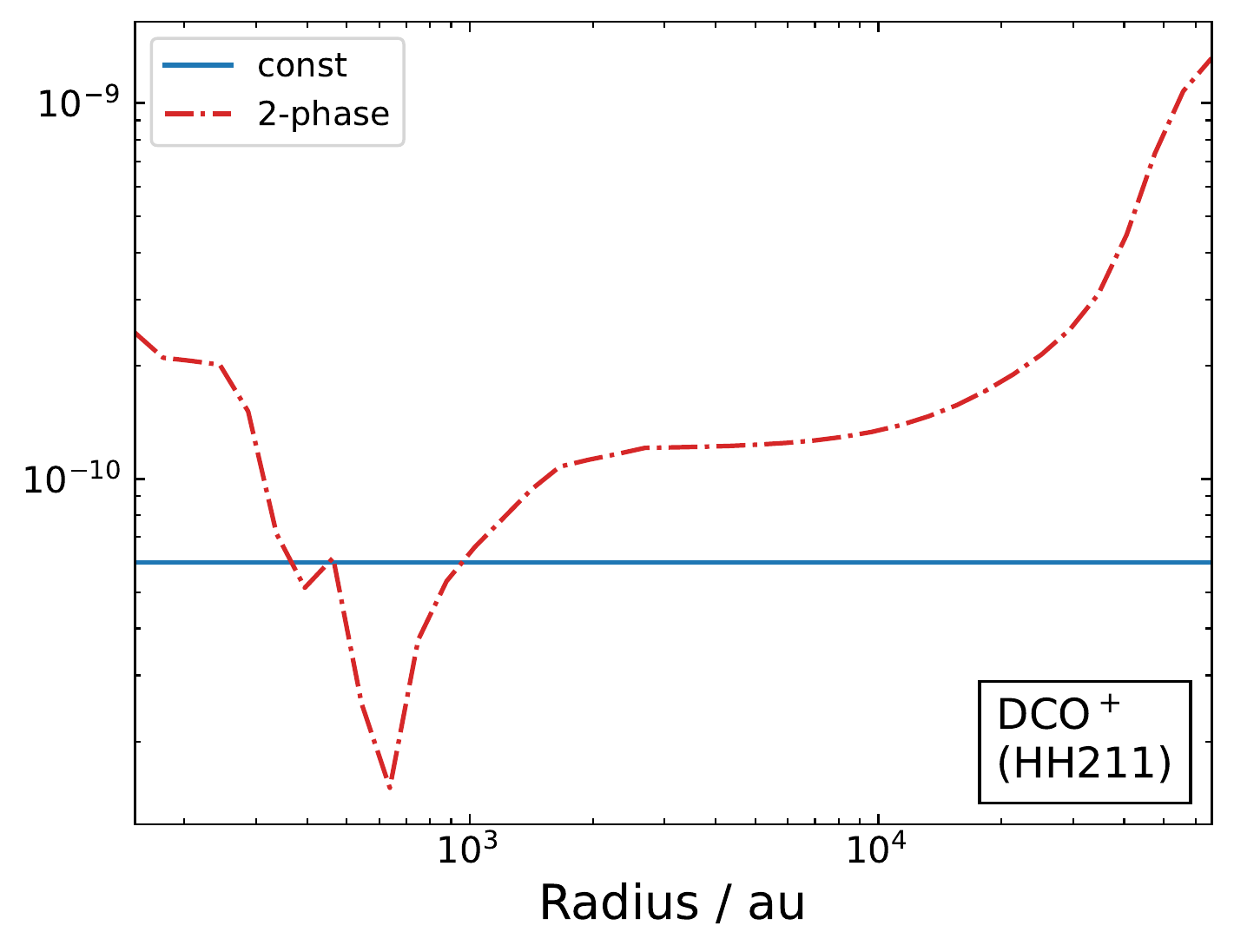}
   \caption{Fractional molecular abundance (with respect to H$_2$) profiles of the best-fit results produced with LOC for the spectra observed towards HH211. In the case of the $^{13}$C and $^{18}$O isotopologues, the abundance profiles correspond to the profiles of the main species, scaled down by the isotopic ratio (68 and 557, respectively).}
              \label{Fig:LOC-abuprofilesHH211}
\end{figure*}

Fig.~\ref{Fig:LOC-abuprofilesL1544} and \ref{Fig:LOC-abuprofilesHH211} show the corresponding molecular abundance profiles of the best-fit modelling results derived with LOC and presented in this work, for L1544 and HH211, respectively.


\end{appendix}

\end{document}